\documentclass[aps,twocolumn,amsmath,amssymb,floatfix,reprint,footinbib,superscriptaddress,longbibliography,showkeys]{revtex4-1}
\usepackage{graphicx}
\usepackage{mathrsfs}
\usepackage{mdwlist}
\usepackage{subfigure}
\usepackage{booktabs}
\usepackage{amsmath}
\usepackage{extarrows}
\usepackage{dsfont}
\usepackage{amstext}
\usepackage{amsbsy}
\usepackage{bbm}
\usepackage{bm}
\usepackage{amssymb,mathrsfs,amsmath}
\usepackage{threeparttable}
\usepackage{booktabs}
\usepackage{bbding,pifont}
\usepackage{amsthm}
\usepackage{graphicx}
\usepackage{textcomp}
\usepackage{multirow}%
\usepackage{pifont}
\usepackage{color}
\usepackage{sansmath}
\usepackage{makecell}
\usepackage{tabularx}
\usepackage[colorlinks,citecolor=blue]{hyperref}
\definecolor{Dgreen}{RGB}{0, 100, 0}
\usepackage{url}
\usepackage[colorlinks]{hyperref}
\hypersetup{%
	plainpages=true,
	breaklinks=true,  
	hypertexnames=false,  
	pageanchor=true,
	colorlinks=true,
	linkcolor={blue},
	citecolor={blue},
	urlcolor={blue},
	anchorcolor={black}
}

\begin{document}

\title{Supervised learning for robust quantum control in composite-pulse systems}

\author{Zhi-Cheng Shi}
\affiliation{Fujian Key Laboratory of Quantum Information and Quantum Optics, Fuzhou University, Fuzhou
350108, China}
\affiliation{Department of Physics, Fuzhou University, Fuzhou 350108, China}

\author{Jun-Tong Ding}
\affiliation{Fujian Key Laboratory of Quantum Information and Quantum Optics, Fuzhou University, Fuzhou
350108, China}
\affiliation{Department of Physics, Fuzhou University, Fuzhou 350108, China}

\author{Ye-Hong Chen}\thanks{cyhong\underline{ }jmu@163.com}
\affiliation{Fujian Key Laboratory of Quantum Information and Quantum Optics, Fuzhou University, Fuzhou
350108, China}
\affiliation{Department of Physics, Fuzhou University, Fuzhou 350108, China}
\affiliation{Theoretical Quantum Physics Laboratory, Cluster for Pioneering Research, RIKEN, Wakoshi, Saitama 351-0198, Japan}
\affiliation{\mbox{Quantum Information Physics Theory Research Team, Center for Quantum Computing, RIKEN, Wakoshi,} Saitama 351-0198, Japan}

\author{Jie Song}
\affiliation{Department of Physics, Harbin Institute of Technology, Harbin 150001, China}

\author{Yan Xia}\thanks{xia-208@163.com}
\affiliation{Fujian Key Laboratory of Quantum Information and Quantum Optics, Fuzhou University, Fuzhou
350108, China}
\affiliation{Department of Physics, Fuzhou University, Fuzhou 350108, China}

\author{X. X. Yi} \thanks{yixx@nenu.edu.cn}
\affiliation{\mbox{Center for Quantum Sciences and School of Physics, Northeast Normal University, Changchun 130024, China}}

\author{Franco Nori}
\affiliation{Theoretical Quantum Physics Laboratory, Cluster for Pioneering Research, RIKEN, Wakoshi, Saitama 351-0198, Japan}
\affiliation{\mbox{Quantum Information Physics Theory Research Team, Center for Quantum Computing, RIKEN, Wakoshi,} Saitama 351-0198, Japan}
\affiliation{Department of Physics, University of Michigan, Ann Arbor, Michigan 48109-1040, USA}

\begin{abstract}
In this work,
we develop a supervised learning model for implementing robust quantum control in composite-pulse systems, where the training parameters can be either phases, detunings, or Rabi frequencies.
This model exhibits great resistance to all kinds of systematic errors, including single, multiple, and time-varying errors.
We propose a modified gradient descent algorithm for adapting the training of phase parameters, and show that different sampling methods result in different robust performances.
In particular, there is a trade-off between high fidelity and robustness for a given number of training parameters, and both can be simultaneously enhanced by increasing the number of training parameters (pulses).
For its applications, we demonstrate that the current model can be used for achieving high-fidelity arbitrary superposition states and universal quantum gates in a robust manner.
This work provides a highly efficient learning model for fault-tolerant quantum computation by training various physical parameters.

\end{abstract}

\maketitle

\section{Introduction}

Quantum control with high precision is an essential prerequisite for the implementation of reliable quantum computation \cite{Nielsen00}.
Here, high precision has two meanings: one for perfect control, with knowledge of given physical parameters, and the other for being as free as possible from the effects of various systematic errors.
The former is readily attainable, for instance, by adopting the familiar resonant pulse technique \cite{osti7365050}. The latter is truly difficult and requires a great deal of attention.
One of the main reasons is that we cannot really capture the complicated nature of systematic errors, such as the inevitable perturbations in parameters and decoherence noises.
Thus far, numerous approaches have been put forward to address them, including: adiabatic passage \cite{physchem.52.1.763,PhysRevLett.95.087001,RevModPhys.79.53,RevModPhys.89.015006}, dynamical decoupling \cite{PhysRevLett.82.2417,PhysRevLett.95.180501,Souza2012}, quantum feedback \cite{PhysRevA.82.062103,Wiseman2011,PhysRevA.92.033812,Zhang2017pr}, single-shot-shaped pulses \cite{PhysRevLett.111.050404,PhysRevA.96.022309}, derivative removal by an adiabatic gate \cite{PhysRevLett.103.110501,PhysRevA.88.052330,PhysRevA.94.032306}, sampling-based algorithms \cite{PhysRevA.89.023402,Dong2015ieee,Dong2020ieee}, geometric optimization \cite{PhysRevA.95.063403,PhysRevLett.125.250403,PhysRevApplied.16.064040,PhysRevApplied.18.034072,PhysRevA.105.042404}, and robust optimal control \cite{PhysRevA.49.2241,PhysRevA.104.042226,PhysRevA.106.052608,PhysRevApplied.17.014036,PhysRevResearch.5.033052}.

The emergence of machine learning \cite{mohri2012} offers another powerful way to tackle the issue of robustness against various errors.
Essentially, machine learning is a process of using data to learn the rules applicable to it, and then utilizing the learned rules to make predictions on new data.
Typical algorithms involved in machine learning are decision trees, neural networks, support vector machines, random forests, and collaborative filtering, to name a few \cite{Bonaccorso2017}.
Broadly speaking, machine learning can be divided into three main categories: unsupervised learning, supervised learning, and reinforcement learning.
Over the last few years, machine learning, especially supervised learning, has achieved great success in many fields of physics \cite{Biamonte2017,Dunjko2018,RevModPhys.91.045002,Arrachea2023,PhysRevA.107.010101}, such as the study of nonequilibrium quantum dynamics problems \cite{PhysRevA.99.042316,Zeng2020,PhysRevB.102.134213,PhysRevA.103.L040401,Ai2022, PhysRevLett.127.140502,PhysRevApplied.17.024040,PhysRevApplied.14.064020,PRXQuantum.2.010316}, the classification of quantum states \cite{Ma2018npj,PhysRevX.8.011006,PhysRevLett.120.240501,Harney2020,PhysRevResearch.3.033278,PhysRevA.107.032421}, and the design of control fields for high-fidelity quantum gates \cite{PhysRevApplied.6.054005,PhysRevA.97.042324,PhysRevA.100.012326,Spiteri2018, Arrazola2019, PhysRevApplied.9.064042,PRXQuantum.2.040324,PhysRevLett.131.050601,PRXQuantum.4.030305}.

To date, a variety of methods, e.g., the Krotov \cite{krotov1995global}, gradient ascent pulse engineering (GRAPE) \cite{Khaneja2005}, evolutionary \cite{PhysRevLett.110.220501,PhysRevLett.114.200502,PhysRevA.105.052414}, and variational quantum algorithms \cite{PRXQuantum.2.010101}, have been raised to update variables in optimization problems.
Among them, the high-efficiency GRAPE algorithm \cite{Khaneja2005} and its variants  \cite{PhysRevA.79.053417,Johansson2013,PhysRevA.99.042327,PhysRevA.101.052317,PhysRevApplied.16.014056,Li2022quantum} are successful in designing the control fields to sharply suppress the influence of various errors, e.g., amplitude uncertainty and frequency drift.
Recently, there has been an increasing interest in the combination of GRAPE with other methods, such as the chopped random basis \cite{PhysRevLett.106.190501,PhysRevA.84.022326,PhysRevA.92.062343}, generally referred to as gradient optimization using the parametrization algorithm \cite{PhysRevA.98.022119}.
In particular, de Fouquieres \emph{et al.} \cite{deFouquieres2011} showed that the gradient accuracy and the convergence rate of the GRAPE algorithm can be effectively improved by introducing the Broyden-Fletcher-Goldfarb-Shanno quasi-Newton algorithm~\cite{Nocedal2006}.

In this work, we provide a systematic methodology for robust quantum control through constructing a supervised learning model in composite-pulse systems.
This model is quite universal and very robust to all kinds of systematic errors, including single error, multiple types of errors, time-varying errors, and so forth.
Specifically, we establish the cost function of the supervised learning model, and then propose a modified version of the  GRAPE algorithm to update the phases. The learned parameters (e.g., phases or detunings) possessing robustness against all kinds of systematic errors are subsequently determined via training the sample set.

In the current model, one of the most critical steps is to ascertain the \emph{sampling} method, reflecting in the sampling distributions. We demonstrate that it is necessary to select a suitable sampling distribution to receive robust and high-fidelity quantum control.
Furthermore, the generalization ability of this model is significantly enhanced by increasing the number of pulses.
We finally extend this model to train any physical parameters to implement robust quantum control.
Our method paves an efficient way toward the establishment of reliable quantum gates for fault-tolerant quantum computation.

The rest of this paper is organized as follows. In Sec.~\ref{ii}, we introduce the physical system and the possible difficulties in composite pulses. In Sec.~\ref{iiia}, we first show how to design the cost function for the supervised learning model.
Next, we propose a modified gradient descent algorithm to train the phase parameters, and then investigate in detail the sampling method and the generalization ability of this model.
In Sec.~\ref{iii}, we provide some applications for this model, including the implementation of arbitrary superposition states and general single-qubit gates, which are robust against all kinds of systematic errors.
In Sec.~\ref{iv}, we show that this model has the ability to train any physical parameters to achieve  robust quantum control in different quantum systems. Conclusions are given in Sec.~\ref{v}.

\section{Physical Model} \label{ii}

Consider a qubit coherently driven by a control field; the general form of the Hamiltonian in the interaction picture reads ($\hbar=1$)
\begin{eqnarray}
H(\theta)=\Delta\sigma_z+ \Omega e^{-i\theta}\sigma_{+}+ \Omega e^{i\theta}\sigma_{-},
\end{eqnarray}
where $\Delta$ represents the detuning between the transition frequency of the qubit and the carrier frequency of the control field, $\Omega$ is the Rabi frequency, $\theta$ denotes the phase, and $\sigma_{\pm}=\frac{1}{2}(\sigma_x\pm i\sigma_y)$, with $\sigma_\nu$ ($\nu=x,y,z$) being Pauli operators.

To achieve general single-qubit gates, the most intuitive and fastest method is to exploit a resonant pulse, i.e., $\Delta=0$. Then, the evolution operator of this system becomes (up to a global phase)
\begin{eqnarray}\label{prpagatorN}
		\mathcal{U}(\theta)=e^{-iH(\theta)t}=\left[
		\begin{array}{cc}
			\cos A & -i e^{i \theta}\sin{A}\\[1.5ex]
			- i e^{-i \theta}\sin{A} & \cos A
		\end{array}
		\right],
\end{eqnarray}
with pulse area $A=\Omega t$.
In the ideal case, one can effortlessly obtain a multitude of perfect single-qubit gates by simply choosing different pulse areas and phases.
Considering the imperfect knowledge of quantum systems, there are two typical types of systematic errors: pulse area errors and off-resonance effects.

Pulse area errors usually originate from incorrect duration and deviation in the Rabi frequency. The former is mainly caused by the inability of selecting a precise operation time. For instance, when an atomic beam passes through a laser field, the interaction times between the atoms and the laser field may be different because of the distinct longitudinal velocities of atoms \cite{PhysRevA.103.033110}.
The dominant sources of the latter are due to the spatial inhomogeneity of control fields or the intensity of control fields deviating from its nominal value. For instance, this error results from fluctuations in the Overhauser field in quantum dot systems \cite{Wang2012,PhysRevA.89.022310}.
In a doped crystal, the typical variation of the amplitude of the drive is about 20\%--30\% \cite{PhysRevA.103.033110}.
Hereafter, we use a dimensionless parameter $\epsilon_A$ to characterize the pulse area error, i.e.,
\begin{eqnarray}\label{epa}
A\rightarrow A(1+\epsilon_A).
\end{eqnarray}

Off-resonance effects, which refer to the mismatch between the control field frequency and the associated transition frequency, arise due the deviation from either of these frequencies.
In practice, the transition frequency can be varied, for instance, due to an energy level shift in atomic systems \cite{scully97} or inhomogeneous broadening of the hyperfine levels in an ion-doped yttrium orthosilicate crystal \cite{PhysRevLett.118.133202,PhysRevA.101.013827}.
Moreover, the control field frequency may also be altered on account of fluctuations in the external environment.
In some integrated photonic circuits, off-resonance effects occur if the waveguides have distinct geometries \cite{PhysRevA.100.032333,PhysRevA.108.042401}.
For the sake of description, we introduce the detuning error $\epsilon_\Delta$ to represent off-resonance effects, i.e.,
\begin{eqnarray}\label{delta}
\Delta\rightarrow \Delta+\epsilon_\Delta,
\end{eqnarray}
where $\Delta=0$ means that the system is in the resonant regime.

It is worth noting that the major disadvantage of the resonant pulse is its high susceptibility to pulse area errors and off-resonance effects. To significantly enhance the robustness against both types of systematic errors, one can turn to the composite pulses \cite{Levitt1979,Levitt1986,Wimperis1994}, a train of pulses with identical amplitudes and relative phases to be addressed.
The total evolution operator for the $N$ pulse can be formulated as
\begin{eqnarray}
\mathcal{U}_{\mathrm{tot}}=\mathcal{U}(\theta_N)\mathcal{U}(\theta_{N-1})\cdots \mathcal{U}(\theta_2)\mathcal{U}(\theta_1),
\end{eqnarray}
where $\mathcal{U}(\theta_n)=\exp[-iH(\theta_n)T]$ is the evolution operator of the $n$th pulse with pulse duration $T$ and Hamiltonian $H(\theta_n)$, $n=1, \dots, N$.

There are numerous methods to determine the value of the phase $\theta_n$ ($n=1,\dots,N$). Among them, the most commonly used is to perform a Taylor expansion in the vicinity of the error $\epsilon=0$ \cite{PhysRevA.67.042308,PhysRevA.70.052318,PhysRevA.84.062311,PhysRevA.87.052317,PhysRevA.93.032340, PhysRevA.101.012321,PhysRevResearch.2.043194, Bulmer2020}, which has been experimentally demonstrated in various physical platforms \cite{Rong2015nc,PhysRevA.92.060301,PhysRevLett.129.240505,PhysRevApplied.18.034062}.
This method has its own limitation.
Whereas the analytical solutions for the phases have been derived in several simple cases (e.g., a single type of error) \cite{PhysRevA.83.053420,PhysRevLett.113.043001,PhysRevA.89.022341,PhysRevA.97.043408}, more often,
the expressions of the Taylor expansion coefficients become particularly cumbersome for multiple types of systematic errors.
As a result, even if the complicated expressions are given, only numerical solutions are accessible for the phases \cite{PhysRevA.93.023830,PhysRevA.103.052612,PhysRevA.104.012609,PhysRevResearch.2.043235}.
Worse still, sometimes the solutions may not exist, and thus one requires the aid of cost functions \cite{Shi2022,PhysRevA.107.023103}.

In this work, we ascertain the phase $\theta_n$ ($n=1,\dots,N$) by training the samples using the supervised learning model, which can effectively settle the aforementioned issues.
Not only that, this model can also solve problems that traditional composite pulses are incapable of handling, e.g., for time-varying errors \cite{PhysRevA.94.022303,PhysRevA.95.062325,PhysRevLett.110.140502,PhysRevB.90.155306, PhysRevA.106.032611,PhysRevApplied.13.024022,PhysRevApplied.19.014014,PhysRevA.107.032609}.
For briefness, from now on, we adopt the vector $\bm{\theta}$ to represent all phase parameters, i.e., $\bm{\theta}=(\theta_1,\dots,\theta_N)$.

\section{Supervised Learning Model} \label{iiia}

\subsection{Cost function}

We begin by constructing the cost function for the supervised learning model.
Given a set of training data $\{\bm{x}_k,y_k\}$, the goal of supervised learning is to learn a mapping from $\bm{x}$ to $y$, where $y_k$ is called the label or target of sample $\bm{x}_k$ (usually a vector) in the data set.
Here, the phases $\bm{\theta}$ are the training parameters to be learned to make quantum control (operations) as immune to errors as possible.

In the absence of errors, one can also acquire perfect quantum operations by composite pulses. Nevertheless, the fidelity of quantum operations declines to varying degrees when exhibiting different types of errors.
Under such circumstances, the fidelity of quantum operations can be regarded as a function of errors.
Therefore, the samples in the training set come from various types of systematic errors, i.e.,
\begin{eqnarray}
\bm{x}_k=\bm{\epsilon}_k=(\epsilon_k^1, \dots, \epsilon_k^\ell, \dots, \epsilon_k^L),
\end{eqnarray}
where $L$ represents the highest dimension of samples, the subscript $k$ denotes the $k$th sample, and the $\ell$th component $\epsilon^\ell_k$ $(\ell=1,\dots,L)$ characterizes a certain type (time interval) of systematic error.

More specifically, if systematic errors are unknown constants, different components of the sample represent different error types. For example, two components of the sample $\bm{x}_k=\{\epsilon^A_{k},\epsilon^\Delta_{k},\dots\}$ are the pulse area error $\epsilon^A_{k}$ and detuning error $\epsilon^\Delta_{k}$.
If the systematic error belongs to one type of time-varying noise, then different components of the sample represent the error in different time intervals, e.g., $\bm{x}_k=\{\epsilon^{A,t_1}_k,\dots,\epsilon^{A,t_\ell}_k,\dots\}$, where $\epsilon^{A,t_\ell}_{k}$ denotes the $k$th sample of the pulse area error in the time interval $[t_\ell,t_{\ell+1}]$.
Of course, one can utilize a mixture of different types and time intervals of errors to generate the samples, e.g., $\bm{x}_k=\{\epsilon^{A,t_1}_k,\dots,\epsilon^{A,t_\ell}_k,\dots, \epsilon^{\Delta,t_1}_k,\dots,\epsilon^{\Delta,t_\ell}_k,\dots\}$.
In other words, the samples can be simultaneously drawn from both time-independent and time-dependent systematic errors to form a training set in this model.

The label we adopt here is the expected fidelity $\hat{F}(\bm{x}_k)$ of quantum operations,
\begin{eqnarray}
y_k=\hat{F}(\bm{x}_k).
\end{eqnarray}
In the ideal situation, we naturally expect the system to be completely resistant to all errors. That is, $\hat{F}(\bm{x}_k)=1$ for all samples $\bm{x}_k$, reflecting the ideal fact that quantum operations are still fully accurate in the presence of a variety of systematic errors.

For each sample $\bm{x}_k$, we define the loss function
\begin{eqnarray}
\mathcal{L}_k=y_k-{F}(\bm{x}_k),
\end{eqnarray}
where ${F}(\bm{x}_k)$ is the actual fidelity of quantum operations when the system exhibits error $\bm{\epsilon}_k$ (equivalent to $\bm{x}_k$). Note that the expressions for ${F}(\bm{x}_k)$ rely on different quantum operations.
For quantum state preparations, ${F}(\bm{x}_k)$ can be given by
\begin{eqnarray} \label{state7}
{F}(\bm{x}_k)=|\langle \Psi_\mathrm{T}|\Psi\rangle|^2,
\end{eqnarray}
where $|\Psi_\mathrm{T}\rangle$ is the target state and $|\Psi\rangle$ is the final state of the system after performing composite pulses.
With regard to quantum gate operations, ${F}(\bm{x}_k)$ can be written as
\begin{eqnarray} \label{gate8}
{F}(\bm{x}_k)=\frac{1}{M}\mathrm{tr}(\mathrm{\mathbf{U}}^{\dag}_{\mathrm{T}}\mathcal{U}_{\mathrm{tot}}),
\end{eqnarray}
where $\mathrm{\mathbf{U}}_\mathrm{T}$ is the target quantum gate with $M$ being its dimension, and $\mathcal{U}_{\mathrm{tot}}$ denotes the total evolution operator for composite pulses.
Then, the cost function of the supervised learning model is defined by
\begin{eqnarray} \label{cost9}
J=\frac{1}{K}\sum_{k=1}^K\mathcal{L}_k=\frac{1}{K}\sum_{k=1}^K\left[y_k- {F}(\bm{x}_k)\right],
\end{eqnarray}
which represents the average of all loss functions, where $K$ is the sample size.
The phase parameters $\bm{\theta}$ are trained in the direction that the cost function~(\ref{cost9}) decreases.

\subsection{Robustness infidelity measure}

In general, one training session on the samples does not necessarily lead to the solution we desire.
Thus, performance evaluation for the solution (i.e., the phase parameters $\bm{\theta}$) is required to determine whether or not to stop the training procedure. To this end, we next introduce the definitions of fidelity and robustness.

Let us start with the case of one-dimensional error. Given a set of the phase parameters $\bm{\theta}$, the definition of the average fidelity $\bar{F}$ is expressed by
\begin{eqnarray} \label{gf14}
\bar{F}=\int \! \rho(\epsilon) \; F_{\bm{\theta}}(\epsilon)\;d\epsilon,
\end{eqnarray}
where $F_{\bm{\theta}}(\epsilon)$ represents the fidelity of quantum operations obtained by using the phase parameters $\bm{\theta}$ when the error is $\epsilon$, and $\rho(\epsilon)$ is the probability density distribution of the error.
In reality, it is difficult to exactly predetermine the specific distribution of $\rho(\epsilon)$. The usual way to tackle it is to use an \emph{a priori} density distribution such as the uniform distribution or the Gaussian distribution.
Here, $\rho(\epsilon)$ is simply taken to be a uniform distribution, i.e., all errors in the interval $[\epsilon_-,\epsilon_+]$ are considered equally weighted. Therefore, the form of $\rho(\epsilon)$ is
\begin{eqnarray}\label{27a}
&\rho(\epsilon)= \left \{
\begin{array}{ll}
    \displaystyle\frac{1}{\epsilon_+-\epsilon_-},           & ~~~   \epsilon\in[\epsilon_-,\epsilon_+], \\[2.8ex]
    0,     & ~~~ \mathrm{otherwise}.~~ \\
\end{array}
\right.
\end{eqnarray}
As a result, Eq.~(\ref{gf14}) becomes
\begin{eqnarray} \label{f14}
\bar{F}(\epsilon_-,\epsilon_+)=\frac{1}{\epsilon_+-\epsilon_-}\int_{\epsilon_-}^{\epsilon_+}F_{\bm{\theta}}(\epsilon)\;d\epsilon.
\end{eqnarray}
Its discrete version reads
\begin{eqnarray}
\bar{F}(\epsilon_-,\epsilon_+)=\frac{1}{\epsilon_+-\epsilon_-}\sum_{j}F_{\bm{\theta}}(\epsilon_j)\;\Delta\epsilon_j.
\end{eqnarray}

Accordingly, we can define the average infidelity by
\begin{eqnarray} \label{g16}
I_f(\epsilon_-,\epsilon_+)=\frac{1}{\epsilon_+-\epsilon_-}\int_{\epsilon_-}^{\epsilon_+}\!\!\big[1\!-\!F_{\bm{\theta}}(\epsilon)\big] \;d\epsilon =1\!-\!\bar{F}(\epsilon_-,\epsilon_+). \cr
\end{eqnarray}
In fact, Eq.~(\ref{g16}) is also the definition of the first-order \emph{robust-infidelity measure} (RIM) \cite{PhysRevA.107.032606} following a uniform distribution.
A small RIM indicates that the learned parameters make the system dynamics very robust and have high-fidelity quantum operations, while a large RIM signifies poor robustness and fidelity.

Although RIM can be used to quantify both robustness and fidelity for a set of learned parameters,
it is difficult to intuitively benchmark how robust the learned parameters are from Eq.~(\ref{g16}).
To describe it more precisely, we define the robust width $W(\xi)$ as 
\begin{eqnarray} \label{w17}
W(\xi)=\epsilon_{\mathrm{max}}-\epsilon_{\mathrm{min}},
\end{eqnarray}
where all errors in the interval $[\epsilon_{\mathrm{min}},\epsilon_{\mathrm{max}}]$ satisfy
\begin{eqnarray}
\max\big\{1-F_{\bm{\theta}}(\epsilon)\big\}\leq\xi, ~~~\epsilon\in[\epsilon_{\mathrm{min}},\epsilon_{\mathrm{max}}],
\end{eqnarray}
with a given threshold $\xi$. It is clear that a large (small) $W(\xi)$ means high (poor) robustness. Remarkably, these definitions can be easily generalized to the case of multidimensional errors.

\subsection{Training and testing}

Given the cost function (\ref{cost9}), we train the phase parameters to yield its minimum value, and then test the robust performance for the learned phase parameters according to Eq.~(\ref{f14}).
The learning procedure is divided into two steps: \emph{training} and \emph{testing}.

In the \emph{training stage}, we first specify a training set by drawing $K$ samples from errors through several common probability distributions, such as the Gaussian distribution, the uniform distribution, the beta distribution, and the exponential distribution.
Once the training set of $K$ samples is obtained, we evaluate the cost function $J$ given by Eq.~(\ref{cost9}) via a random set of initial phase parameters, and then update the phase parameters using the modified gradient descent algorithm, which is introduced in the next section.
The cost function $J$ gradually diminishes with the passage of iterations, and the training does not end until seeking out the optimal value or reaching a certain threshold of this cost function.

In the \emph{testing stage}, we apply the phase parameters learned from the training stage to assess the robust performance for this composite pulse sequence using the average fidelity.
Specifically, we first calculate the fidelity $F_{\bm{\theta}}(\epsilon)$ for all possible errors rather than resampling them, and then the average fidelity $\bar{F}(\epsilon_-,\epsilon_+)$ in the interval $[\epsilon_-,\epsilon_+]$ according to Eq.~(\ref{f14}).
When the average fidelity reaches the expected value, we accept the learned phase parameters and terminate the whole learning procedure. Otherwise, the learning procedure is a failure and we must retrain the phase parameters to obtain another solution, i.e., returning to the training stage with a new set of samples or initial phase parameters.

\begin{figure}[t]\centering
\includegraphics[width=0.47\textwidth]{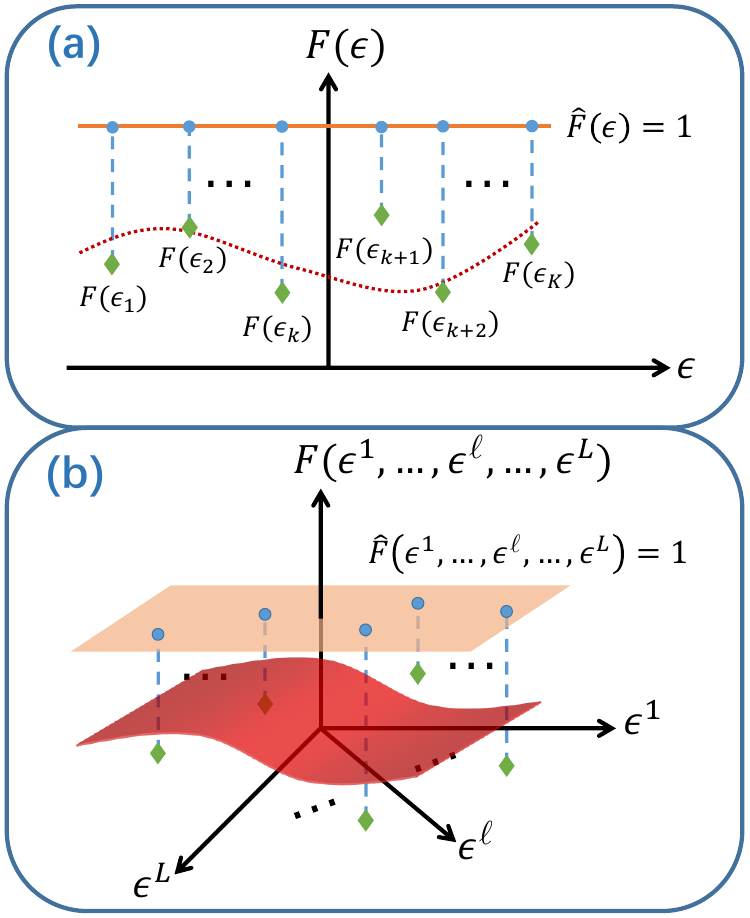}
\caption{Schematic diagram of the data space. (a) Two-dimensional space composed of error $\epsilon$ and fidelity $F(\epsilon)$, where the filled circles and rhomboids represent the expected and actual sample fidelities, respectively. According to the labeled training set $\{\epsilon_k,\hat{F}(\epsilon_k)\}$, the phase parameters $\bm{\theta}$ are trained so that the red dotted curve fit by the data points is as close as possible to the orange solid line, where $k=1,\dots, K$ with $K$ the sample size. (b) The $(L+1)$-dimensional space, where the $L$-dimensional hyperplane $\hat{F}(\epsilon^1, \dots, \epsilon^\ell, \dots, \epsilon^L)=1$ represents an ideal situation in which the fidelity of quantum operations is completely unaffected by all types (time intervals) of errors. We have to train the phase parameters $\bm{\theta}$ so that the red hypersurface fit by the filled rhomboids approaches as much as possible the orange hyperplane $\hat{F}(\epsilon^1, \dots, \epsilon^\ell, \dots, \epsilon^L)=1$. }
	\label{super}
\end{figure}

In Fig.~\ref{super}, the blue dots represent the samples in the training set. Note that all blue dots have a value of 1, which implies that the quantum operation we desire is perfect however large the systematic errors.
The green rhomboids are the actual fidelities of the samples, and their positions vary depending on the values of phase parameters.
The red dotted curve (hypersurface) is plotted according to the phase parameters learned from the training stage, and each point on the curve (hypersurface) corresponds to the actual fidelity of the quantum operation at specific values of the systematic errors.

From a physical point of view, the closer the curve (hypersurface) is to the line $\hat{F}(\epsilon)=1$ [the hyperplane $\hat{F}(\epsilon^1, \dots, \epsilon^\ell, \dots, \epsilon^L)=1$], the less susceptible the quantum operation is to errors.
Obviously, the shapes of the curves (hypersurfaces) are quite different for distinct phase parameters, and the objective is to attain an optimal solution of the phase parameters $\bm{\theta}$ by training these rhomboids in Fig.~\ref{super}.

\subsection{Modified gradient descent algorithm}

In the original GRAPE algorithm \cite{Khaneja2005}, the control variables are \emph{amplitudes} of control fields, but they become \emph{phases} here. As a result, the original GRAPE algorithm cannot be directly applied in the current model, and a modified version is urgently required to accommodate this situation. We next explain the details of this modified GRAPE (mGRAPE) algorithm.

\begin{figure}[t]\centering
\includegraphics[width=0.44\textwidth]{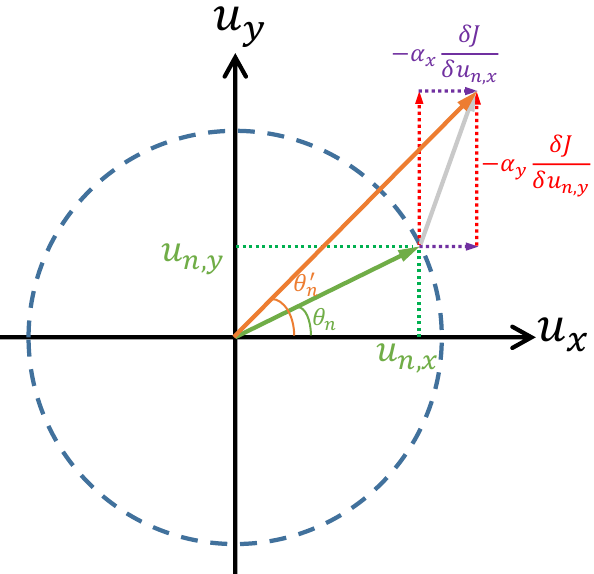}
\caption{Geometric interpretation of the modified GRAPE algorithm. First, the phase $\theta_n$ ($n=1, \dots, N$) is mapped onto the unit circle to yield a unit vector (the green line).
By orthogonal decomposition of this vector, we obtain two variables $u_{n,x}$ and $u_{n,y}$. Then, we use the gradient descent algorithm to calculate the increment in both directions, and thus acquire the increment vector (the gray line). According to the parallelogram law of vectors, we obtain a new vector (the orange line) whose direction is actually the gradient descent direction. As a result, we have the new phase $\theta_n'$ in the next iteration. The iteration process continues until the gradient threshold is reached. For detailed steps, see Algorithm 1.}
	\label{ita}
\end{figure}

At first, we need to rewrite the Hamiltonian of the $n$th pulse as ($n=1,\dots,N$)
\begin{eqnarray} \label{h10}
H(\theta_n)&=&e^{-i\theta_n}\sigma_{+}+e^{i\theta_n}\sigma_{-} \nonumber\\[0.5ex]
           &=&\cos\theta_n\sigma_x+\sin\theta_n\sigma_y \nonumber\\[0.5ex]
           &=&u_{n,x}H_x+u_{n,y}H_y,
\end{eqnarray}
where $H_{x(y)}=\sigma_{x(y)}$, and $\Omega$ is regarded as a unit.
In Eq.~(\ref{h10}), two virtual variables $u_{n,x}$ and $u_{n,y}$ are exploited to substitute for the phase parameter $\theta_n$, and they satisfy the relation
\begin{eqnarray} \label{12theta}
\theta_n=\arctan\frac{u_{n,y}}{u_{n,x}}.
\end{eqnarray}
After this substitution, we can use the gradient descent algorithm to update the virtual variables $u_{n,x}$ and $u_{n,y}$, i.e.,
\begin{eqnarray}
u'_{n,x}&=&u_{n,x}-\alpha_x\frac{\delta J}{\delta u_{n,x}},    \\[0.5ex]
u'_{n,y}&=&u_{n,y}-\alpha_y\frac{\delta J}{\delta u_{n,y}},
\end{eqnarray}
where both $\alpha_x$ and $\alpha_y$ are some prescribed step sizes.
Then, the new phase parameter in the next iteration reads
\begin{eqnarray}  \label{22theta}
\theta'_n=\arctan\frac{u'_{n,y}}{u'_{n,x}}.
\end{eqnarray}
It is worth stressing that $u_{n,x}$ and $u_{n,y}$ are only introduced as a mathematical tool, and they do not have to possess physical meaning. Figure~\ref{ita} illustrates the geometric interpretation of the gradient descent principle.
Here, the updated formulas for the phases given by Eqs.~(\ref{12theta})--(\ref{22theta}) are quite different from those proposed in Ref.~\cite{bhole2020a}. Specifically, we evaluate the gradient of the cost function with the aid of two types of virtual variables, while the gradient with respect to the phase is directly calculated after performing the rotating-wave approximation in a rotating frame in Ref.~\cite{bhole2020a}.

\renewcommand\arraystretch{1.2}
\begin{table}[htbp] 
 \begin{tabular}{l}
  \hline
  \multirow{2}{*}{\textbf{ALGORITHM 1. Modified GRAPE algorithm}}\\\\
  \hline
1. Randomly input the initial phase parameters $\theta_n$. \cr
2. Calculate the virtual variables $u_{n,x}=\cos\theta_n$ and  \cr
~~~~$u_{n,y}=\sin\theta_n$. \cr
3. Evaluate the increments and update the virtual variables, \cr
~~~~labeled as $u'_{n,x}$ and $u'_{n,y}$. \cr
4. Calculate the parameters $\theta'_n=\arctan u'_{n,y}/u'_{n,x}$. \cr
5. Regard the $\theta'_n$ as new phase parameters. \cr
6. Go to step 2 until the threshold is reached. \cr
  \hline
 \end{tabular}
\end{table}

It is important to note that the modified GRAPE algorithm is very crude, not because we cannot find a solution of the training parameters $\bm{\theta}$, but because there are very many locally optimal solutions.
To search global optimal solutions of the training parameters (probably more than one), two common methods of ``escaping'' or ``restarting'' are usually adopted, namely, randomly searching the neighborhood of the current solution in parameter space, or ignoring the current solution, and exploring a new one in parameter space.
Here, we adopt the latter method, which can take full advantage of parallel training to improve computational efficiency. Specifically, we increase the number of training groups large enough, and each group is trained individually.
After accomplishing the training, we measure the performance of each group using the average fidelity defined by Eq.~(\ref{f14}).
Then, the phase parameters that make the average fidelity reach the desired value are left, and the rest are simply discarded.
As for the former, we can introduce the simulated annealing method into the modified GRAPE algorithm, similar to that used in Refs.~\cite{PhysRevA.105.042437,Mahesh2022}. For details, we refer the reader to Appendix~\ref{appendixa}.

\subsection{Sampling methods} \label{methods}

In the current model, one of the most important parts is to acquire enough representative samples.
If the sample size is too small, it is easy to observe underfitting, where the parameter law of the model may not be well learned.
How to efficiently draw enough samples becomes very critical in supervised learning.
In particular,
the usage of different training samples will produce different solutions of training parameters, and such discrepancies can significantly impact the quality of robust quantum control.

\begin{figure}[t]
\includegraphics[width=0.48\textwidth]{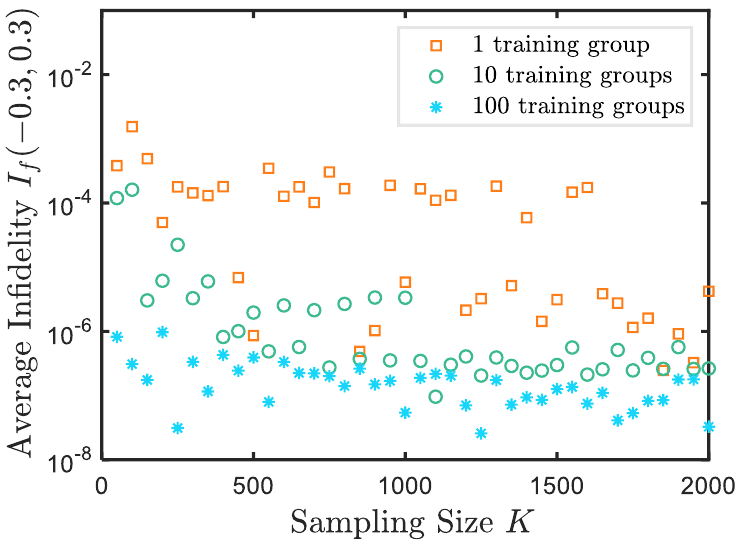}
\caption{ Average infidelity $I_f(-0.3,0.3)$ versus the sampling size $K$. The objective is to robustly attain population inversion in the presence of pulse area error, where we uniformly draw the samples from the interval $[0, 0.3]$, set the pulse area of each pulse to $\pi/2$, choose the pulse number $N=7$, adopt a step size of 0.001, and randomly select initial values of the training parameters $\bm{\theta}$ from the range of $[-\pi,\pi]$.  } \label{dvsF}
\end{figure}

Figure~\ref{dvsF} shows the relationship between the average infidelity $I_f(-0.3,0.3)$ and the total number of samples, where the objective is to achieve population inversion in a robust manner, and the samples come from the pulse area error given by Eq.~(\ref{epa}). For cases in which the sample size is particularly large, we split them into multiple groups for independent training, and then pick out the optimal solutions from these groups. The purpose of this is that we can train the parameters in parallel, thus improving the computational efficiency. It is readily found that the average infidelity is relatively large and oscillates when the sample size $K$ is not large enough. With increasing the sampling size, the average infidelity gradually decreases and stabilizes at a relatively low level.
This implies that having sufficient samples is imperative for successfully training the parameters. Furthermore, increasing the number of training groups also facilitates parameter training, making it easier to find the optimal solutions.

\begin{figure}[t]
\includegraphics[width=0.47\textwidth]{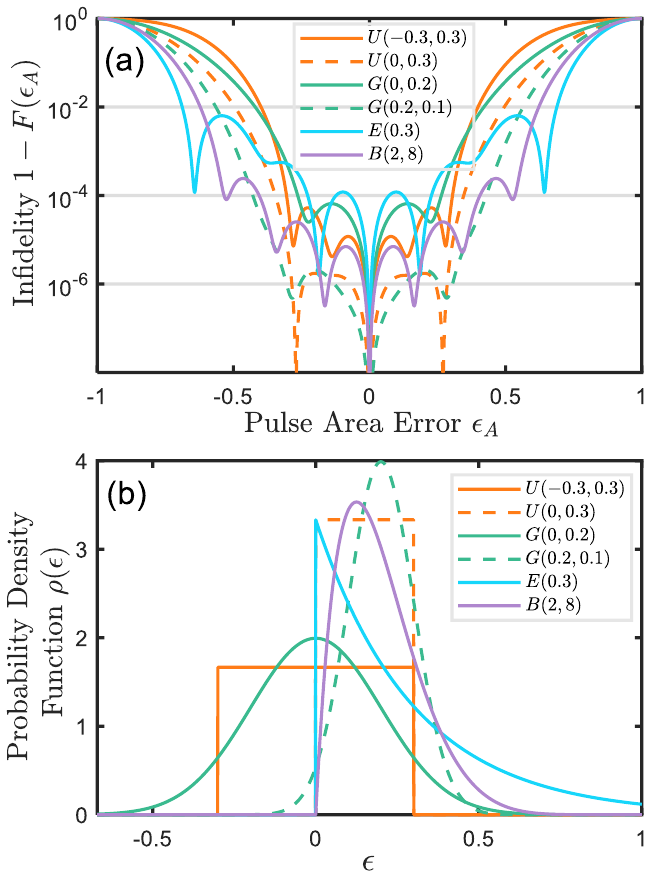}
\caption{ (a) Infidelity $1-F(\epsilon_A)$ of population inversion versus the pulse area error $\epsilon_A$ by using different sampling distributions. The learned phase parameters are given in Table~\ref{tab1}. (b) Probability density function of different distributions.} \label{diffsamplings}
\end{figure}

We next investigate the influence of different sampling distributions on the robust performance of quantum control.
We plot in Fig.~\ref{diffsamplings}(a) the robust performances of population inversion for various learned phase parameters $\bm{\theta}$, which have been trained by several common sampling distributions, including the uniform distribution $U(\epsilon_-, \epsilon_+)$ with sampling interval $[\epsilon_-, \epsilon_+]$, the Gaussian distribution $G(\mu, \nu)$ with expectation $\mu$ and variance $\nu$, the exponential distribution $E(\lambda)$ with rate parameter $\lambda$, and the Beta distribution $B(\alpha,\beta)$ with parameters $\alpha$ and $\beta$.
The specific expressions are given in Appendix~\ref{appendixb} and the corresponding probability density functions are plotted in Fig.~\ref{diffsamplings}(b).

An inspection of Fig.~\ref{diffsamplings}(a) shows that the infidelity is almost much less than $10^{-4}$ in the vicinity of the pulse area error $\epsilon=0$ for all sampling distributions, indicating the success of these trainings. A closer look at Fig.~\ref{diffsamplings}(a) reveals that different sampling distributions lead to different robust performances of population inversion.
For example, the phase parameters learned via the exponential distribution $E(0.3)$ have very good robustness with respect to errors, e.g., see the robust width $W(\xi)$ with the threshold $\xi$ being $10^{-2}$ in Fig.~\ref{diffsamplings}(a).
Nevertheless, the fidelity is not particularly high in the presence of errors when comparing to other distributions.
The reason behind these results is that a small amount of samples with large magnitude are considered in the exponential distribution, as shown in Fig.~\ref{diffsamplings}(b).

\begin{figure}[t]
\includegraphics[width=0.48\textwidth]{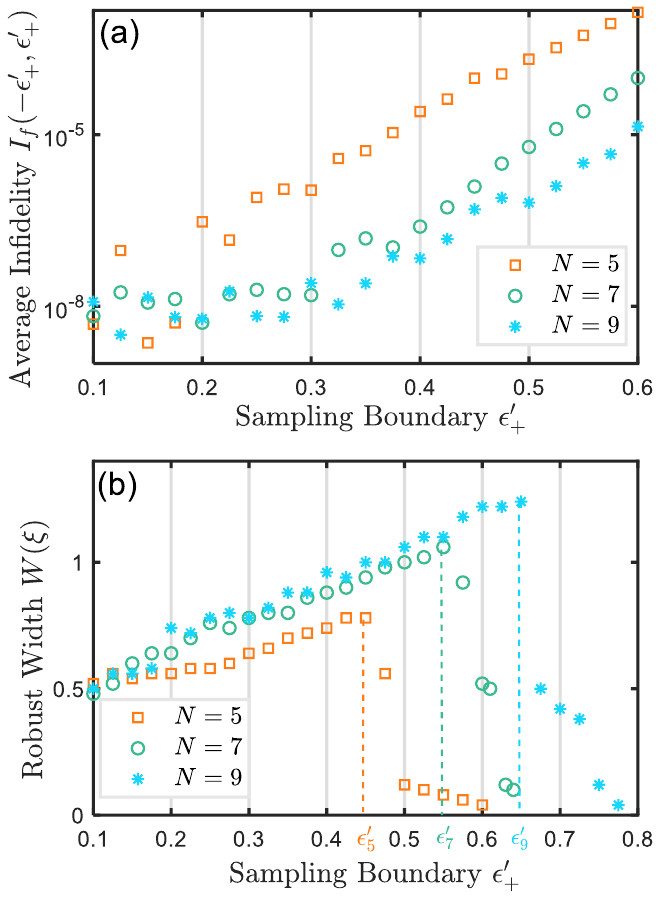}
\caption{(a) Average infidelity $I_f(-\epsilon'_+,\epsilon'_+)$ and (b) robust width $W(\xi)$ versus the sampling boundary $\epsilon'_+$ for different pulse numbers $N$. The objective and the parameters are the same as in Fig.~\ref{dvsF}. } \label{generalization_width}
\end{figure}

On the contrary, the fidelity is high in the presence of errors with small magnitude but the robust width is not very large for the uniform distribution $U(0,0.3)$.
Therefore, during the training process, the samples with small magnitude are in favor of improving high fidelity, while the samples with large magnitude facilitate the promotion of robustness.

\renewcommand\arraystretch{1.3}
\begin{table*}[htbp]
\caption{Learned phase parameters $\bm{\theta}$ in Figs.~\ref{diffsamplings}--\ref{Fvstimevaryingerror} and Videos \ref{bloch} and \ref{bloch2}. We set the pulse area of each pulse to $\pi/2$, adopt a step size of $0.001$, train 500 groups of samples with a sample size of $K=1000$, and randomly select the initial values of the phase parameters $\bm{\theta}$ from the range of $[-\pi,\pi]$. }\label{tab1}
\setlength\tabcolsep{4.4pt}
 \centering
 \begin{tabular}{llcrrrrrrrrr}
\hline
  \hline \multirow{2}{*}{}&\multirow{2}{*}{Distribution}&\multirow{2}{*}{$N$}&\multirow{2}{*}{$\theta_1~~~$}&\multirow{2}{*}{$\theta_2~~~$}& \multirow{2}{*}{$\theta_3~~~$}&\multirow{2}{*}{$\theta_4~~~$}&\multirow{2}{*}{$\theta_5~~~$}&\multirow{2}{*}{$\theta_6~~~$} &\multirow{2}{*}{$\theta_7~~~$}&\multirow{2}{*}{$\theta_8~~~$} &\multirow{2}{*}{$\theta_9~~~$}\\\\
  \hline\\[-2ex]
  Fig.~\ref{diffsamplings}&$U(-0.3,0.3)$&7&$1.1349 $ & $ 0.3521 $ & $-1.8097$  & $ 2.3882 $ & $-1.4894$ &  $-2.2752$  &  $2.9204$&&\\
  Fig.~\ref{diffsamplings}&$U(0,0.3)$&7&$-0.0890$ &  $-0.0883$  &  $2.8061$ &   $1.8183$ &  $ 0.3086$ &  $-1.7302$ &   $0.9139$&&\\
  Fig.~\ref{diffsamplings}&$G(0,0.2)$&7&$0.3267$ & $ -2.2901$  &  $1.7353$ &  $-1.4601$ & $ -1.3068$ &  $-2.7500$ &  $-0.3568$&&\\
  Fig.~\ref{diffsamplings}&$G(0.2,0.1)$&7&$-1.1718$  & $-1.6562$  &  $2.9098$   & $0.2857$   & $2.8737$  & $-1.9269$ &  $-0.9110$&&\\
  Fig.~\ref{diffsamplings}&$E(0.3)$&7&$-1.7314$  &$ -0.9706 $ & $-0.0454$  &  $2.3807 $ & $ 0.0650$  & $-1.3612$  & $-1.5465$&&\\
  Fig.~\ref{diffsamplings}&$B(2,8)$&7&$1.2869$   & $0.8791$  & $-0.6698$  &  $3.1261$ &  $-0.6310$  &  $0.6509$  &  $1.4951$&&\\
  Fig.~\ref{FvsDetuning}&$U(0,0.3)$&5&$2.8622$  &  $2.4234$  &  $0.0425$  &  $0.6872$  & $-0.8656$&&&&\\
  Fig.~\ref{FvsDetuning}&$U(0,0.5)$&7&$1.1002$  &  $2.7012$  &  $1.4171$  & $-1.9374$  &  $0.9740$  &  $1.8036$  & $-0.2582$&&\\
  Fig.~\ref{FvsDetuning}&$U(0,0.6)$&9&$0.7598$  &  $1.1145$  & $-0.8872$  & $-0.0167$  & $-1.5056$  &  $0.4233$  & $-0.0248$  &  $2.9724$ &  $-3.0918$\\
  Fig.~\ref{gate_iterations}&$U(-0.1,0.3)$&9& $1.1516$  & $ 0.3675$  & $-2.2800$  & $ 0.5743$  & $ 2.0933$   & $0.2061$  &$ -1.6437$  & $-0.7061$  &  $0.9273$\\
  Fig.~\ref{gate_range}&$U(-0.25,0)$&9&$0.6220 $ & $ 0.6251$  & $ 1.1929 $  &$ 1.6412$ &  $-0.8985 $&  $-1.5013$  & $-0.5969$ &  $-0.6176 $ &  $1.2448$\\
  Fig.~\ref{gate_range}&$U(-0.15,0.15)$&9&$2.7879$  & $ 1.8734$  &$ -2.8629 $  & $1.2522$  &  $2.4905$  &$ -1.1821$ & $ -1.8967 $ & $-2.1757$   &$ 2.4733$\\
  Fig.~\ref{gate_range}&$U(0,0.25)$&9&$-2.8126$  &$ -2.7701$ &  $-1.0354 $& $ -2.8256 $  &$ 2.3944 $&  $-1.7776$ & $ -0.7125 $ &  $0.5963 $ & $-1.4869$\\
  Fig.~\ref{FvsAreaDetuning}(a)&$U(0,0.22)$&7&$1.8826$  & $-0.5939$  &  $0.1365$  & $-1.2262$  & $-1.8446$  &  $1.7622$  &  $2.2986$&&\\
  Fig.~\ref{FvsAreaDetuning}(b)&$U(0,0.22)$&9&$-1.2035$ &  $ 1.7050$  &  $2.0577$  & $-0.7816$  &  $0.7853$ &  $-1.3623$  & $ 0.9122$  &  $0.0378 $ & $ 2.7522$\\
  Fig.~\ref{Fvstimevaryingerror}&G(0.1,0.02)&5 &1.2993  &  1.5066 &  -0.4582  &  1.6285  &  2.6998&&&&\\
  Video~\ref{bloch}&$U(-0.1,0.3)$&7&$2.2802$  &  $2.6146$  &  $0.0101$  & $-0.3657$   & $0.9080$  & $-2.3146$  &  $1.3232$&&\\
  Video~\ref{bloch2}&$U(-0.1,0.3)$&7&$1.8992$  &  $2.4589$  & $-0.0234$  & $-0.5775$ &  $ 0.5731$  &  $2.5166 $  & $0.1152$&&\\
  \hline
  \hline
 \end{tabular}
\end{table*}

\subsection{Generalization ability}\label{iif}

The generalization ability is a key indicator to measure the quality of supervised learning models. It refers to the ability to generalize from the training set to the testing set.
Generally speaking, the supervised learning model possesses strong generalization ability when executing well in predicting samples that have never been seen before.
For learned phase parameters $\bm{\theta}$,  the generalization ability can be quantified by the average infidelity $I_f(\epsilon_-,\epsilon_+)$ in Eq.~(\ref{g16}).
Specifically, a strong generalization ability is reflected in a small average infidelity $I_f(\epsilon_-,\epsilon_+)$, while a large $I_f(\epsilon_-,\epsilon_+)$ indicates that the fidelity of quantum control is not very high and also has poor robustness with respect to the errors.
Note that the average infidelity $I_f(\epsilon_-,\epsilon_+)$ is difficult to accurately characterize within what error range the learned phase parameters $\bm{\theta}$ have strong robustness. Therefore, we adopt the robust width $W(\xi)$ in Eq.~(\ref{w17}). Apparently, the robustness is better if $W(\xi)$ is larger.

Figure~\ref{generalization_width}(a) shows the average infidelity of population inversion as a function of the sampling boundary for different pulse numbers, where the samples are uniformly drawn from the interval $[0,\epsilon'_+]$ with the prime notation added to $\epsilon_+$ to distinguish the sampling boundary from the error boundary.
We can observe that the average infidelity is particularly small when the sampling range is small; see $\epsilon'_+<0.2$.
This indicates that the generalization ability is very strong within the small sampling range.
As the sampling boundary increases, the average infidelity tends to increase on the whole, signifying that a large sampling range is not conducive to obtaining high fidelity of quantum operations.

In Fig.~\ref{generalization_width}(b), we show the relationship between the robust width $W(\xi)$ and the sampling boundary $\epsilon'_+$, where the threshold $\xi$ is equal to $10^{-4}$.
First, within a small sampling boundary, e.g., $\epsilon'_+<0.2$ in Fig.~\ref{generalization_width}(b), there is not much improvement in the robust width when increasing the pulse number or sampling range, implying that the phase parameters are not properly trained.
This is mainly due to the fact that the sampling range is too narrow to effectively represent all errors.
As a result, although complex models may possess strong generalization ability, they are overlearning on the training set.
To avoid this, we can either expand the sampling range or change the sampling method [e.g., employing the Gaussian distribution to consider a small amount of errors (samples) with large magnitude].

For a given pulse number, the robust width mainly exhibits a positive correlation when increasing the sampling range; see Fig.~\ref{generalization_width}(b).
While the sampling boundary exceeds a certain threshold, the robust width drops sharply, because the average infidelity becomes relatively large in such circumstance.
Finally, the threshold of the sampling range varies depending on the number of pulses; see $\epsilon'_5$, $\epsilon'_7$, and $\epsilon'_9$ in Fig.~\ref{generalization_width}(b). In particular, increasing the pulse number leads to an increase in both the threshold range and robust width.

Therefore, there is a trade-off between high fidelity and strong robustness at a given number of pulses in our model, because both the average infidelity and the robust width grow when increasing the sampling range; see Figs.~\ref{generalization_width}(a) and \ref{generalization_width}(b).
High fidelity usually leads to a reduction in robust width, and vice versa. Thereby, for a specific number of pulses, an appropriate sampling range favors balancing high fidelity and a large robust width.

To simultaneously promote fidelity and robustness for quantum control, a feasible way is to increase the pulse number such that more phase parameters are involved during the training.
The reason is that the rule to be learned is already known in the current model, i.e., the hyperplane $\hat{F}(\epsilon^1, \dots, \epsilon^\ell, \dots, \epsilon^L)=1$ in Fig.~\ref{super}(b).
This is a little different from traditional supervised learning \cite{Bishop2006} in which the rule also needs to be learned in training. In some sense, our model can also be regarded as a multiple linear regression, where the surfaces made up of training samples and learned by researchers differ from the traditional ones.
To be specific, in traditional supervised learning, the shape composed from training samples is usually a complicated hypersurface, and one always intends to fit it with a simple hyperplane determined by learned parameters. Here, a simple hyperplane is formed by the training samples $\{\bm{\epsilon}_k,\hat{F}(\bm{\epsilon}_k)\}$, and we require to yield a hypersurface by training physical parameters to fit it. In particular, the more phase parameters, the closer the hypersurface will be to the hyperplane.

\begin{figure}[t]
\includegraphics[width=0.47\textwidth]{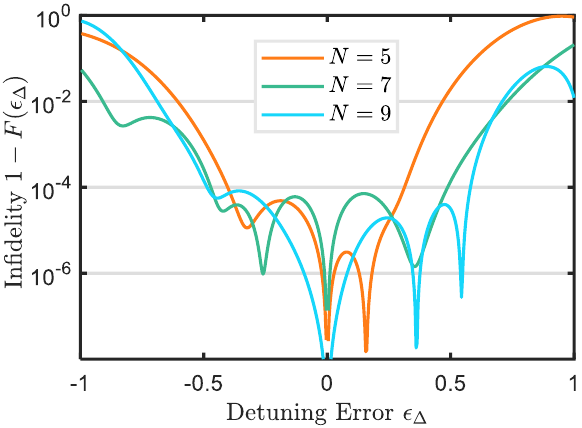}
\caption{ Infidelity $1-F(\epsilon_\Delta)$ of population inversion versus the detuning error $\epsilon_\Delta$ for different pulse numbers $N$. Each profile is plotted according to the learned phase parameters $\bm{\theta}$ offered in Table~\ref{tab1}. } \label{FvsDetuning}
\end{figure}

\section{Applications: Robust Quantum Control}  \label{iii}

\subsection{Single type of error}

We begin by illustrating a scenario in which the system exhibits only one type of error. That is, the samples are one dimensional.
Indeed, we have explained in detail in Sec.~\ref{ii} how to successfully train the phase parameters $\bm{\theta}$ to achieve robust population inversion in the presence of the pulse area error $\epsilon_A$ given by Eq.~(\ref{epa}).
We now demonstrate that this model is equally applicable to suppressing the influence of other types of errors, such as the detuning error $\epsilon_\Delta$ given by Eq.~(\ref{delta}).

Figure~\ref{FvsDetuning} shows the infidelity of population inversion in the presence of detuning error by the learned phase parameters, where the samples $\epsilon^\Delta_k$ are drawn from the uniform distribution $U(0,0.3)$.
The ultralow infidelity ranges reveal that it is also feasible to make the system robust with respect to detuning error by training on the phase parameters. Furthermore, the robustness becomes stronger as the number of pulses increases, as also indicated in Fig.~\ref{FvsDetuning}.

\subsubsection{Arbitrary superposition states}

\begin{video}[t]
\includegraphics[width=0.24\textwidth]{bloch.pdf}
\caption{Visualization of robustly preparing the superposition state $1/\sqrt{2}(|0\rangle+|1\rangle)$ on the Bloch sphere. The learned phase parameters $\bm{\theta}$ are presented in Table~\ref{tab1}, and the blue (red) arrow denotes the Bloch vector pointing to the initial (final) state.
The solid, dashed, and dash-dot curves represent the evolution trajectories of the system state in the presence of the pulse area error $\epsilon_A=0$, $+0.1$, and $-0.1$, respectively. } \label{bloch}
\end{video}

\begin{video}[t]
\includegraphics[width=0.24\textwidth]{bloch2.pdf}
\caption{Visualization of robustly preparing the superposition state $1/2(|0\rangle+\sqrt{3}|1\rangle)$ on the Bloch sphere. The learned phase parameters $\bm{\theta}$ are given in Table~\ref{tab1}.} \label{bloch2}
\end{video}

\begin{figure*}[htbp]
\includegraphics[width=0.98\textwidth]{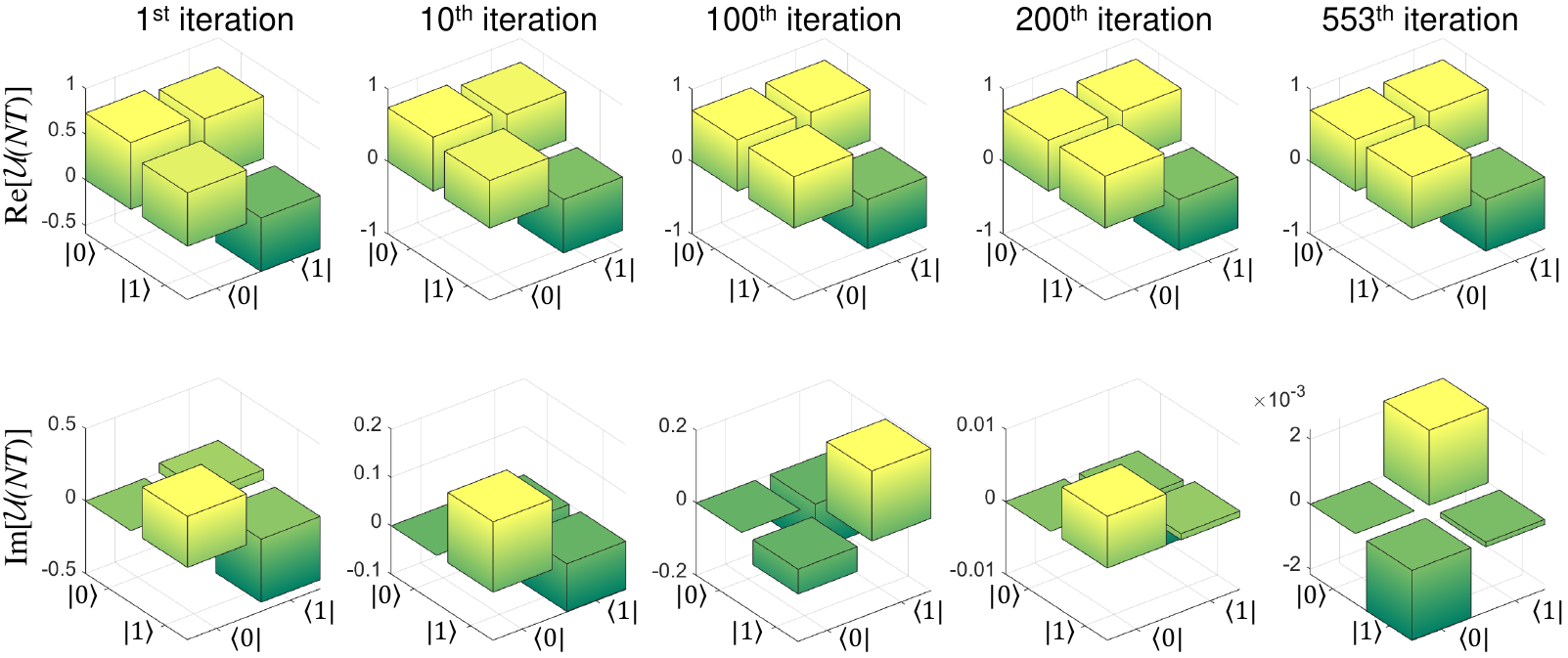}
\caption{Real and imaginary parts of the evolution operator $\mathcal{U}(NT)$ at different iterations during the training process. The objective is to robustly implement the Hadamard gate given by Eq.~(\ref{haN}); the learned parameters are given in Table~\ref{tab1}.} \label{gate_iterations}
\end{figure*}

So far, the supervised learning model is merely illustrated to realize robust population inversion. Actually, this model is suitable for implementing arbitrary population transfer in a robust manner as well.
To this end, it is imperative to transform the target state into a superposition state we desire to prepare, whose general form is given by
\begin{eqnarray} \label{24s}
|\psi_T\rangle=\cos\phi|0\rangle+e^{i\varphi}\sin\phi|1\rangle,
\end{eqnarray}
where $\phi$ and $\varphi$ can be arbitrary.
After establishing the target state, we then employ Eq.~(\ref{state7}) to train the phase parameters to accomplish this goal.

Videos~\ref{bloch} and \ref{bloch2} show the time-evolution trajectories for robust preparation of different superposition states through the learned phase parameters $\bm{\theta}$.
We can see that, when the system exhibits pulse area error, the trajectories (the dashed and dash-dot curves) gradually deviate from the exact one (the solid curve) during the evolution process. Despite this, these three trajectories eventually come together. Therefore, the learned phase parameters $\bm{\theta}$ do ensure that the target state $|\psi_T\rangle$ is still obtained with an ultrahigh fidelity even in the presence of systematic errors.

\subsubsection{General single-qubit gates}

This supervised learning model can be used not only to prepare arbitrary superposition states, but also to implement general single-qubit gates. The expression for a general single-qubit gate can be written as
\begin{eqnarray}\label{unitaryN}
\mathrm{\mathbf{U}}(\phi,\varphi,\Phi)=\left[
		\begin{array}{cc}
			e^{i\Phi}\cos\phi &~~ -e^{-i(\varphi+\Phi)}\sin\phi\\[1.5ex]
			e^{i(\varphi+\Phi)}\sin\phi & e^{-i\Phi}\cos\phi
		\end{array}
		\right],~~~
\end{eqnarray}
where $\phi,\varphi$, and $\Phi$ are arbitrary real numbers based on the desired quantum gate.
In the current model, there are two methods to robustly obtain the general single-qubit gate given by Eq.~(\ref{unitaryN}). Next, we present the details of these two methods.

(i) \emph{State-based method}. Given that the initial state of the system is $|0\rangle$, we first train the phase parameters $\bm{\theta}$ to robustly obtain a superposition state
$|\psi_T\rangle=\cos\phi|0\rangle+e^{i\varphi}\sin\phi|1\rangle$, where the values of $\phi$ and $\varphi$ are determined by the desired quantum gate.
Then, we shift the learned phase parameters by an appropriate angle:
\begin{eqnarray}
\bm{\theta}\leftarrow\bm{\theta}+\Phi
\end{eqnarray}
with the shift angle $\Phi$ dependent on the desired quantum gate as well.
We emphasize that this phase shift manipulation does not change the probability amplitude of the matrix elements in Eq.~(\ref{unitaryN}); rather, it modifies only the relative phase between different matrix elements.
Therefore, the final learned phase parameters for implementing the general single-qubit gate in a robust manner become $\bm{\theta}+\Phi$.
As an example, the learned phase parameters $\bm{\theta}$ adopted in Video~\ref{bloch} can also be used to implement the Hadamard gate with a specific phase that can be properly modulated by the phase shift $\Phi$.

(ii) \emph{Evolution operator-based method}.
We directly solve the Schr\"{o}dinger equation for the evolution operator $\mathcal{U}(t)$ of this system,
\begin{eqnarray}\label{unitary11}
\dot{\mathcal{U}}(t)=-iH(t)\mathcal{U}(t),
\end{eqnarray}
where $H(t)$ is the system Hamiltonian, and $\mathcal{U}(0)={\mathbb{I}}$ is an identity operator at the initial time.
The target quantum gate $\mathrm{\mathbf{U}}(\phi,\varphi,\Phi)$ can now be received via the evolution operator after $N$ pulses, i.e.,
\begin{eqnarray}
\mathcal{U}(NT)=\mathrm{\mathbf{U}}(\phi,\varphi,\Phi).
\end{eqnarray}
In this situation, we adopt Eq.~(\ref{gate8}) to calculate the corresponding gradients.

Suppose now that we intend to execute a Hadamard gate:
\begin{eqnarray}\label{haN}
\mathrm{\mathbf{H}}=\frac{1}{\sqrt{2}}\left[
		\begin{array}{cc}
			1 & 1\\[1.5ex]
			1 & -1
		\end{array}
		\right].
\end{eqnarray}
To this end, we proceed to train the phase parameters $\bm{\theta}$ by minimizing the cost function given in Eq.~(\ref{cost9}).
Figure~\ref{gate_iterations} shows the variation of tomography results for the evolution operator $\mathcal{U}(NT)$ at different iterations during the training stage. We can see that the evolution operator of the system gradually approaches the Hadamard gate as the number of iterations increases, and eventually reaches the goal with ultrahigh fidelity.

\begin{figure}[t]
\includegraphics[width=0.48\textwidth]{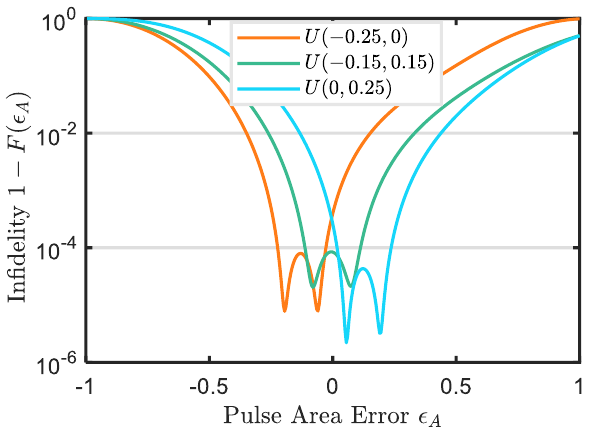}
\caption{Robust performances for the phase parameters $\bm{\theta}$ learned from different ranges of the uniform distribution. The objective is the same as in Fig.~\ref{gate_iterations}, and the learned parameters are given in Table~\ref{tab1}. } \label{gate_range}
\end{figure}

Different from the case of population inversion, the robust performance of quantum gates is strongly correlated with the sampling distribution here. In other words, when training on the samples with different ranges, we learn different phase parameters. These phase parameters exhibit different robust performances for systematic errors.

In Fig.~\ref{gate_range}, we plot the infidelity of the Hadamard gate versus the pulse area error using various phase parameters that are learned from the training set drawn from distinct sampling ranges. It is clearly shown that the learned phase parameters manifest varying degrees of robustness to the pulse area error under different sampling distributions.

\subsection{Multiple types of errors}

\begin{figure}[t]
\includegraphics[width=0.48\textwidth]{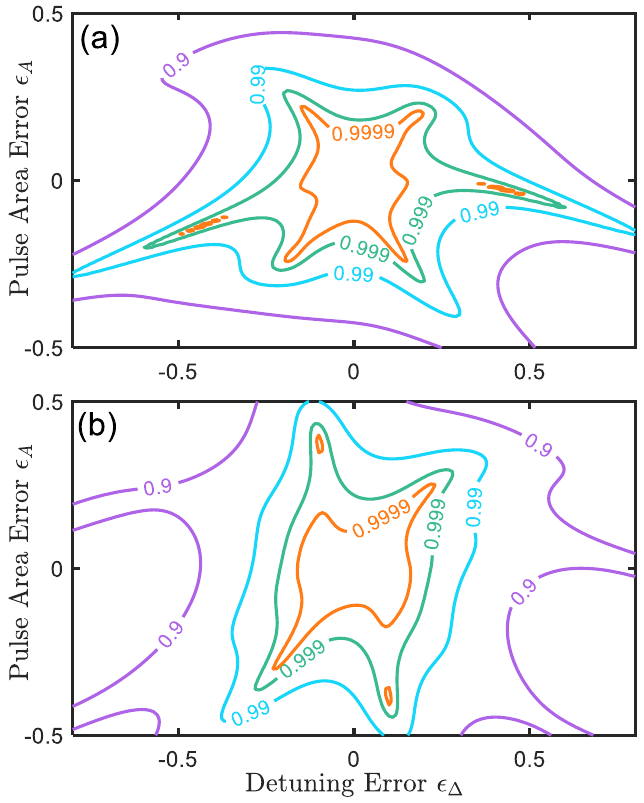}
\caption{Fidelity $F(\epsilon_A,\epsilon_\Delta)$ of population inversion versus the pulse area error $\epsilon_A$ and detuning error $\epsilon_\Delta$ for different pulse numbers: (a) $N=7$, (b) $N=9$. The learned parameters are given in Table~\ref{tab1}. } \label{FvsAreaDetuning}
\end{figure}

When the system simultaneously exhibits different types of errors (e.g., pulse area and detuning errors), it is also convenient to employ this model to implement robust quantum control.
In such a situation, the sample space becomes multidimensional, where each dimension represents a specific type of error.
As a result, the training set can be represented as
$\{\epsilon^A_{k},\epsilon^\Delta_{k},\dots,\hat{F}(\epsilon^A_{k},\epsilon^\Delta_{k},\dots)\}$,
where $\hat{F}(\epsilon^A_{k},\epsilon^\Delta_{k},\dots)=1$, $k=1,\dots, K.$
In Fig.~\ref{FvsAreaDetuning}, we plot the fidelity of population inversion as a function of the pulse area and detuning errors by training different numbers of phase parameters. The results demonstrate that the training methodology is able to maintain its efficacy even when confronted with multiple types of errors.

\subsection{Time-varying errors}

In this section, we manage to train the phase parameters when the errors are time dependent. It is worth mentioning that one rarely obtains the exact Taylor series expression for time-varying errors by using traditional composite pulse technologies \cite{Freeman97}, because the Hamiltonian generally cannot commute with itself at different times.
The popular method of tackling time-varying errors is to make use of an appropriate unitary transformation to formally divide the evolution operator of the system into error-free and error terms, and then design related parameters to minimize the influence of the error term \cite{Green2013,PhysRevA.90.012316,PhysRevLett.113.250501,Zhen2016,Haas2019,Li2021,PhysRevA.104.052625}. While this method is effective, it is quite cumbersome.

Here we show that it is still feasible to use the supervised learning model to address the case of time-varying errors.
To this end, the training set needs to be modified as  $\{\epsilon^{t_1}_k,\dots,\epsilon^{t_L}_k,\hat{F}(\epsilon^{t_1}_k,\dots,\epsilon^{t_L}_k)\}$, where $\hat{F}(\epsilon^{t_1}_k,\dots,\epsilon^{t_L}_k)=1$, and $\epsilon^{t_\ell}_k$ denotes the error magnitude in the time interval $[t_\ell,t_{\ell+1}]$, $\ell=1, \dots, L$, $k=1,\dots, K$.
Namely, the time-dependent nature of errors is reflected in distinct dimensions of the samples.

\begin{figure}[t]
\includegraphics[width=0.48\textwidth]{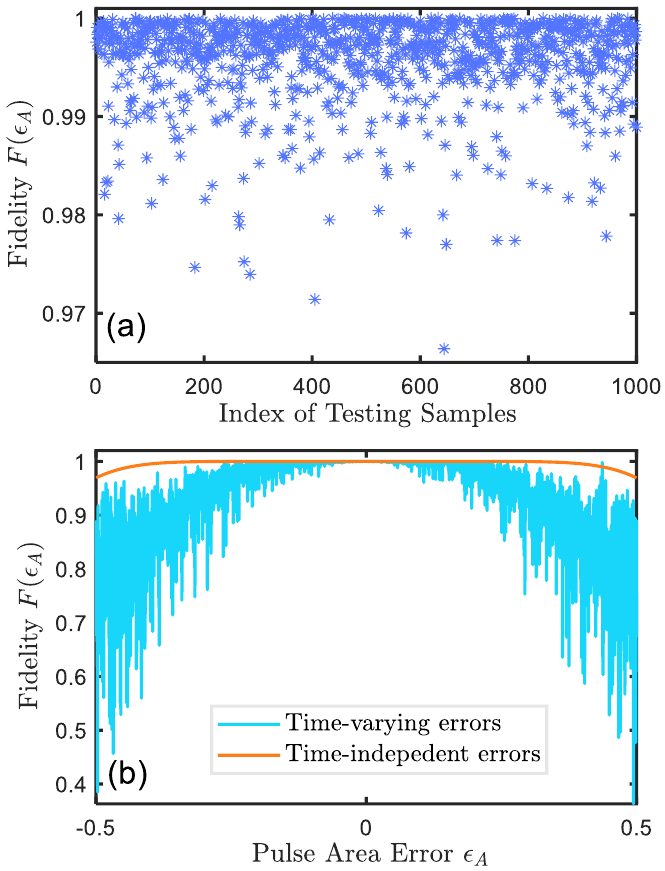}
\caption{(a) Testing performance for the learned phase parameters $\bm{\theta}$, where the average fidelity of population inversion is about 99.55\% for 1000 testing samples. The time interval is $t_{\ell+1}-t_\ell=T$, with the other parameters given in Table~\ref{tab1}. (b) Fidelity $F(\epsilon_A)$ of population inversion versus the  pulse area error $\epsilon_A$, where the pulse area error suffers from time-dependent Gaussian noises with expectation $0.1\epsilon_A$ and variance $0.1\epsilon_A$.
} \label{Fvstimevaryingerror}
\end{figure}

Unlike the case of time-independent errors, it is difficult to account for all time-varying errors in the testing stage. To validate the phase parameters $\bm{\theta}$ learned from the training stage, we randomly draw enough testing samples from time-varying errors for different time intervals. The fidelity of each testing sample is plotted in Fig.~\ref{Fvstimevaryingerror}(a). It is clearly shown that the learned phase parameters $\bm{\theta}$ perform very well in the vast majority of testing samples, where a total average fidelity of 99.55\% is achieved.
Furthermore, we demonstrate in Fig.~\ref{Fvstimevaryingerror}(b) how to successfully obtain robust population inversion when the pulse area error suffers from time-dependent Gaussian noises. For a comparison, Fig.~\ref{Fvstimevaryingerror}(b) also shows the fidelity as a function of the time-independent errors using the learned phase parameters. We can observe that the robustness with respect to time-independent errors is much better than that of time-varying errors.

\section{Further Extensions}   \label{iv}

\subsection{Training of detuning parameters}

Up to now, we have regarded the phases as training parameters in the supervised learning model.
Actually, the phases do not have unique specifications, and thus other physical parameters (e.g., detunings or Rabi frequencies) are also able to serve as training parameters instead.
In the following, we investigate the feasibility of utilizing the detuning $\Delta_n$ ($n=1, \dots, N$) of each pulse as training parameters to implement robust quantum control.
Undoubtedly, the training of Rabi frequencies can be accomplished as well through a comparable procedure.
To simplify the notation, all detunings $\Delta_n$ are collectively represented by a vector $\bm{\Delta}$, i.e., $\bm{\Delta}=(\Delta_1, \dots, \Delta_N)$.

\begin{figure}[t]
\includegraphics[width=0.48\textwidth]{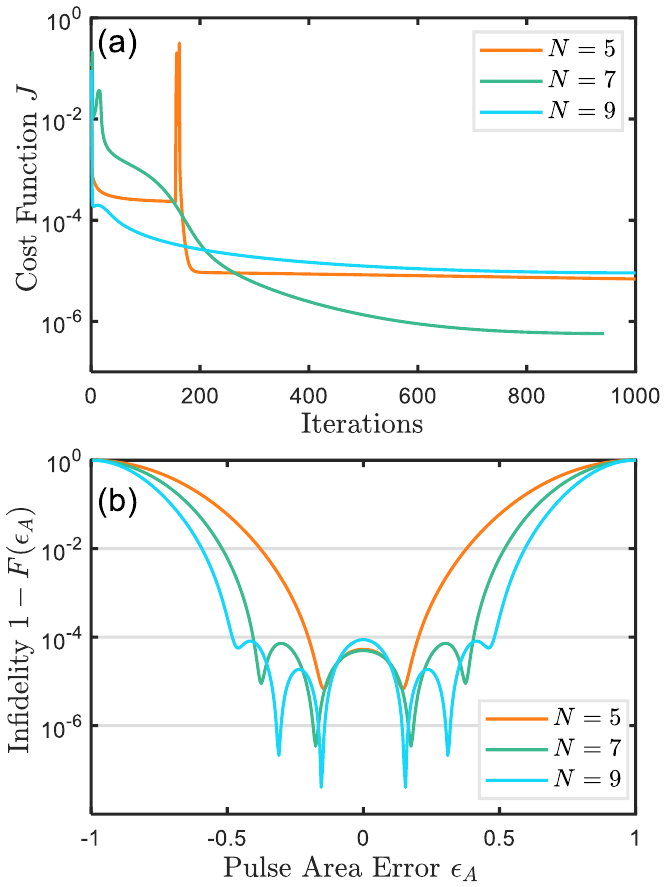}
\caption{(a) Cost function versus the number of iterations for different numbers of detuning parameters. The sudden increase in the cost function is due to the reinitialization of the large detuning.
(b) Infidelity $1-F(\epsilon_A)$ of population inversion versus the pulse area error $\epsilon_A$ by using distinct learned detuning parameters $\bm{\Delta}$ given in Table~\ref{tab2}. } \label{detuning_var}
\end{figure}

Figure~\ref{detuning_var}(a) shows the relationship between the cost function and the number of iterations for different numbers of detuning parameters in the training process,in order to achieve robust population inversion.
During the procedure, the detuning $\Delta_n$ is reinitialized if $\Delta_n/\Omega_n>10$. The reason for doing this can be found as follows. Physically, the pulse becomes highly detuned and primarily influences the phase rather than probability amplitude of quantum states when $\Delta_n$ is excessively large. As a result, the highly detuned pulse makes little contribution to population transfer, and thus needs to be reinitialized.

\renewcommand\arraystretch{1.3}
\begin{table*}[htbp]
\caption{Learned detuning parameters $\bm{\Delta}$ in Fig.~\ref{detuning_var}. We set the duration of each pulse to $T_n=\pi/(2\sqrt{\Omega_n^2+\Delta_n^2})$, adopt a step size of 0.1, train 500 groups of samples with a sample size of $K=1000$, and randomly select the initial values of the detuning parameters $\bm{\Delta}$ from the range of $[-3,3]$.}\label{tab2}
\setlength\tabcolsep{4.6pt}
 \centering
 \begin{tabular}{llcrrrrrrrrr}
\hline
  \hline \multirow{2}{*}{Distribution}&\multirow{2}{*}{$N$}&\multirow{2}{*}{$~~~\Delta_1~~~$}&\multirow{2}{*}{$~\Delta_2~~~$}& \multirow{2}{*}{$\Delta_3~~~$}&\multirow{2}{*}{$\Delta_4~~~$}&\multirow{2}{*}{$\Delta_5~~~$}&\multirow{2}{*}{$\Delta_6~~~$} &\multirow{2}{*}{$\Delta_7~~~$}&\multirow{2}{*}{$\Delta_8~~~$} &\multirow{2}{*}{$\Delta_9~~~$}\\\\
  \hline\\[-2ex]
  $U(0,0.2)$&5&$-4.8860$  &  $0.2452$  & $ 4.0916$ &  $-5.0015$  & $-1.9768$&&&&\\
  $U(0,0.3)$&7&$-5.8333$ &  $-1.4134$  & $ 0.4753$  & $-2.6560$  &  $9.5803$  &  $7.9718$ & $-2.7511$&&\\
  $U(0,0.4)$&9&$-9.2604$ & $ -3.2210$  &$ -9.8686$  & $-1.7511$  & $ 6.9721$   & $7.0765$ &  $-1.0804$ &  $-0.2291$  &  $2.7464$\\
  \hline
  \hline
 \end{tabular}
\end{table*}

An inspection of Fig.~\ref{detuning_var}(a) demonstrates that the value of the cost function for $N=9$ is larger than that of other two situations. This is because the sampling ranges of the uniform distribution are different for different pulse numbers. As explained in Sec.~\ref{methods}, a large sampling range leads to a large value of the cost function.  Thus, what we care about is not the value of the cost function, but its convergence.  When the cost function converges to a small steady value, we can claim that a set of possible solutions for the training parameters is obtained.
It is clearly found from Fig.~\ref{detuning_var}(a) that the cost functions do gradually diminish with each subsequent iteration and finally tend to extremely small values, verifying the validity of the gradient descent algorithm and the success of the trainings.

We can also see in Fig.~\ref{detuning_var}(b) that the detuning parameters $\bm{\Delta}$ learned from the training stage perform particularly well in terms of robustness against pulse area errors around $\epsilon_A=0$.
Furthermore, similar to the case of the phase parameters, the system becomes more resilient to pulse area error when increasing the number of detuning parameters, as illustrated by different robust widths in Fig.~\ref{detuning_var}(b).

\subsection{Supervised learning in high-dimension spaces}

So far, the supervised learning model has mainly been demonstrated to work well in the two-dimension Hilbert space. Indeed, this model can also be applied in higher-dimension Hilbert spaces. Next, we take four dimensions as an illustrative example to briefly show how to perform a high-fidelity two-qubit gate in error environments; the treatment in higher dimensions is similar.

\begin{figure}[t]
\includegraphics[width=0.48\textwidth]{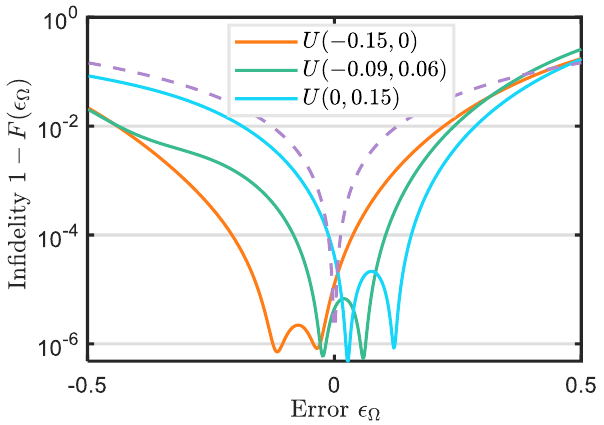}
\caption{ Infidelity $1-F(\epsilon_\Omega)$ of the two-qubit gate $\mathrm{\mathbf{U}}_\mathrm{M}$ versus error $\epsilon_\Omega$. Each solid profile is plotted according to the learned detuning parameters $\bm{\Delta}$ given in Table~\ref{tab3}, while the dashed curve is plotted by setting $T=\pi/(4\Omega)$ and $\Delta^a=\Delta^b=0$.} \label{twogates}
\end{figure}

Assume that the Hamiltonian of a two-qubit system can be written as
\begin{eqnarray}
H(\Delta^a,\Delta^b)=\Omega(1+\epsilon_\Omega)\sigma_x^a\sigma_x^b+\Delta^a\sigma_z^a+\Delta^b\sigma_z^b,
\end{eqnarray}
where $\Omega$ is the coupling strength between qubits $a$ and $b$, $\Delta^a$ and $\Delta^b$ are the corresponding detunings, and $\epsilon_\Omega$ denotes a possible error in this system.
The objective is to implement a two-qubit gate $\mathrm{\mathbf{U}}_\mathrm{M}=\exp(-i\pi/4\sigma_x^a\sigma_x^b)$, which is actually equivalent to the controlled-NOT (CNOT) gate after performing several single-qubit operations \cite{PhysRevA.95.032314}.

In the ideal case, i.e., $\epsilon_\Omega=0$, the target two-qubit gate $\mathrm{\mathbf{U}}_\mathrm{M}$ is readily attained by setting the operation time $T=\pi/(4\Omega)$ and $\Delta^a=\Delta^b=0$.
When taking account of the presence of error $\epsilon_\Omega$, we will obtain a faulty gate $\mathrm{\mathbf{U}}^f_\mathrm{M}=\exp[-i\pi/4(1+\epsilon_\Omega)\sigma_x^a\sigma_x^b]$, which is highly sensitive to $\epsilon_\Omega$ (see the dashed curves in Fig.~\ref{twogates}).
To improve the robustness with respect to error $\epsilon_\Omega$, we adopt a composite pulse sequence and regard the detunings $\Delta^a_n$ and $\Delta^b_n$ ($n=1,2,\dots,N$) as training parameters, i.e., $\bm{\Delta}=(\Delta^a_1,\Delta^b_1,\dots,\Delta^a_N,\Delta^b_N)$.
As shown in Fig.~\ref{twogates}, the infidelity is extremely small around $\epsilon_\Omega=0$ according to the learned detuning parameters, verifying that robust quantum control is also accessible in high-dimension Hilbert spaces by using this model.

It should be noted that the dimension of the Hilbert space grows exponentially as the number of qubits increases. At this point, successfully training the parameters becomes progressively more difficult for multiple-qubit systems, because the amount of computation will explode. Fortunately, it has been shown that any multi-qubit quantum gate may be constituted from CNOT and single-qubit gates \cite{Nielsen00}, and both types of logic gates can be robustly trained via the current model. Therefore, it is possible to achieve any multi-qubit quantum gate in a robust manner using a universal gate set formed by high-fidelity CNOT and single-qubit gates.

\renewcommand\arraystretch{1.3}
\begin{table*}[htbp]
\caption{Learned detuning parameters $\bm{\Delta}$ for implementing the two-qubit gate $\mathrm{\mathbf{U}}_\mathrm{M}$ in a robust manner. We set the duration of each pulse to $T=\pi/(4\Omega)$, adopt a step size of 0.001, choose the pulse number as $N=9$, train 1000 groups of samples with a sample size of $K=3000$, and randomly select the initial values of the detuning parameters $\bm{\Delta}$ from the range of $[-1,1]$.}\label{tab3}
\setlength\tabcolsep{4.6pt}
 \centering
 \begin{tabular}{lrrrrrrrrrrr}
  \hline
  \hline \multirow{2}{*}{Distribution}&\multirow{2}{*}{$~~~\Delta^a_1~~~$}&\multirow{2}{*}{$~~\Delta^a_2~~~$}& \multirow{2}{*}{$~\Delta^a_3~~~$}&\multirow{2}{*}{$\Delta^a_4~~~$}&\multirow{2}{*}{$\Delta^a_5~~~$}&\multirow{2}{*}{$\Delta^a_6~~~$} &\multirow{2}{*}{$\Delta^a_7~~~$}&\multirow{2}{*}{$\Delta^a_8~~~$} &\multirow{2}{*}{$\Delta^a_9~~~$}\\\\
  \hline\\[-2ex]
  $U(-0.15,0)$&$-0.8094$  & $-0.3266$ &  $-0.4048$  &  $1.0508$  & $ 1.1646$  &  $0.1250$  &  $0.1357$ &  $-1.5451$  &  $0.0231$\\
  $U(-0.09,0.06)$&$1.5060$  & $-1.2594$  & $-0.2565$  &  $0.6741$  & $-1.6983$  &  $1.5649$   & $0.4478$  & $-0.9407$  &  $0.5922$\\
  $U(0,0.15)$&$-0.3514$  &  $1.2741$ &  $-0.0107$  &  $1.7152$  &  $0.2579$ & $ -1.9542$ & $ -0.4414$  & $-0.9180$  & $ 0.4095$\\
  \hline \multirow{2}{*}{Distribution}&\multirow{2}{*}{$\Delta^b_1~~~$}&\multirow{2}{*}{$~~\Delta^b_2~~~$}& \multirow{2}{*}{$~\Delta^b_3~~~$}&\multirow{2}{*}{$\Delta^b_4~~~$}&\multirow{2}{*}{$\Delta^b_5~~~$}&\multirow{2}{*}{$\Delta^b_6~~~$} &\multirow{2}{*}{$\Delta^b_7~~~$}&\multirow{2}{*}{$\Delta^b_8~~~$} &\multirow{2}{*}{$\Delta^b_9~~~$}\\\\
  \hline\\[-2ex]
  $U(-0.15,0)$&$1.1007$ &  $-1.6323$  &  $0.5713$  &  $1.0148$ &  $-1.6145$  & $ 1.6170$  & $-0.8991$  &$ -0.5304 $ &  $0.8847$\\
  $U(-0.09,0.06)$&$-0.2990$  &$ -0.9496$  &$ -0.6132$   & $0.9016$  & $ 0.5801$  & $ 0.5451$  &  $0.2672 $ & $-0.7618$  & $-0.5137$\\
  $U(0,0.15)$& $-1.4707$ &  $-0.8018 $ &  $0.7620 $ & $-0.8676 $ &  $1.6910$  &  $0.4098$ &  $-0.1400$  &  $1.2832$  & $-1.0190$\\
  \hline
  \hline
 \end{tabular}
\end{table*}

\section{Conclusion}  \label{v}

In conclusion, we have proposed a supervised learning model used for implementing robust quantum control in composite-pulse systems.
We first constructed the cost function for this model, and demonstrated how this model works. By introducing two types of virtual variables, we put forward a modified gradient descent algorithm to train phase parameters.
We found that the robustness depends heavily on sampling methods.
To be specific, the average infidelity tends to enlarge as the sampling range increases, and different sampling distributions result in different robust performances.

{Afterwards, we characterized the generalization ability of this model.
The results demonstrate that robustness is not greatly improved when the sampling range is too small, whereas robustness is poor once the sampling range exceeds a certain threshold.
In particular, there is a corresponding strengthening of the generalization ability when increasing the model complexity (i.e., increasing the pulse number).
Therefore, for a given pulse number, a suitable sampling range is conducive to maximizing the generalization ability of the model.}

Finally, we demonstrated a wide range of applications for the current model, e.g., robust realization of arbitrary superposition states as well as general single-qubit gates.
Specifically, the phase parameters learned from this model perform very well in robustness with respect to various systematic errors, such as single error, multiple errors, and time-varying errors.
It is particularly important that all kinds of situations (including single, multiple, and time-varying errors) can be well tackled in the same model, and the only difference lies in the dimension of the samples.
To implement general single-qubit gates in a robust manner, we proposed two different methods, the state-based and evolution operator-based methods, to train the phase parameters and its robust property relies on the sampling distribution.
Moreover, this model can be applied in higher-dimensional Hilbert spaces, and is suitable for training any physical parameters to attain robust quantum control.
It is believed that this supervised learning model provides a universal and high-efficiency plateau for realizing reliable quantum information processing in different quantum systems.

\begin{acknowledgments}
We would like to thank the valuable suggestions of Dr.~Ye-Xiong Zeng, Dr.~Yanming Che, and Dr.~Clemens Gneiting.
This work is supported by the Natural Science Foundation of Fujian Province under
Grant No. 2021J01575, the Natural Science Funds
for Distinguished Young Scholars of Fujian Province
under Grant No. 2020J06011, and the Project from Fuzhou
University under Grant No. JG202001-2. F.N. is supported in part by
Nippon Telegraph and Telephone Corporation (NTT) Research,
the Japan Science and Technology Agency (JST)
[via the Quantum Leap Flagship Program (Q-LEAP), and the Moonshot R\&D Grant No. JPMJMS2061],
the Asian Office of Aerospace Research and Development (AOARD) (via Grant No. FA2386-20-1-4069),
the Office of Naval Research (ONR Global),
and the Foundational Questions Institute Fund (FQXi) via Grant No. FQXi-IAF19-06.
\end{acknowledgments}

\begin{appendix}

\section{Search for optimal solutions by the escaping method} \label{appendixa}

In this appendix, we mainly demonstrate how to search the optimal solutions by the escaping method, a synthesis of the modified GRAPE and simulated annealing \cite{PhysRevA.105.042437,Mahesh2022}.
First, we employ the modified GRAPE algorithm for learning $M$ groups of phase parameters $\bm{\theta}_m$ as a starting point, where $m=1, \dots, M$.
To escape the locally optimal solution, we artificially add a random value to each phase so as to generate new input variables.
By making use of the modified GRAPE algorithm again, we obtain the new learned parameters $\bm{\theta}^{\prime}_m$ and the corresponding average fidelity $\bar{F}'_m(\epsilon_-,\epsilon_+)$, where the subscript $m$ represents the $m$th group of training sets.
If $\bar{F}'_m(\epsilon_-,\epsilon_+)\leq\bar{F}_m(\epsilon_-,\epsilon_+)$ then the new learned parameters $\bm{\theta}'_m$ are not as good as the old ones. However, we accept these new learned parameters $\bm{\theta}^{\prime}_m$ with probability
\begin{eqnarray} \label{pc}
p_c=\exp\left[-\frac{\bar{F}_m(\epsilon_-,\epsilon_+)-\bar{F}'_m(\epsilon_-,\epsilon_+)}{T_{j}}\right],
\end{eqnarray}
where the sequence $\{T_j\}$ ($j=1,\dots,J$) is monotonically decreasing, and $T_j>0$ denotes the temperature in simulated annealing \cite{PhysRevA.105.042437,Mahesh2022}.
On the other hand, if $\bar{F}'_m(\epsilon_-,\epsilon_+)>\bar{F}_m(\epsilon_-,\epsilon_+)$, it indicates that the solution has jumped out of locally optimal regions, and thus we accept the new ones.
After updating each training group $J$ times, we next calculate the total average fidelity for all training groups,
\begin{eqnarray}
\bar{F}_\mathrm{tot}=\frac{1}{M}\sum_{m=1}^M \bar{F}_m(\epsilon_-,\epsilon_+).
\end{eqnarray}
Then, the learned parameters whose average fidelities are lower than $\bar{F}_\mathrm{tot}$ are discarded.
This iteration process continues until the threshold of the total average fidelity is reached; see Algorithm 2 for details.
At the end, the optimal solutions for robust quantum control are left.

\renewcommand\arraystretch{1.2}
\begin{table}[htbp]
 \begin{tabular}{l}
  \hline
  \multirow{2}{*}{\textbf{ALGORITHM 2. Escape-based GRAPE algorithm}}\\\\
  \hline
1. Use the modified GRAPE algorithm to learn $M$ group \cr
~~~~parameters $\bm{\theta}_m$ and calculate the average fidelity  \cr
~~~~$\bar{F}_m(\epsilon_-,\epsilon_+)$.\cr
2. Add random values $\delta\theta^m_n$ to the phase parameters: \cr ~~~~$\theta^m_n\leftarrow\theta^m_n+\delta\theta^m_n$, the new input variables. \cr
3. Reuse the modified GRAPE algorithm to learn new\cr
~~~~parameters $\bm{\theta}'_m$ and the average fidelity $\bar{F}'_m(\epsilon_-,\epsilon_+)$. \cr
4. If $\bar{F}_m(\epsilon_-,\epsilon_+)<\bar{F}'_m(\epsilon_-,\epsilon_+)$, \cr
~~~~then $\bm{\theta}_m\leftarrow\bm{\theta}'_m$ and $\bar{F}_m(\epsilon_-,\epsilon_+)\leftarrow\bar{F}'_m (\epsilon_-,\epsilon_+)$. \cr
~~~~Otherwise, accept $\bm{\theta}'_m$ with probability $p_c$ in Eq.~(\ref{pc}).\cr
5. Repeat steps 2 through 4 $J$ times.  \cr
6. Calculate the total average fidelity $\bar{F}_\mathrm{tot}$. \cr
7. Discard learned parameters $\bm{\theta}_m$ when $\bar{F}_m(\epsilon_-,\epsilon_+)<\bar{F}_\mathrm{tot}$. \cr
8. Go to step 2 until $\big(\max\{\bar{F}_m(\epsilon_-,\epsilon_+)\}-\bar{F}_\mathrm{tot}\big)<\varepsilon$.  \cr
  \hline
 \end{tabular}
\end{table}

\section{Common sampling distributions} \label{appendixb}

This appendix briefly introduces several common sampling distributions.

($i$) Uniform distribution $U(\epsilon_-, \epsilon_+)$. The probability density function of the uniform distribution reads
\begin{eqnarray}\label{bu}
&\rho_u(\epsilon)= \left \{
\begin{array}{ll}
    \displaystyle\frac{1}{\epsilon_+-\epsilon_-},           & ~~~   \epsilon\in[\epsilon_-,\epsilon_+], \\[2.8ex]
    0,     & ~~~ \mathrm{others},~~ \\
\end{array}
\right.
\end{eqnarray}
with sampling interval $[\epsilon_-, \epsilon_+]$.
The expectation and variance are
\begin{eqnarray}
\mathcal{E}(\epsilon)=\frac{\epsilon_++\epsilon_-}{2}, ~~\mathcal{D}(\epsilon)=\frac{(\epsilon_+-\epsilon_-)^2}{12}.
\end{eqnarray}

($ii$) Gaussian distribution $G(\mu, \nu)$. The probability density function of the Gaussian distribution reads
\begin{eqnarray}\label{bg}
\rho_g(\epsilon)= \frac{1}{\sqrt{2\pi}\nu}\exp\left[-\frac{(\epsilon-\mu)^2}{2\nu^2}\right].
\end{eqnarray}
The expectation and variance are
\begin{eqnarray}
\mathcal{E}(\epsilon)=\mu, ~~\mathcal{D}(\epsilon)=\nu^2.
\end{eqnarray}

($iii$) Exponential distribution $E(\lambda)$. The probability density function of the exponential distribution reads
\begin{eqnarray}\label{bu}
&\rho_e(\epsilon)= \left \{
\begin{array}{ll}
    \displaystyle\frac{1}{\lambda} \exp\left({-\frac{\epsilon}{\lambda}}\right),           & ~~~   \epsilon\geq0, \\[1.8ex]
    0,     & ~~~ \epsilon<0,~~ \\
\end{array}
\right.
\end{eqnarray}
with rate parameter $\lambda$.
The expectation and variance are
\begin{eqnarray}
\mathcal{E}(\epsilon)={\lambda}, ~~\mathcal{D}(\epsilon)={\lambda^2}.
\end{eqnarray}

($iv$) Beta distribution $B(\alpha,\beta)$. The probability density function of the Beta distribution reads
\begin{eqnarray}\label{bb}
\rho_b(\epsilon)= \frac{\Gamma(\alpha+\beta)}{\Gamma(\alpha)\Gamma(\beta)}\epsilon^{\alpha-1}(1-\epsilon)^{\beta-1},
\end{eqnarray}
where $\Gamma(\cdot)$ is the Gamma function.
The expectation and variance are
\begin{eqnarray}
\mathcal{E}(\epsilon)=\frac{\alpha}{\alpha+\beta}, ~~\mathcal{D}(\epsilon)=\frac{\alpha\beta}{(\alpha+\beta)^2(\alpha+\beta+1)}.
\end{eqnarray}

\end{appendix}

\bibliographystyle{apsrev4-1}
\bibliography{references}

\begin{thebibliography}{135}%
\makeatletter
\providecommand \@ifxundefined [1]{%
 \@ifx{#1\undefined}
}%
\providecommand \@ifnum [1]{%
 \ifnum #1\expandafter \@firstoftwo
 \else \expandafter \@secondoftwo
 \fi
}%
\providecommand \@ifx [1]{%
 \ifx #1\expandafter \@firstoftwo
 \else \expandafter \@secondoftwo
 \fi
}%
\providecommand \natexlab [1]{#1}%
\providecommand \enquote  [1]{``#1''}%
\providecommand \bibnamefont  [1]{#1}%
\providecommand \bibfnamefont [1]{#1}%
\providecommand \citenamefont [1]{#1}%
\providecommand \href@noop [0]{\@secondoftwo}%
\providecommand \href [0]{\begingroup \@sanitize@url \@href}%
\providecommand \@href[1]{\@@startlink{#1}\@@href}%
\providecommand \@@href[1]{\endgroup#1\@@endlink}%
\providecommand \@sanitize@url [0]{\catcode `\\12\catcode `\$12\catcode
  `\&12\catcode `\#12\catcode `\^12\catcode `\_12\catcode `\%12\relax}%
\providecommand \@@startlink[1]{}%
\providecommand \@@endlink[0]{}%
\providecommand \url  [0]{\begingroup\@sanitize@url \@url }%
\providecommand \@url [1]{\endgroup\@href {#1}{\urlprefix }}%
\providecommand \urlprefix  [0]{URL }%
\providecommand \Eprint [0]{\href }%
\providecommand \doibase [0]{http://dx.doi.org/}%
\providecommand \selectlanguage [0]{\@gobble}%
\providecommand \bibinfo  [0]{\@secondoftwo}%
\providecommand \bibfield  [0]{\@secondoftwo}%
\providecommand \translation [1]{[#1]}%
\providecommand \BibitemOpen [0]{}%
\providecommand \bibitemStop [0]{}%
\providecommand \bibitemNoStop [0]{.\EOS\space}%
\providecommand \EOS [0]{\spacefactor3000\relax}%
\providecommand \BibitemShut  [1]{\csname bibitem#1\endcsname}%
\let\auto@bib@innerbib\@empty
\bibitem [{\citenamefont {Nielsen}\ and\ \citenamefont
  {Chuang}(2000)}]{Nielsen00}%
  \BibitemOpen
  \bibfield  {author} {\bibinfo {author} {\bibfnamefont {M.~A.}\ \bibnamefont
  {Nielsen}}\ and\ \bibinfo {author} {\bibfnamefont {I.~L.}\ \bibnamefont
  {Chuang}},\ }\href@noop {} {\emph {\bibinfo {title} {Quantum {C}omputation
  and {Q}uantum {I}nformation}}}\ (\bibinfo  {publisher} {Cambridge University
  Press, Cambridge},\ \bibinfo {year} {2000})\BibitemShut {NoStop}%
\bibitem [{\citenamefont {Allen}\ and\ \citenamefont
  {Eberly}(1975)}]{osti7365050}%
  \BibitemOpen
  \bibfield  {author} {\bibinfo {author} {\bibfnamefont {L.}~\bibnamefont
  {Allen}}\ and\ \bibinfo {author} {\bibfnamefont {J.~H.}\ \bibnamefont
  {Eberly}},\ }\href@noop {} {\emph {\bibinfo {title} {Optical {R}esonance and
  {T}wo-{L}evel {A}toms}}}\ (\bibinfo  {publisher} {New York: Dover
  Publications},\ \bibinfo {year} {1975})\BibitemShut {NoStop}%
\bibitem [{\citenamefont {Vitanov}\ \emph {et~al.}(2001)\citenamefont
  {Vitanov}, \citenamefont {Halfmann}, \citenamefont {Shore},\ and\
  \citenamefont {Bergmann}}]{physchem.52.1.763}%
  \BibitemOpen
  \bibfield  {author} {\bibinfo {author} {\bibfnamefont {N.~V.}\ \bibnamefont
  {Vitanov}}, \bibinfo {author} {\bibfnamefont {T.}~\bibnamefont {Halfmann}},
  \bibinfo {author} {\bibfnamefont {B.~W.}\ \bibnamefont {Shore}}, \ and\
  \bibinfo {author} {\bibfnamefont {K.}~\bibnamefont {Bergmann}},\ }\bibinfo
  {title} {Laser-induced population transfer by adiabatic passage techniques},\
  \href {\doibase 10.1146/annurev.physchem.52.1.763} {\bibfield  {journal}
  {\bibinfo  {journal} {Annu. Rev. Phys. Chem.}\ }\textbf {\bibinfo {volume}
  {52}},\ \bibinfo {pages} {763} (\bibinfo {year} {2001})}\BibitemShut
  {NoStop}%
\bibitem [{\citenamefont {Liu}\ \emph {et~al.}(2005)\citenamefont {Liu},
  \citenamefont {You}, \citenamefont {Wei}, \citenamefont {Sun},\ and\
  \citenamefont {Nori}}]{PhysRevLett.95.087001}%
  \BibitemOpen
  \bibfield  {author} {\bibinfo {author} {\bibfnamefont {Y.-X.}\ \bibnamefont
  {Liu}}, \bibinfo {author} {\bibfnamefont {J.~Q.}\ \bibnamefont {You}},
  \bibinfo {author} {\bibfnamefont {L.~F.}\ \bibnamefont {Wei}}, \bibinfo
  {author} {\bibfnamefont {C.~P.}\ \bibnamefont {Sun}}, \ and\ \bibinfo
  {author} {\bibfnamefont {F.}~\bibnamefont {Nori}},\ }\bibinfo {title}
  {Optical Selection Rules and Phase-Dependent Adiabatic State Control in a
  Superconducting Quantum Circuit},\ \href {\doibase
  10.1103/PhysRevLett.95.087001} {\bibfield  {journal} {\bibinfo  {journal}
  {Phys. Rev. Lett.}\ }\textbf {\bibinfo {volume} {95}},\ \bibinfo {pages}
  {087001} (\bibinfo {year} {2005})}\BibitemShut {NoStop}%
\bibitem [{\citenamefont {Kr\'al}\ \emph {et~al.}(2007)\citenamefont {Kr\'al},
  \citenamefont {Thanopulos},\ and\ \citenamefont
  {Shapiro}}]{RevModPhys.79.53}%
  \BibitemOpen
  \bibfield  {author} {\bibinfo {author} {\bibfnamefont {P.}~\bibnamefont
  {Kr\'al}}, \bibinfo {author} {\bibfnamefont {I.}~\bibnamefont {Thanopulos}},
  \ and\ \bibinfo {author} {\bibfnamefont {M.}~\bibnamefont {Shapiro}},\
  }\bibinfo {title} {Colloquium: Coherently controlled adiabatic passage},\
  \href {\doibase 10.1103/RevModPhys.79.53} {\bibfield  {journal} {\bibinfo
  {journal} {Rev. Mod. Phys.}\ }\textbf {\bibinfo {volume} {79}},\ \bibinfo
  {pages} {53} (\bibinfo {year} {2007})}\BibitemShut {NoStop}%
\bibitem [{\citenamefont {Vitanov}\ \emph {et~al.}(2017)\citenamefont
  {Vitanov}, \citenamefont {Rangelov}, \citenamefont {Shore},\ and\
  \citenamefont {Bergmann}}]{RevModPhys.89.015006}%
  \BibitemOpen
  \bibfield  {author} {\bibinfo {author} {\bibfnamefont {N.~V.}\ \bibnamefont
  {Vitanov}}, \bibinfo {author} {\bibfnamefont {A.~A.}\ \bibnamefont
  {Rangelov}}, \bibinfo {author} {\bibfnamefont {B.~W.}\ \bibnamefont {Shore}},
  \ and\ \bibinfo {author} {\bibfnamefont {K.}~\bibnamefont {Bergmann}},\
  }\bibinfo {title} {Stimulated {R}aman adiabatic passage in physics,
  chemistry, and beyond},\ \href {\doibase 10.1103/RevModPhys.89.015006}
  {\bibfield  {journal} {\bibinfo  {journal} {Rev. Mod. Phys.}\ }\textbf
  {\bibinfo {volume} {89}},\ \bibinfo {pages} {015006} (\bibinfo {year}
  {2017})}\BibitemShut {NoStop}%
\bibitem [{\citenamefont {Viola}\ \emph {et~al.}(1999)\citenamefont {Viola},
  \citenamefont {Knill},\ and\ \citenamefont {Lloyd}}]{PhysRevLett.82.2417}%
  \BibitemOpen
  \bibfield  {author} {\bibinfo {author} {\bibfnamefont {L.}~\bibnamefont
  {Viola}}, \bibinfo {author} {\bibfnamefont {E.}~\bibnamefont {Knill}}, \ and\
  \bibinfo {author} {\bibfnamefont {S.}~\bibnamefont {Lloyd}},\ }\bibinfo
  {title} {Dynamical Decoupling of Open Quantum Systems},\ \href {\doibase
  10.1103/PhysRevLett.82.2417} {\bibfield  {journal} {\bibinfo  {journal}
  {Phys. Rev. Lett.}\ }\textbf {\bibinfo {volume} {82}},\ \bibinfo {pages}
  {2417} (\bibinfo {year} {1999})}\BibitemShut {NoStop}%
\bibitem [{\citenamefont {Khodjasteh}\ and\ \citenamefont
  {Lidar}(2005)}]{PhysRevLett.95.180501}%
  \BibitemOpen
  \bibfield  {author} {\bibinfo {author} {\bibfnamefont {K.}~\bibnamefont
  {Khodjasteh}}\ and\ \bibinfo {author} {\bibfnamefont {D.~A.}\ \bibnamefont
  {Lidar}},\ }\bibinfo {title} {Fault-Tolerant Quantum Dynamical Decoupling},\
  \href {\doibase 10.1103/PhysRevLett.95.180501} {\bibfield  {journal}
  {\bibinfo  {journal} {Phys. Rev. Lett.}\ }\textbf {\bibinfo {volume} {95}},\
  \bibinfo {pages} {180501} (\bibinfo {year} {2005})}\BibitemShut {NoStop}%
\bibitem [{\citenamefont {Souza}\ \emph {et~al.}(2012)\citenamefont {Souza},
  \citenamefont {{\'{A}}lvarez},\ and\ \citenamefont {Suter}}]{Souza2012}%
  \BibitemOpen
  \bibfield  {author} {\bibinfo {author} {\bibfnamefont {A.~M.}\ \bibnamefont
  {Souza}}, \bibinfo {author} {\bibfnamefont {G.~A.}\ \bibnamefont
  {{\'{A}}lvarez}}, \ and\ \bibinfo {author} {\bibfnamefont {D.}~\bibnamefont
  {Suter}},\ }\bibinfo {title} {Robust dynamical decoupling},\ \href {\doibase
  10.1098/rsta.2011.0355} {\bibfield  {journal} {\bibinfo  {journal} {Phil.
  Trans. R. Soc. A}\ }\textbf {\bibinfo {volume} {370}},\ \bibinfo {pages}
  {4748} (\bibinfo {year} {2012})}\BibitemShut {NoStop}%
\bibitem [{\citenamefont {Ashhab}\ and\ \citenamefont
  {Nori}(2010)}]{PhysRevA.82.062103}%
  \BibitemOpen
  \bibfield  {author} {\bibinfo {author} {\bibfnamefont {S.}~\bibnamefont
  {Ashhab}}\ and\ \bibinfo {author} {\bibfnamefont {F.}~\bibnamefont {Nori}},\
  }\bibinfo {title} {Control-free control: Manipulating a quantum system using
  only a limited set of measurements},\ \href {\doibase
  10.1103/PhysRevA.82.062103} {\bibfield  {journal} {\bibinfo  {journal} {Phys.
  Rev. A}\ }\textbf {\bibinfo {volume} {82}},\ \bibinfo {pages} {062103}
  (\bibinfo {year} {2010})}\BibitemShut {NoStop}%
\bibitem [{\citenamefont {Wiseman}(2011)}]{Wiseman2011}%
  \BibitemOpen
  \bibfield  {author} {\bibinfo {author} {\bibfnamefont {H.~M.}\ \bibnamefont
  {Wiseman}},\ }\bibinfo {title} {Squinting at quantum systems},\ \href
  {\doibase 10.1038/470178a} {\bibfield  {journal} {\bibinfo  {journal}
  {Nature}\ }\textbf {\bibinfo {volume} {470}},\ \bibinfo {pages} {178}
  (\bibinfo {year} {2011})}\BibitemShut {NoStop}%
\bibitem [{\citenamefont {Yang}\ \emph {et~al.}(2015)\citenamefont {Yang},
  \citenamefont {Zhang}, \citenamefont {Wang}, \citenamefont {Liu},
  \citenamefont {Wu}, \citenamefont {Liu}, \citenamefont {Li},\ and\
  \citenamefont {Nori}}]{PhysRevA.92.033812}%
  \BibitemOpen
  \bibfield  {author} {\bibinfo {author} {\bibfnamefont {N.}~\bibnamefont
  {Yang}}, \bibinfo {author} {\bibfnamefont {J.}~\bibnamefont {Zhang}},
  \bibinfo {author} {\bibfnamefont {H.}~\bibnamefont {Wang}}, \bibinfo {author}
  {\bibfnamefont {Y.-X.}\ \bibnamefont {Liu}}, \bibinfo {author} {\bibfnamefont
  {R.-B.}\ \bibnamefont {Wu}}, \bibinfo {author} {\bibfnamefont {L.-Q.}\
  \bibnamefont {Liu}}, \bibinfo {author} {\bibfnamefont {C.-W.}\ \bibnamefont
  {Li}}, \ and\ \bibinfo {author} {\bibfnamefont {F.}~\bibnamefont {Nori}},\
  }\bibinfo {title} {Noise suppression of on-chip mechanical resonators by
  chaotic coherent feedback},\ \href {\doibase 10.1103/PhysRevA.92.033812}
  {\bibfield  {journal} {\bibinfo  {journal} {Phys. Rev. A}\ }\textbf {\bibinfo
  {volume} {92}},\ \bibinfo {pages} {033812} (\bibinfo {year}
  {2015})}\BibitemShut {NoStop}%
\bibitem [{\citenamefont {Zhang}\ \emph {et~al.}(2017)\citenamefont {Zhang},
  \citenamefont {Liu}, \citenamefont {Wu}, \citenamefont {Jacobs},\ and\
  \citenamefont {Nori}}]{Zhang2017pr}%
  \BibitemOpen
  \bibfield  {author} {\bibinfo {author} {\bibfnamefont {J.}~\bibnamefont
  {Zhang}}, \bibinfo {author} {\bibfnamefont {Y.-X.}\ \bibnamefont {Liu}},
  \bibinfo {author} {\bibfnamefont {R.-B.}\ \bibnamefont {Wu}}, \bibinfo
  {author} {\bibfnamefont {K.}~\bibnamefont {Jacobs}}, \ and\ \bibinfo {author}
  {\bibfnamefont {F.}~\bibnamefont {Nori}},\ }\bibinfo {title} {Quantum
  feedback: Theory, experiments, and applications},\ \href {\doibase
  10.1016/j.physrep.2017.02.003} {\bibfield  {journal} {\bibinfo  {journal}
  {Phys. Rep.}\ }\textbf {\bibinfo {volume} {679}},\ \bibinfo {pages} {1}
  (\bibinfo {year} {2017})}\BibitemShut {NoStop}%
\bibitem [{\citenamefont {Daems}\ \emph {et~al.}(2013)\citenamefont {Daems},
  \citenamefont {Ruschhaupt}, \citenamefont {Sugny},\ and\ \citenamefont
  {Gu\'erin}}]{PhysRevLett.111.050404}%
  \BibitemOpen
  \bibfield  {author} {\bibinfo {author} {\bibfnamefont {D.}~\bibnamefont
  {Daems}}, \bibinfo {author} {\bibfnamefont {A.}~\bibnamefont {Ruschhaupt}},
  \bibinfo {author} {\bibfnamefont {D.}~\bibnamefont {Sugny}}, \ and\ \bibinfo
  {author} {\bibfnamefont {S.}~\bibnamefont {Gu\'erin}},\ }\bibinfo {title}
  {Robust Quantum Control by a Single-Shot Shaped Pulse},\ \href {\doibase
  10.1103/PhysRevLett.111.050404} {\bibfield  {journal} {\bibinfo  {journal}
  {Phys. Rev. Lett.}\ }\textbf {\bibinfo {volume} {111}},\ \bibinfo {pages}
  {050404} (\bibinfo {year} {2013})}\BibitemShut {NoStop}%
\bibitem [{\citenamefont {Van-Damme}\ \emph {et~al.}(2017)\citenamefont
  {Van-Damme}, \citenamefont {Schraft}, \citenamefont {Genov}, \citenamefont
  {Sugny}, \citenamefont {Halfmann},\ and\ \citenamefont
  {Gu\'erin}}]{PhysRevA.96.022309}%
  \BibitemOpen
  \bibfield  {author} {\bibinfo {author} {\bibfnamefont {L.}~\bibnamefont
  {Van-Damme}}, \bibinfo {author} {\bibfnamefont {D.}~\bibnamefont {Schraft}},
  \bibinfo {author} {\bibfnamefont {G.~T.}\ \bibnamefont {Genov}}, \bibinfo
  {author} {\bibfnamefont {D.}~\bibnamefont {Sugny}}, \bibinfo {author}
  {\bibfnamefont {T.}~\bibnamefont {Halfmann}}, \ and\ \bibinfo {author}
  {\bibfnamefont {S.}~\bibnamefont {Gu\'erin}},\ }\bibinfo {title} {Robust
  {NOT} gate by single-shot-shaped pulses: Demonstration of the efficiency of
  the pulses in rephasing atomic coherences},\ \href {\doibase
  10.1103/PhysRevA.96.022309} {\bibfield  {journal} {\bibinfo  {journal} {Phys.
  Rev. A}\ }\textbf {\bibinfo {volume} {96}},\ \bibinfo {pages} {022309}
  (\bibinfo {year} {2017})}\BibitemShut {NoStop}%
\bibitem [{\citenamefont {Motzoi}\ \emph {et~al.}(2009)\citenamefont {Motzoi},
  \citenamefont {Gambetta}, \citenamefont {Rebentrost},\ and\ \citenamefont
  {Wilhelm}}]{PhysRevLett.103.110501}%
  \BibitemOpen
  \bibfield  {author} {\bibinfo {author} {\bibfnamefont {F.}~\bibnamefont
  {Motzoi}}, \bibinfo {author} {\bibfnamefont {J.~M.}\ \bibnamefont
  {Gambetta}}, \bibinfo {author} {\bibfnamefont {P.}~\bibnamefont
  {Rebentrost}}, \ and\ \bibinfo {author} {\bibfnamefont {F.~K.}\ \bibnamefont
  {Wilhelm}},\ }\bibinfo {title} {Simple Pulses for Elimination of Leakage in
  Weakly Nonlinear Qubits},\ \href {\doibase 10.1103/PhysRevLett.103.110501}
  {\bibfield  {journal} {\bibinfo  {journal} {Phys. Rev. Lett.}\ }\textbf
  {\bibinfo {volume} {103}},\ \bibinfo {pages} {110501} (\bibinfo {year}
  {2009})}\BibitemShut {NoStop}%
\bibitem [{\citenamefont {Schutjens}\ \emph {et~al.}(2013)\citenamefont
  {Schutjens}, \citenamefont {Dagga}, \citenamefont {Egger},\ and\
  \citenamefont {Wilhelm}}]{PhysRevA.88.052330}%
  \BibitemOpen
  \bibfield  {author} {\bibinfo {author} {\bibfnamefont {R.}~\bibnamefont
  {Schutjens}}, \bibinfo {author} {\bibfnamefont {F.~A.}\ \bibnamefont
  {Dagga}}, \bibinfo {author} {\bibfnamefont {D.~J.}\ \bibnamefont {Egger}}, \
  and\ \bibinfo {author} {\bibfnamefont {F.~K.}\ \bibnamefont {Wilhelm}},\
  }\bibinfo {title} {Single-qubit gates in frequency-crowded transmon
  systems},\ \href {\doibase 10.1103/PhysRevA.88.052330} {\bibfield  {journal}
  {\bibinfo  {journal} {Phys. Rev. A}\ }\textbf {\bibinfo {volume} {88}},\
  \bibinfo {pages} {052330} (\bibinfo {year} {2013})}\BibitemShut {NoStop}%
\bibitem [{\citenamefont {Theis}\ \emph {et~al.}(2016)\citenamefont {Theis},
  \citenamefont {Motzoi}, \citenamefont {Wilhelm},\ and\ \citenamefont
  {Saffman}}]{PhysRevA.94.032306}%
  \BibitemOpen
  \bibfield  {author} {\bibinfo {author} {\bibfnamefont {L.~S.}\ \bibnamefont
  {Theis}}, \bibinfo {author} {\bibfnamefont {F.}~\bibnamefont {Motzoi}},
  \bibinfo {author} {\bibfnamefont {F.~K.}\ \bibnamefont {Wilhelm}}, \ and\
  \bibinfo {author} {\bibfnamefont {M.}~\bibnamefont {Saffman}},\ }\bibinfo
  {title} {High-fidelity Rydberg-blockade entangling gate using shaped,
  analytic pulses},\ \href {\doibase 10.1103/PhysRevA.94.032306} {\bibfield
  {journal} {\bibinfo  {journal} {Phys. Rev. A}\ }\textbf {\bibinfo {volume}
  {94}},\ \bibinfo {pages} {032306} (\bibinfo {year} {2016})}\BibitemShut
  {NoStop}%
\bibitem [{\citenamefont {Chen}\ \emph {et~al.}(2014)\citenamefont {Chen},
  \citenamefont {Dong}, \citenamefont {Long}, \citenamefont {Petersen},\ and\
  \citenamefont {Rabitz}}]{PhysRevA.89.023402}%
  \BibitemOpen
  \bibfield  {author} {\bibinfo {author} {\bibfnamefont {C.}~\bibnamefont
  {Chen}}, \bibinfo {author} {\bibfnamefont {D.}~\bibnamefont {Dong}}, \bibinfo
  {author} {\bibfnamefont {R.}~\bibnamefont {Long}}, \bibinfo {author}
  {\bibfnamefont {I.~R.}\ \bibnamefont {Petersen}}, \ and\ \bibinfo {author}
  {\bibfnamefont {H.~A.}\ \bibnamefont {Rabitz}},\ }\bibinfo {title}
  {Sampling-based learning control of inhomogeneous quantum ensembles},\ \href
  {\doibase 10.1103/PhysRevA.89.023402} {\bibfield  {journal} {\bibinfo
  {journal} {Phys. Rev. A}\ }\textbf {\bibinfo {volume} {89}},\ \bibinfo
  {pages} {023402} (\bibinfo {year} {2014})}\BibitemShut {NoStop}%
\bibitem [{\citenamefont {Dong}\ \emph {et~al.}(2015)\citenamefont {Dong},
  \citenamefont {Mabrok}, \citenamefont {Petersen}, \citenamefont {Qi},
  \citenamefont {Chen},\ and\ \citenamefont {Rabitz}}]{Dong2015ieee}%
  \BibitemOpen
  \bibfield  {author} {\bibinfo {author} {\bibfnamefont {D.}~\bibnamefont
  {Dong}}, \bibinfo {author} {\bibfnamefont {M.~A.}\ \bibnamefont {Mabrok}},
  \bibinfo {author} {\bibfnamefont {I.~R.}\ \bibnamefont {Petersen}}, \bibinfo
  {author} {\bibfnamefont {B.}~\bibnamefont {Qi}}, \bibinfo {author}
  {\bibfnamefont {C.}~\bibnamefont {Chen}}, \ and\ \bibinfo {author}
  {\bibfnamefont {H.}~\bibnamefont {Rabitz}},\ }\bibinfo {title}
  {Sampling-Based Learning Control for Quantum Systems With Uncertainties},\
  \href {\doibase 10.1109/tcst.2015.2404292} {\bibfield  {journal} {\bibinfo
  {journal} {{IEEE} Trans. Control Syst. Technol.}\ }\textbf {\bibinfo {volume}
  {23}},\ \bibinfo {pages} {2155} (\bibinfo {year} {2015})}\BibitemShut
  {NoStop}%
\bibitem [{\citenamefont {Dong}\ \emph {et~al.}(2020)\citenamefont {Dong},
  \citenamefont {Xing}, \citenamefont {Ma}, \citenamefont {Chen}, \citenamefont
  {Liu},\ and\ \citenamefont {Rabitz}}]{Dong2020ieee}%
  \BibitemOpen
  \bibfield  {author} {\bibinfo {author} {\bibfnamefont {D.}~\bibnamefont
  {Dong}}, \bibinfo {author} {\bibfnamefont {X.}~\bibnamefont {Xing}}, \bibinfo
  {author} {\bibfnamefont {H.}~\bibnamefont {Ma}}, \bibinfo {author}
  {\bibfnamefont {C.}~\bibnamefont {Chen}}, \bibinfo {author} {\bibfnamefont
  {Z.}~\bibnamefont {Liu}}, \ and\ \bibinfo {author} {\bibfnamefont
  {H.}~\bibnamefont {Rabitz}},\ }\bibinfo {title} {Learning-Based Quantum
  Robust Control: Algorithm, Applications, and Experiments},\ \href {\doibase
  10.1109/tcyb.2019.2921424} {\bibfield  {journal} {\bibinfo  {journal} {{IEEE}
  Trans. Cybern.}\ }\textbf {\bibinfo {volume} {50}},\ \bibinfo {pages} {3581}
  (\bibinfo {year} {2020})}\BibitemShut {NoStop}%
\bibitem [{\citenamefont {Van~Damme}\ \emph {et~al.}(2017)\citenamefont
  {Van~Damme}, \citenamefont {Ansel}, \citenamefont {Glaser},\ and\
  \citenamefont {Sugny}}]{PhysRevA.95.063403}%
  \BibitemOpen
  \bibfield  {author} {\bibinfo {author} {\bibfnamefont {L.}~\bibnamefont
  {Van~Damme}}, \bibinfo {author} {\bibfnamefont {Q.}~\bibnamefont {Ansel}},
  \bibinfo {author} {\bibfnamefont {S.~J.}\ \bibnamefont {Glaser}}, \ and\
  \bibinfo {author} {\bibfnamefont {D.}~\bibnamefont {Sugny}},\ }\bibinfo
  {title} {Robust optimal control of two-level quantum systems},\ \href
  {\doibase 10.1103/PhysRevA.95.063403} {\bibfield  {journal} {\bibinfo
  {journal} {Phys. Rev. A}\ }\textbf {\bibinfo {volume} {95}},\ \bibinfo
  {pages} {063403} (\bibinfo {year} {2017})}\BibitemShut {NoStop}%
\bibitem [{\citenamefont {Dridi}\ \emph
  {et~al.}(2020{\natexlab{a}})\citenamefont {Dridi}, \citenamefont {Liu},\ and\
  \citenamefont {Gu\'erin}}]{PhysRevLett.125.250403}%
  \BibitemOpen
  \bibfield  {author} {\bibinfo {author} {\bibfnamefont {G.}~\bibnamefont
  {Dridi}}, \bibinfo {author} {\bibfnamefont {K.}~\bibnamefont {Liu}}, \ and\
  \bibinfo {author} {\bibfnamefont {S.}~\bibnamefont {Gu\'erin}},\ }\bibinfo
  {title} {Optimal Robust Quantum Control by Inverse Geometric Optimization},\
  \href {\doibase 10.1103/PhysRevLett.125.250403} {\bibfield  {journal}
  {\bibinfo  {journal} {Phys. Rev. Lett.}\ }\textbf {\bibinfo {volume} {125}},\
  \bibinfo {pages} {250403} (\bibinfo {year} {2020}{\natexlab{a}})}\BibitemShut
  {NoStop}%
\bibitem [{\citenamefont {Sun}\ \emph {et~al.}(2021)\citenamefont {Sun},
  \citenamefont {Yan}, \citenamefont {Su},\ and\ \citenamefont
  {Jia}}]{PhysRevApplied.16.064040}%
  \BibitemOpen
  \bibfield  {author} {\bibinfo {author} {\bibfnamefont {L.-N.}\ \bibnamefont
  {Sun}}, \bibinfo {author} {\bibfnamefont {L.-L.}\ \bibnamefont {Yan}},
  \bibinfo {author} {\bibfnamefont {S.-L.}\ \bibnamefont {Su}}, \ and\ \bibinfo
  {author} {\bibfnamefont {Y.}~\bibnamefont {Jia}},\ }\bibinfo {title}
  {One-Step Implementation of Time-Optimal-Control Three-Qubit Nonadiabatic
  Holonomic Controlled Gates in Rydberg Atoms},\ \href {\doibase
  10.1103/PhysRevApplied.16.064040} {\bibfield  {journal} {\bibinfo  {journal}
  {Phys. Rev. Appl.}\ }\textbf {\bibinfo {volume} {16}},\ \bibinfo {pages}
  {064040} (\bibinfo {year} {2021})}\BibitemShut {NoStop}%
\bibitem [{\citenamefont {Yu}\ \emph {et~al.}(2022)\citenamefont {Yu},
  \citenamefont {Wang}, \citenamefont {Liu}, \citenamefont {Su}, \citenamefont
  {Qian},\ and\ \citenamefont {Zhang}}]{PhysRevApplied.18.034072}%
  \BibitemOpen
  \bibfield  {author} {\bibinfo {author} {\bibfnamefont {D.}~\bibnamefont
  {Yu}}, \bibinfo {author} {\bibfnamefont {H.}~\bibnamefont {Wang}}, \bibinfo
  {author} {\bibfnamefont {J.-M.}\ \bibnamefont {Liu}}, \bibinfo {author}
  {\bibfnamefont {S.-L.}\ \bibnamefont {Su}}, \bibinfo {author} {\bibfnamefont
  {J.}~\bibnamefont {Qian}}, \ and\ \bibinfo {author} {\bibfnamefont
  {W.}~\bibnamefont {Zhang}},\ }\bibinfo {title} {Multiqubit Toffoli Gates and
  Optimal Geometry with Rydberg Atoms},\ \href {\doibase
  10.1103/PhysRevApplied.18.034072} {\bibfield  {journal} {\bibinfo  {journal}
  {Phys. Rev. Appl.}\ }\textbf {\bibinfo {volume} {18}},\ \bibinfo {pages}
  {034072} (\bibinfo {year} {2022})}\BibitemShut {NoStop}%
\bibitem [{\citenamefont {Wei}\ \emph {et~al.}(2022)\citenamefont {Wei},
  \citenamefont {Guo}, \citenamefont {Wang}, \citenamefont {Jia}, \citenamefont
  {Yan}, \citenamefont {Feng},\ and\ \citenamefont {Su}}]{PhysRevA.105.042404}%
  \BibitemOpen
  \bibfield  {author} {\bibinfo {author} {\bibfnamefont {J.-F.}\ \bibnamefont
  {Wei}}, \bibinfo {author} {\bibfnamefont {F.-Q.}\ \bibnamefont {Guo}},
  \bibinfo {author} {\bibfnamefont {D.-Y.}\ \bibnamefont {Wang}}, \bibinfo
  {author} {\bibfnamefont {Y.}~\bibnamefont {Jia}}, \bibinfo {author}
  {\bibfnamefont {L.-L.}\ \bibnamefont {Yan}}, \bibinfo {author} {\bibfnamefont
  {M.}~\bibnamefont {Feng}}, \ and\ \bibinfo {author} {\bibfnamefont {S.-L.}\
  \bibnamefont {Su}},\ }\bibinfo {title} {Fast multiqubit Rydberg geometric
  fan-out gates with optimal control technology},\ \href {\doibase
  10.1103/PhysRevA.105.042404} {\bibfield  {journal} {\bibinfo  {journal}
  {Phys. Rev. A}\ }\textbf {\bibinfo {volume} {105}},\ \bibinfo {pages}
  {042404} (\bibinfo {year} {2022})}\BibitemShut {NoStop}%
\bibitem [{\citenamefont {Zhang}\ and\ \citenamefont
  {Rabitz}(1994)}]{PhysRevA.49.2241}%
  \BibitemOpen
  \bibfield  {author} {\bibinfo {author} {\bibfnamefont {H.}~\bibnamefont
  {Zhang}}\ and\ \bibinfo {author} {\bibfnamefont {H.}~\bibnamefont {Rabitz}},\
  }\bibinfo {title} {Robust optimal control of quantum molecular systems in the
  presence of disturbances and uncertainties},\ \href {\doibase
  10.1103/PhysRevA.49.2241} {\bibfield  {journal} {\bibinfo  {journal} {Phys.
  Rev. A}\ }\textbf {\bibinfo {volume} {49}},\ \bibinfo {pages} {2241}
  (\bibinfo {year} {1994})}\BibitemShut {NoStop}%
\bibitem [{\citenamefont {Van~Damme}\ \emph {et~al.}(2021)\citenamefont
  {Van~Damme}, \citenamefont {Sugny},\ and\ \citenamefont
  {Glaser}}]{PhysRevA.104.042226}%
  \BibitemOpen
  \bibfield  {author} {\bibinfo {author} {\bibfnamefont {L.}~\bibnamefont
  {Van~Damme}}, \bibinfo {author} {\bibfnamefont {D.}~\bibnamefont {Sugny}}, \
  and\ \bibinfo {author} {\bibfnamefont {S.~J.}\ \bibnamefont {Glaser}},\
  }\bibinfo {title} {Application of the small-tip-angle approximation in the
  toggling frame for the design of analytic robust pulses in quantum control},\
  \href {\doibase 10.1103/PhysRevA.104.042226} {\bibfield  {journal} {\bibinfo
  {journal} {Phys. Rev. A}\ }\textbf {\bibinfo {volume} {104}},\ \bibinfo
  {pages} {042226} (\bibinfo {year} {2021})}\BibitemShut {NoStop}%
\bibitem [{\citenamefont {Laforgue}\ \emph {et~al.}(2022)\citenamefont
  {Laforgue}, \citenamefont {Dridi},\ and\ \citenamefont
  {Gu\'erin}}]{PhysRevA.106.052608}%
  \BibitemOpen
  \bibfield  {author} {\bibinfo {author} {\bibfnamefont {X.}~\bibnamefont
  {Laforgue}}, \bibinfo {author} {\bibfnamefont {G.}~\bibnamefont {Dridi}}, \
  and\ \bibinfo {author} {\bibfnamefont {S.}~\bibnamefont {Gu\'erin}},\
  }\bibinfo {title} {Optimal quantum control robust against pulse
  inhomogeneities: Analytic solutions},\ \href {\doibase
  10.1103/PhysRevA.106.052608} {\bibfield  {journal} {\bibinfo  {journal}
  {Phys. Rev. A}\ }\textbf {\bibinfo {volume} {106}},\ \bibinfo {pages}
  {052608} (\bibinfo {year} {2022})}\BibitemShut {NoStop}%
\bibitem [{\citenamefont {Propson}\ \emph {et~al.}(2022)\citenamefont
  {Propson}, \citenamefont {Jackson}, \citenamefont {Koch}, \citenamefont
  {Manchester},\ and\ \citenamefont {Schuster}}]{PhysRevApplied.17.014036}%
  \BibitemOpen
  \bibfield  {author} {\bibinfo {author} {\bibfnamefont {T.}~\bibnamefont
  {Propson}}, \bibinfo {author} {\bibfnamefont {B.~E.}\ \bibnamefont
  {Jackson}}, \bibinfo {author} {\bibfnamefont {J.}~\bibnamefont {Koch}},
  \bibinfo {author} {\bibfnamefont {Z.}~\bibnamefont {Manchester}}, \ and\
  \bibinfo {author} {\bibfnamefont {D.~I.}\ \bibnamefont {Schuster}},\
  }\bibinfo {title} {Robust Quantum Optimal Control with Trajectory
  Optimization},\ \href {\doibase 10.1103/PhysRevApplied.17.014036} {\bibfield
  {journal} {\bibinfo  {journal} {Phys. Rev. Appl.}\ }\textbf {\bibinfo
  {volume} {17}},\ \bibinfo {pages} {014036} (\bibinfo {year}
  {2022})}\BibitemShut {NoStop}%
\bibitem [{\citenamefont {Mohan}\ \emph {et~al.}(2023)\citenamefont {Mohan},
  \citenamefont {de~Keijzer},\ and\ \citenamefont
  {Kokkelmans}}]{PhysRevResearch.5.033052}%
  \BibitemOpen
  \bibfield  {author} {\bibinfo {author} {\bibfnamefont {M.}~\bibnamefont
  {Mohan}}, \bibinfo {author} {\bibfnamefont {R.}~\bibnamefont {de~Keijzer}}, \
  and\ \bibinfo {author} {\bibfnamefont {S.}~\bibnamefont {Kokkelmans}},\
  }\bibinfo {title} {Robust control and optimal Rydberg states for neutral atom
  two-qubit gates},\ \href {\doibase 10.1103/PhysRevResearch.5.033052}
  {\bibfield  {journal} {\bibinfo  {journal} {Phys. Rev. Res.}\ }\textbf
  {\bibinfo {volume} {5}},\ \bibinfo {pages} {033052} (\bibinfo {year}
  {2023})}\BibitemShut {NoStop}%
\bibitem [{\citenamefont {Mohri}\ \emph {et~al.}(2012)\citenamefont {Mohri},
  \citenamefont {Rostamizadeh},\ and\ \citenamefont {Talwalkar}}]{mohri2012}%
  \BibitemOpen
  \bibfield  {author} {\bibinfo {author} {\bibfnamefont {M.}~\bibnamefont
  {Mohri}}, \bibinfo {author} {\bibfnamefont {A.}~\bibnamefont {Rostamizadeh}},
  \ and\ \bibinfo {author} {\bibfnamefont {A.}~\bibnamefont {Talwalkar}},\
  }\href@noop {} {\emph {\bibinfo {title} {Foundations of {M}achine
  {L}earning}}}\ (\bibinfo  {publisher} {MIT Press, Cambridge},\ \bibinfo
  {year} {2012})\BibitemShut {NoStop}%
\bibitem [{\citenamefont {Bonaccorso}(2017)}]{Bonaccorso2017}%
  \BibitemOpen
  \bibfield  {author} {\bibinfo {author} {\bibfnamefont {G.}~\bibnamefont
  {Bonaccorso}},\ }\href@noop {} {\emph {\bibinfo {title} {Machine learning
  algorithms}}}\ (\bibinfo  {publisher} {Packt Publishing, Birmingham},\
  \bibinfo {year} {2017})\BibitemShut {NoStop}%
\bibitem [{\citenamefont {Biamonte}\ \emph {et~al.}(2017)\citenamefont
  {Biamonte}, \citenamefont {Wittek}, \citenamefont {Pancotti}, \citenamefont
  {Rebentrost}, \citenamefont {Wiebe},\ and\ \citenamefont
  {Lloyd}}]{Biamonte2017}%
  \BibitemOpen
  \bibfield  {author} {\bibinfo {author} {\bibfnamefont {J.}~\bibnamefont
  {Biamonte}}, \bibinfo {author} {\bibfnamefont {P.}~\bibnamefont {Wittek}},
  \bibinfo {author} {\bibfnamefont {N.}~\bibnamefont {Pancotti}}, \bibinfo
  {author} {\bibfnamefont {P.}~\bibnamefont {Rebentrost}}, \bibinfo {author}
  {\bibfnamefont {N.}~\bibnamefont {Wiebe}}, \ and\ \bibinfo {author}
  {\bibfnamefont {S.}~\bibnamefont {Lloyd}},\ }\bibinfo {title} {Quantum
  machine learning},\ \href {\doibase 10.1038/nature23474} {\bibfield
  {journal} {\bibinfo  {journal} {Nature}\ }\textbf {\bibinfo {volume} {549}},\
  \bibinfo {pages} {195} (\bibinfo {year} {2017})}\BibitemShut {NoStop}%
\bibitem [{\citenamefont {Dunjko}\ and\ \citenamefont
  {Briegel}(2018)}]{Dunjko2018}%
  \BibitemOpen
  \bibfield  {author} {\bibinfo {author} {\bibfnamefont {V.}~\bibnamefont
  {Dunjko}}\ and\ \bibinfo {author} {\bibfnamefont {H.~J.}\ \bibnamefont
  {Briegel}},\ }\bibinfo {title} {Machine learning {\&} artificial intelligence
  in the quantum domain: a review of recent progress},\ \href {\doibase
  10.1088/1361-6633/aab406} {\bibfield  {journal} {\bibinfo  {journal} {Rep.
  Prog. Phys.}\ }\textbf {\bibinfo {volume} {81}},\ \bibinfo {pages} {074001}
  (\bibinfo {year} {2018})}\BibitemShut {NoStop}%
\bibitem [{\citenamefont {Carleo}\ \emph {et~al.}(2019)\citenamefont {Carleo},
  \citenamefont {Cirac}, \citenamefont {Cranmer}, \citenamefont {Daudet},
  \citenamefont {Schuld}, \citenamefont {Tishby}, \citenamefont
  {Vogt-Maranto},\ and\ \citenamefont {Zdeborov\'a}}]{RevModPhys.91.045002}%
  \BibitemOpen
  \bibfield  {author} {\bibinfo {author} {\bibfnamefont {G.}~\bibnamefont
  {Carleo}}, \bibinfo {author} {\bibfnamefont {I.}~\bibnamefont {Cirac}},
  \bibinfo {author} {\bibfnamefont {K.}~\bibnamefont {Cranmer}}, \bibinfo
  {author} {\bibfnamefont {L.}~\bibnamefont {Daudet}}, \bibinfo {author}
  {\bibfnamefont {M.}~\bibnamefont {Schuld}}, \bibinfo {author} {\bibfnamefont
  {N.}~\bibnamefont {Tishby}}, \bibinfo {author} {\bibfnamefont
  {L.}~\bibnamefont {Vogt-Maranto}}, \ and\ \bibinfo {author} {\bibfnamefont
  {L.}~\bibnamefont {Zdeborov\'a}},\ }\bibinfo {title} {Machine learning and
  the physical sciences},\ \href {\doibase 10.1103/RevModPhys.91.045002}
  {\bibfield  {journal} {\bibinfo  {journal} {Rev. Mod. Phys.}\ }\textbf
  {\bibinfo {volume} {91}},\ \bibinfo {pages} {045002} (\bibinfo {year}
  {2019})}\BibitemShut {NoStop}%
\bibitem [{\citenamefont {Arrachea}(2023)}]{Arrachea2023}%
  \BibitemOpen
  \bibfield  {author} {\bibinfo {author} {\bibfnamefont {L.}~\bibnamefont
  {Arrachea}},\ }\bibinfo {title} {Energy dynamics, heat production and
  heat{\textendash}work conversion with qubits: toward the development of
  quantum machines},\ \href {\doibase 10.1088/1361-6633/acb06b} {\bibfield
  {journal} {\bibinfo  {journal} {Rep. Prog. Phys.}\ }\textbf {\bibinfo
  {volume} {86}},\ \bibinfo {pages} {036501} (\bibinfo {year}
  {2023})}\BibitemShut {NoStop}%
\bibitem [{\citenamefont {Krenn}\ \emph {et~al.}(2023)\citenamefont {Krenn},
  \citenamefont {Landgraf}, \citenamefont {Foesel},\ and\ \citenamefont
  {Marquardt}}]{PhysRevA.107.010101}%
  \BibitemOpen
  \bibfield  {author} {\bibinfo {author} {\bibfnamefont {M.}~\bibnamefont
  {Krenn}}, \bibinfo {author} {\bibfnamefont {J.}~\bibnamefont {Landgraf}},
  \bibinfo {author} {\bibfnamefont {T.}~\bibnamefont {Foesel}}, \ and\ \bibinfo
  {author} {\bibfnamefont {F.}~\bibnamefont {Marquardt}},\ }\bibinfo {title}
  {Artificial intelligence and machine learning for quantum technologies},\
  \href {\doibase 10.1103/PhysRevA.107.010101} {\bibfield  {journal} {\bibinfo
  {journal} {Phys. Rev. A}\ }\textbf {\bibinfo {volume} {107}},\ \bibinfo
  {pages} {010101} (\bibinfo {year} {2023})}\BibitemShut {NoStop}%
\bibitem [{\citenamefont {Zhang}\ and\ \citenamefont
  {Wang}(2019)}]{PhysRevA.99.042316}%
  \BibitemOpen
  \bibfield  {author} {\bibinfo {author} {\bibfnamefont {C.}~\bibnamefont
  {Zhang}}\ and\ \bibinfo {author} {\bibfnamefont {X.}~\bibnamefont {Wang}},\
  }\bibinfo {title} {Spin-qubit noise spectroscopy from randomized benchmarking
  by supervised learning},\ \href {\doibase 10.1103/PhysRevA.99.042316}
  {\bibfield  {journal} {\bibinfo  {journal} {Phys. Rev. A}\ }\textbf {\bibinfo
  {volume} {99}},\ \bibinfo {pages} {042316} (\bibinfo {year}
  {2019})}\BibitemShut {NoStop}%
\bibitem [{\citenamefont {Zeng}\ \emph {et~al.}(2020)\citenamefont {Zeng},
  \citenamefont {Shen}, \citenamefont {Hou}, \citenamefont {Gebremariam},\ and\
  \citenamefont {Li}}]{Zeng2020}%
  \BibitemOpen
  \bibfield  {author} {\bibinfo {author} {\bibfnamefont {Y.}~\bibnamefont
  {Zeng}}, \bibinfo {author} {\bibfnamefont {J.}~\bibnamefont {Shen}}, \bibinfo
  {author} {\bibfnamefont {S.}~\bibnamefont {Hou}}, \bibinfo {author}
  {\bibfnamefont {T.}~\bibnamefont {Gebremariam}}, \ and\ \bibinfo {author}
  {\bibfnamefont {C.}~\bibnamefont {Li}},\ }\bibinfo {title} {Quantum control
  based on machine learning in an open quantum system},\ \href {\doibase
  10.1016/j.physleta.2020.126886} {\bibfield  {journal} {\bibinfo  {journal}
  {Phys. Lett. A}\ }\textbf {\bibinfo {volume} {384}},\ \bibinfo {pages}
  {126886} (\bibinfo {year} {2020})}\BibitemShut {NoStop}%
\bibitem [{\citenamefont {Che}\ \emph {et~al.}(2020)\citenamefont {Che},
  \citenamefont {Gneiting}, \citenamefont {Liu},\ and\ \citenamefont
  {Nori}}]{PhysRevB.102.134213}%
  \BibitemOpen
  \bibfield  {author} {\bibinfo {author} {\bibfnamefont {Y.}~\bibnamefont
  {Che}}, \bibinfo {author} {\bibfnamefont {C.}~\bibnamefont {Gneiting}},
  \bibinfo {author} {\bibfnamefont {T.}~\bibnamefont {Liu}}, \ and\ \bibinfo
  {author} {\bibfnamefont {F.}~\bibnamefont {Nori}},\ }\bibinfo {title}
  {Topological quantum phase transitions retrieved through unsupervised machine
  learning},\ \href {\doibase 10.1103/PhysRevB.102.134213} {\bibfield
  {journal} {\bibinfo  {journal} {Phys. Rev. B}\ }\textbf {\bibinfo {volume}
  {102}},\ \bibinfo {pages} {134213} (\bibinfo {year} {2020})}\BibitemShut
  {NoStop}%
\bibitem [{\citenamefont {Ding}\ \emph
  {et~al.}(2021{\natexlab{a}})\citenamefont {Ding}, \citenamefont {Ban},
  \citenamefont {Mart\'{\i}n-Guerrero}, \citenamefont {Solano}, \citenamefont
  {Casanova},\ and\ \citenamefont {Chen}}]{PhysRevA.103.L040401}%
  \BibitemOpen
  \bibfield  {author} {\bibinfo {author} {\bibfnamefont {Y.}~\bibnamefont
  {Ding}}, \bibinfo {author} {\bibfnamefont {Y.}~\bibnamefont {Ban}}, \bibinfo
  {author} {\bibfnamefont {J.~D.}\ \bibnamefont {Mart\'{\i}n-Guerrero}},
  \bibinfo {author} {\bibfnamefont {E.}~\bibnamefont {Solano}}, \bibinfo
  {author} {\bibfnamefont {J.}~\bibnamefont {Casanova}}, \ and\ \bibinfo
  {author} {\bibfnamefont {X.}~\bibnamefont {Chen}},\ }\bibinfo {title}
  {Breaking adiabatic quantum control with deep learning},\ \href {\doibase
  10.1103/PhysRevA.103.L040401} {\bibfield  {journal} {\bibinfo  {journal}
  {Phys. Rev. A}\ }\textbf {\bibinfo {volume} {103}},\ \bibinfo {pages}
  {L040401} (\bibinfo {year} {2021}{\natexlab{a}})}\BibitemShut {NoStop}%
\bibitem [{\citenamefont {Ai}\ \emph {et~al.}(2022)\citenamefont {Ai},
  \citenamefont {Ding}, \citenamefont {Ban}, \citenamefont
  {Mart{\'{\i}}n-Guerrero}, \citenamefont {Casanova}, \citenamefont {Cui},
  \citenamefont {Huang}, \citenamefont {Chen}, \citenamefont {Li},\ and\
  \citenamefont {Guo}}]{Ai2022}%
  \BibitemOpen
  \bibfield  {author} {\bibinfo {author} {\bibfnamefont {M.-Z.}\ \bibnamefont
  {Ai}}, \bibinfo {author} {\bibfnamefont {Y.}~\bibnamefont {Ding}}, \bibinfo
  {author} {\bibfnamefont {Y.}~\bibnamefont {Ban}}, \bibinfo {author}
  {\bibfnamefont {J.~D.}\ \bibnamefont {Mart{\'{\i}}n-Guerrero}}, \bibinfo
  {author} {\bibfnamefont {J.}~\bibnamefont {Casanova}}, \bibinfo {author}
  {\bibfnamefont {J.-M.}\ \bibnamefont {Cui}}, \bibinfo {author} {\bibfnamefont
  {Y.-F.}\ \bibnamefont {Huang}}, \bibinfo {author} {\bibfnamefont
  {X.}~\bibnamefont {Chen}}, \bibinfo {author} {\bibfnamefont {C.-F.}\
  \bibnamefont {Li}}, \ and\ \bibinfo {author} {\bibfnamefont {G.-C.}\
  \bibnamefont {Guo}},\ }\bibinfo {title} {Experimentally realizing efficient
  quantum control with reinforcement learning},\ \href {\doibase
  10.1007/s11433-021-1841-2} {\bibfield  {journal} {\bibinfo  {journal} {Sci.
  China Phys. Mech. Astron.}\ }\textbf {\bibinfo {volume} {65}},\ \bibinfo
  {pages} {250312} (\bibinfo {year} {2022})}\BibitemShut {NoStop}%
\bibitem [{\citenamefont {Ahmed}\ \emph
  {et~al.}(2021{\natexlab{a}})\citenamefont {Ahmed}, \citenamefont {S\'anchez
  Mu\~noz}, \citenamefont {Nori},\ and\ \citenamefont
  {Kockum}}]{PhysRevLett.127.140502}%
  \BibitemOpen
  \bibfield  {author} {\bibinfo {author} {\bibfnamefont {S.}~\bibnamefont
  {Ahmed}}, \bibinfo {author} {\bibfnamefont {C.}~\bibnamefont {S\'anchez
  Mu\~noz}}, \bibinfo {author} {\bibfnamefont {F.}~\bibnamefont {Nori}}, \ and\
  \bibinfo {author} {\bibfnamefont {A.~F.}\ \bibnamefont {Kockum}},\ }\bibinfo
  {title} {Quantum State Tomography with Conditional Generative Adversarial
  Networks},\ \href {\doibase 10.1103/PhysRevLett.127.140502} {\bibfield
  {journal} {\bibinfo  {journal} {Phys. Rev. Lett.}\ }\textbf {\bibinfo
  {volume} {127}},\ \bibinfo {pages} {140502} (\bibinfo {year}
  {2021}{\natexlab{a}})}\BibitemShut {NoStop}%
\bibitem [{\citenamefont {Huang}\ \emph
  {et~al.}(2022{\natexlab{a}})\citenamefont {Huang}, \citenamefont {Ban},
  \citenamefont {Sherman},\ and\ \citenamefont
  {Chen}}]{PhysRevApplied.17.024040}%
  \BibitemOpen
  \bibfield  {author} {\bibinfo {author} {\bibfnamefont {T.}~\bibnamefont
  {Huang}}, \bibinfo {author} {\bibfnamefont {Y.}~\bibnamefont {Ban}}, \bibinfo
  {author} {\bibfnamefont {E.~Y.}\ \bibnamefont {Sherman}}, \ and\ \bibinfo
  {author} {\bibfnamefont {X.}~\bibnamefont {Chen}},\ }\bibinfo {title}
  {Machine-Learning-Assisted Quantum Control in a Random Environment},\ \href
  {\doibase 10.1103/PhysRevApplied.17.024040} {\bibfield  {journal} {\bibinfo
  {journal} {Phys. Rev. Appl.}\ }\textbf {\bibinfo {volume} {17}},\ \bibinfo
  {pages} {024040} (\bibinfo {year} {2022}{\natexlab{a}})}\BibitemShut
  {NoStop}%
\bibitem [{\citenamefont {Wu}\ \emph {et~al.}(2020)\citenamefont {Wu},
  \citenamefont {Cao}, \citenamefont {Xie},\ and\ \citenamefont
  {Liu}}]{PhysRevApplied.14.064020}%
  \BibitemOpen
  \bibfield  {author} {\bibinfo {author} {\bibfnamefont {R.-B.}\ \bibnamefont
  {Wu}}, \bibinfo {author} {\bibfnamefont {X.}~\bibnamefont {Cao}}, \bibinfo
  {author} {\bibfnamefont {P.}~\bibnamefont {Xie}}, \ and\ \bibinfo {author}
  {\bibfnamefont {Y.-X.}\ \bibnamefont {Liu}},\ }\bibinfo {title} {End-To-End
  Quantum Machine Learning Implemented with Controlled Quantum Dynamics},\
  \href {\doibase 10.1103/PhysRevApplied.14.064020} {\bibfield  {journal}
  {\bibinfo  {journal} {Phys. Rev. Appl.}\ }\textbf {\bibinfo {volume} {14}},\
  \bibinfo {pages} {064020} (\bibinfo {year} {2020})}\BibitemShut {NoStop}%
\bibitem [{\citenamefont {Wise}\ \emph {et~al.}(2021)\citenamefont {Wise},
  \citenamefont {Morton},\ and\ \citenamefont {Dhomkar}}]{PRXQuantum.2.010316}%
  \BibitemOpen
  \bibfield  {author} {\bibinfo {author} {\bibfnamefont {D.~F.}\ \bibnamefont
  {Wise}}, \bibinfo {author} {\bibfnamefont {J.~J.}\ \bibnamefont {Morton}}, \
  and\ \bibinfo {author} {\bibfnamefont {S.}~\bibnamefont {Dhomkar}},\
  }\bibinfo {title} {Using Deep Learning to Understand and Mitigate the Qubit
  Noise Environment},\ \href {\doibase 10.1103/PRXQuantum.2.010316} {\bibfield
  {journal} {\bibinfo  {journal} {PRX Quantum}\ }\textbf {\bibinfo {volume}
  {2}},\ \bibinfo {pages} {010316} (\bibinfo {year} {2021})}\BibitemShut
  {NoStop}%
\bibitem [{\citenamefont {Ma}\ and\ \citenamefont {Yung}(2018)}]{Ma2018npj}%
  \BibitemOpen
  \bibfield  {author} {\bibinfo {author} {\bibfnamefont {Y.-C.}\ \bibnamefont
  {Ma}}\ and\ \bibinfo {author} {\bibfnamefont {M.-H.}\ \bibnamefont {Yung}},\
  }\bibinfo {title} {Transforming Bell's inequalities into state classifiers
  with machine learning},\ \href {\doibase 10.1038/s41534-018-0081-3}
  {\bibfield  {journal} {\bibinfo  {journal} {npj Quantum Inf.}\ }\textbf
  {\bibinfo {volume} {4}},\ \bibinfo {pages} {34} (\bibinfo {year}
  {2018})}\BibitemShut {NoStop}%
\bibitem [{\citenamefont {Glasser}\ \emph {et~al.}(2018)\citenamefont
  {Glasser}, \citenamefont {Pancotti}, \citenamefont {August}, \citenamefont
  {Rodriguez},\ and\ \citenamefont {Cirac}}]{PhysRevX.8.011006}%
  \BibitemOpen
  \bibfield  {author} {\bibinfo {author} {\bibfnamefont {I.}~\bibnamefont
  {Glasser}}, \bibinfo {author} {\bibfnamefont {N.}~\bibnamefont {Pancotti}},
  \bibinfo {author} {\bibfnamefont {M.}~\bibnamefont {August}}, \bibinfo
  {author} {\bibfnamefont {I.~D.}\ \bibnamefont {Rodriguez}}, \ and\ \bibinfo
  {author} {\bibfnamefont {J.~I.}\ \bibnamefont {Cirac}},\ }\bibinfo {title}
  {Neural-Network Quantum States, String-Bond States, and Chiral Topological
  States},\ \href {\doibase 10.1103/PhysRevX.8.011006} {\bibfield  {journal}
  {\bibinfo  {journal} {Phys. Rev. X}\ }\textbf {\bibinfo {volume} {8}},\
  \bibinfo {pages} {011006} (\bibinfo {year} {2018})}\BibitemShut {NoStop}%
\bibitem [{\citenamefont {Gao}\ \emph {et~al.}(2018)\citenamefont {Gao},
  \citenamefont {Qiao}, \citenamefont {Jiao}, \citenamefont {Ma}, \citenamefont
  {Hu}, \citenamefont {Ren}, \citenamefont {Yang}, \citenamefont {Tang},
  \citenamefont {Yung},\ and\ \citenamefont {Jin}}]{PhysRevLett.120.240501}%
  \BibitemOpen
  \bibfield  {author} {\bibinfo {author} {\bibfnamefont {J.}~\bibnamefont
  {Gao}}, \bibinfo {author} {\bibfnamefont {L.-F.}\ \bibnamefont {Qiao}},
  \bibinfo {author} {\bibfnamefont {Z.-Q.}\ \bibnamefont {Jiao}}, \bibinfo
  {author} {\bibfnamefont {Y.-C.}\ \bibnamefont {Ma}}, \bibinfo {author}
  {\bibfnamefont {C.-Q.}\ \bibnamefont {Hu}}, \bibinfo {author} {\bibfnamefont
  {R.-J.}\ \bibnamefont {Ren}}, \bibinfo {author} {\bibfnamefont {A.-L.}\
  \bibnamefont {Yang}}, \bibinfo {author} {\bibfnamefont {H.}~\bibnamefont
  {Tang}}, \bibinfo {author} {\bibfnamefont {M.-H.}\ \bibnamefont {Yung}}, \
  and\ \bibinfo {author} {\bibfnamefont {X.-M.}\ \bibnamefont {Jin}},\
  }\bibinfo {title} {Experimental Machine Learning of Quantum States},\ \href
  {\doibase 10.1103/PhysRevLett.120.240501} {\bibfield  {journal} {\bibinfo
  {journal} {Phys. Rev. Lett.}\ }\textbf {\bibinfo {volume} {120}},\ \bibinfo
  {pages} {240501} (\bibinfo {year} {2018})}\BibitemShut {NoStop}%
\bibitem [{\citenamefont {Harney}\ \emph {et~al.}(2020)\citenamefont {Harney},
  \citenamefont {Pirandola}, \citenamefont {Ferraro},\ and\ \citenamefont
  {Paternostro}}]{Harney2020}%
  \BibitemOpen
  \bibfield  {author} {\bibinfo {author} {\bibfnamefont {C.}~\bibnamefont
  {Harney}}, \bibinfo {author} {\bibfnamefont {S.}~\bibnamefont {Pirandola}},
  \bibinfo {author} {\bibfnamefont {A.}~\bibnamefont {Ferraro}}, \ and\
  \bibinfo {author} {\bibfnamefont {M.}~\bibnamefont {Paternostro}},\ }\bibinfo
  {title} {Entanglement classification via neural network quantum states},\
  \href {\doibase 10.1088/1367-2630/ab783d} {\bibfield  {journal} {\bibinfo
  {journal} {New J. Phys.}\ }\textbf {\bibinfo {volume} {22}},\ \bibinfo
  {pages} {045001} (\bibinfo {year} {2020})}\BibitemShut {NoStop}%
\bibitem [{\citenamefont {Ahmed}\ \emph
  {et~al.}(2021{\natexlab{b}})\citenamefont {Ahmed}, \citenamefont {S\'anchez
  Mu\~noz}, \citenamefont {Nori},\ and\ \citenamefont
  {Kockum}}]{PhysRevResearch.3.033278}%
  \BibitemOpen
  \bibfield  {author} {\bibinfo {author} {\bibfnamefont {S.}~\bibnamefont
  {Ahmed}}, \bibinfo {author} {\bibfnamefont {C.}~\bibnamefont {S\'anchez
  Mu\~noz}}, \bibinfo {author} {\bibfnamefont {F.}~\bibnamefont {Nori}}, \ and\
  \bibinfo {author} {\bibfnamefont {A.~F.}\ \bibnamefont {Kockum}},\ }\bibinfo
  {title} {Classification and reconstruction of optical quantum states with
  deep neural networks},\ \href {\doibase 10.1103/PhysRevResearch.3.033278}
  {\bibfield  {journal} {\bibinfo  {journal} {Phys. Rev. Res.}\ }\textbf
  {\bibinfo {volume} {3}},\ \bibinfo {pages} {033278} (\bibinfo {year}
  {2021}{\natexlab{b}})}\BibitemShut {NoStop}%
\bibitem [{\citenamefont {Vintskevich}\ \emph {et~al.}(2023)\citenamefont
  {Vintskevich}, \citenamefont {Bao}, \citenamefont {Nomerotski}, \citenamefont
  {Stankus},\ and\ \citenamefont {Grigoriev}}]{PhysRevA.107.032421}%
  \BibitemOpen
  \bibfield  {author} {\bibinfo {author} {\bibfnamefont {S.~V.}\ \bibnamefont
  {Vintskevich}}, \bibinfo {author} {\bibfnamefont {N.}~\bibnamefont {Bao}},
  \bibinfo {author} {\bibfnamefont {A.}~\bibnamefont {Nomerotski}}, \bibinfo
  {author} {\bibfnamefont {P.}~\bibnamefont {Stankus}}, \ and\ \bibinfo
  {author} {\bibfnamefont {D.~A.}\ \bibnamefont {Grigoriev}},\ }\bibinfo
  {title} {Classification of four-qubit entangled states via machine
  learning},\ \href {\doibase 10.1103/PhysRevA.107.032421} {\bibfield
  {journal} {\bibinfo  {journal} {Phys. Rev. A}\ }\textbf {\bibinfo {volume}
  {107}},\ \bibinfo {pages} {032421} (\bibinfo {year} {2023})}\BibitemShut
  {NoStop}%
\bibitem [{\citenamefont {Zahedinejad}\ \emph {et~al.}(2016)\citenamefont
  {Zahedinejad}, \citenamefont {Ghosh},\ and\ \citenamefont
  {Sanders}}]{PhysRevApplied.6.054005}%
  \BibitemOpen
  \bibfield  {author} {\bibinfo {author} {\bibfnamefont {E.}~\bibnamefont
  {Zahedinejad}}, \bibinfo {author} {\bibfnamefont {J.}~\bibnamefont {Ghosh}},
  \ and\ \bibinfo {author} {\bibfnamefont {B.~C.}\ \bibnamefont {Sanders}},\
  }\bibinfo {title} {Designing High-Fidelity Single-Shot Three-Qubit Gates: A
  Machine-Learning Approach},\ \href {\doibase 10.1103/PhysRevApplied.6.054005}
  {\bibfield  {journal} {\bibinfo  {journal} {Phys. Rev. Appl.}\ }\textbf
  {\bibinfo {volume} {6}},\ \bibinfo {pages} {054005} (\bibinfo {year}
  {2016})}\BibitemShut {NoStop}%
\bibitem [{\citenamefont {Yang}\ \emph {et~al.}(2018)\citenamefont {Yang},
  \citenamefont {Yung},\ and\ \citenamefont {Wang}}]{PhysRevA.97.042324}%
  \BibitemOpen
  \bibfield  {author} {\bibinfo {author} {\bibfnamefont {X.-C.}\ \bibnamefont
  {Yang}}, \bibinfo {author} {\bibfnamefont {M.-H.}\ \bibnamefont {Yung}}, \
  and\ \bibinfo {author} {\bibfnamefont {X.}~\bibnamefont {Wang}},\ }\bibinfo
  {title} {Neural-network-designed pulse sequences for robust control of
  singlet-triplet qubits},\ \href {\doibase 10.1103/PhysRevA.97.042324}
  {\bibfield  {journal} {\bibinfo  {journal} {Phys. Rev. A}\ }\textbf {\bibinfo
  {volume} {97}},\ \bibinfo {pages} {042324} (\bibinfo {year}
  {2018})}\BibitemShut {NoStop}%
\bibitem [{\citenamefont {Sabapathy}\ \emph {et~al.}(2019)\citenamefont
  {Sabapathy}, \citenamefont {Qi}, \citenamefont {Izaac},\ and\ \citenamefont
  {Weedbrook}}]{PhysRevA.100.012326}%
  \BibitemOpen
  \bibfield  {author} {\bibinfo {author} {\bibfnamefont {K.~K.}\ \bibnamefont
  {Sabapathy}}, \bibinfo {author} {\bibfnamefont {H.}~\bibnamefont {Qi}},
  \bibinfo {author} {\bibfnamefont {J.}~\bibnamefont {Izaac}}, \ and\ \bibinfo
  {author} {\bibfnamefont {C.}~\bibnamefont {Weedbrook}},\ }\bibinfo {title}
  {Production of photonic universal quantum gates enhanced by machine
  learning},\ \href {\doibase 10.1103/PhysRevA.100.012326} {\bibfield
  {journal} {\bibinfo  {journal} {Phys. Rev. A}\ }\textbf {\bibinfo {volume}
  {100}},\ \bibinfo {pages} {012326} (\bibinfo {year} {2019})}\BibitemShut
  {NoStop}%
\bibitem [{\citenamefont {Spiteri}\ \emph {et~al.}(2018)\citenamefont
  {Spiteri}, \citenamefont {Schmidt}, \citenamefont {Ghosh}, \citenamefont
  {Zahedinejad},\ and\ \citenamefont {Sanders}}]{Spiteri2018}%
  \BibitemOpen
  \bibfield  {author} {\bibinfo {author} {\bibfnamefont {R.~J.}\ \bibnamefont
  {Spiteri}}, \bibinfo {author} {\bibfnamefont {M.}~\bibnamefont {Schmidt}},
  \bibinfo {author} {\bibfnamefont {J.}~\bibnamefont {Ghosh}}, \bibinfo
  {author} {\bibfnamefont {E.}~\bibnamefont {Zahedinejad}}, \ and\ \bibinfo
  {author} {\bibfnamefont {B.~C.}\ \bibnamefont {Sanders}},\ }\bibinfo {title}
  {Quantum control for high-fidelity multi-qubit gates},\ \href {\doibase
  10.1088/1367-2630/aae79a} {\bibfield  {journal} {\bibinfo  {journal} {New J.
  Phys.}\ }\textbf {\bibinfo {volume} {20}},\ \bibinfo {pages} {113009}
  (\bibinfo {year} {2018})}\BibitemShut {NoStop}%
\bibitem [{\citenamefont {Arrazola}\ \emph {et~al.}(2019)\citenamefont
  {Arrazola}, \citenamefont {Bromley}, \citenamefont {Izaac}, \citenamefont
  {Myers}, \citenamefont {Br{\'{a}}dler},\ and\ \citenamefont
  {Killoran}}]{Arrazola2019}%
  \BibitemOpen
  \bibfield  {author} {\bibinfo {author} {\bibfnamefont {J.~M.}\ \bibnamefont
  {Arrazola}}, \bibinfo {author} {\bibfnamefont {T.~R.}\ \bibnamefont
  {Bromley}}, \bibinfo {author} {\bibfnamefont {J.}~\bibnamefont {Izaac}},
  \bibinfo {author} {\bibfnamefont {C.~R.}\ \bibnamefont {Myers}}, \bibinfo
  {author} {\bibfnamefont {K.}~\bibnamefont {Br{\'{a}}dler}}, \ and\ \bibinfo
  {author} {\bibfnamefont {N.}~\bibnamefont {Killoran}},\ }\bibinfo {title}
  {Machine learning method for state preparation and gate synthesis on photonic
  quantum computers},\ \href {\doibase 10.1088/2058-9565/aaf59e} {\bibfield
  {journal} {\bibinfo  {journal} {Quantum Sci. Technol.}\ }\textbf {\bibinfo
  {volume} {4}},\ \bibinfo {pages} {024004} (\bibinfo {year}
  {2019})}\BibitemShut {NoStop}%
\bibitem [{\citenamefont {Gupta}\ and\ \citenamefont
  {Biercuk}(2018)}]{PhysRevApplied.9.064042}%
  \BibitemOpen
  \bibfield  {author} {\bibinfo {author} {\bibfnamefont {R.~S.}\ \bibnamefont
  {Gupta}}\ and\ \bibinfo {author} {\bibfnamefont {M.~J.}\ \bibnamefont
  {Biercuk}},\ }\bibinfo {title} {Machine Learning for Predictive Estimation of
  Qubit Dynamics Subject to Dephasing},\ \href {\doibase
  10.1103/PhysRevApplied.9.064042} {\bibfield  {journal} {\bibinfo  {journal}
  {Phys. Rev. Appl.}\ }\textbf {\bibinfo {volume} {9}},\ \bibinfo {pages}
  {064042} (\bibinfo {year} {2018})}\BibitemShut {NoStop}%
\bibitem [{\citenamefont {Baum}\ \emph {et~al.}(2021)\citenamefont {Baum},
  \citenamefont {Amico}, \citenamefont {Howell}, \citenamefont {Hush},
  \citenamefont {Liuzzi}, \citenamefont {Mundada}, \citenamefont {Merkh},
  \citenamefont {Carvalho},\ and\ \citenamefont
  {Biercuk}}]{PRXQuantum.2.040324}%
  \BibitemOpen
  \bibfield  {author} {\bibinfo {author} {\bibfnamefont {Y.}~\bibnamefont
  {Baum}}, \bibinfo {author} {\bibfnamefont {M.}~\bibnamefont {Amico}},
  \bibinfo {author} {\bibfnamefont {S.}~\bibnamefont {Howell}}, \bibinfo
  {author} {\bibfnamefont {M.}~\bibnamefont {Hush}}, \bibinfo {author}
  {\bibfnamefont {M.}~\bibnamefont {Liuzzi}}, \bibinfo {author} {\bibfnamefont
  {P.}~\bibnamefont {Mundada}}, \bibinfo {author} {\bibfnamefont
  {T.}~\bibnamefont {Merkh}}, \bibinfo {author} {\bibfnamefont {A.~R.}\
  \bibnamefont {Carvalho}}, \ and\ \bibinfo {author} {\bibfnamefont {M.~J.}\
  \bibnamefont {Biercuk}},\ }\bibinfo {title} {Experimental Deep Reinforcement
  Learning for Error-Robust Gate-Set Design on a Superconducting Quantum
  Computer},\ \href {\doibase 10.1103/PRXQuantum.2.040324} {\bibfield
  {journal} {\bibinfo  {journal} {PRX Quantum}\ }\textbf {\bibinfo {volume}
  {2}},\ \bibinfo {pages} {040324} (\bibinfo {year} {2021})}\BibitemShut
  {NoStop}%
\bibitem [{\citenamefont {Zeng}\ \emph {et~al.}(2023)\citenamefont {Zeng},
  \citenamefont {Zhou}, \citenamefont {Rinaldi}, \citenamefont {Gneiting},\
  and\ \citenamefont {Nori}}]{PhysRevLett.131.050601}%
  \BibitemOpen
  \bibfield  {author} {\bibinfo {author} {\bibfnamefont {Y.}~\bibnamefont
  {Zeng}}, \bibinfo {author} {\bibfnamefont {Z.-Y.}\ \bibnamefont {Zhou}},
  \bibinfo {author} {\bibfnamefont {E.}~\bibnamefont {Rinaldi}}, \bibinfo
  {author} {\bibfnamefont {C.}~\bibnamefont {Gneiting}}, \ and\ \bibinfo
  {author} {\bibfnamefont {F.}~\bibnamefont {Nori}},\ }\bibinfo {title}
  {Approximate Autonomous Quantum Error Correction with Reinforcement
  Learning},\ \href {\doibase 10.1103/PhysRevLett.131.050601} {\bibfield
  {journal} {\bibinfo  {journal} {Phys. Rev. Lett.}\ }\textbf {\bibinfo
  {volume} {131}},\ \bibinfo {pages} {050601} (\bibinfo {year}
  {2023})}\BibitemShut {NoStop}%
\bibitem [{\citenamefont {Porotti}\ \emph {et~al.}(2023)\citenamefont
  {Porotti}, \citenamefont {Peano},\ and\ \citenamefont
  {Marquardt}}]{PRXQuantum.4.030305}%
  \BibitemOpen
  \bibfield  {author} {\bibinfo {author} {\bibfnamefont {R.}~\bibnamefont
  {Porotti}}, \bibinfo {author} {\bibfnamefont {V.}~\bibnamefont {Peano}}, \
  and\ \bibinfo {author} {\bibfnamefont {F.}~\bibnamefont {Marquardt}},\
  }\bibinfo {title} {Gradient-Ascent Pulse Engineering with Feedback},\ \href
  {\doibase 10.1103/PRXQuantum.4.030305} {\bibfield  {journal} {\bibinfo
  {journal} {PRX Quantum}\ }\textbf {\bibinfo {volume} {4}},\ \bibinfo {pages}
  {030305} (\bibinfo {year} {2023})}\BibitemShut {NoStop}%
\bibitem [{\citenamefont {Krotov}(1996)}]{krotov1995global}%
  \BibitemOpen
  \bibfield  {author} {\bibinfo {author} {\bibfnamefont {V.~F.}\ \bibnamefont
  {Krotov}},\ }\href@noop {} {\emph {\bibinfo {title} {Global Methods in
  Optimal Control Theory}}}\ (\bibinfo  {publisher} {Dekker, New York},\
  \bibinfo {year} {1996})\BibitemShut {NoStop}%
\bibitem [{\citenamefont {Khaneja}\ \emph {et~al.}(2005)\citenamefont
  {Khaneja}, \citenamefont {Reiss}, \citenamefont {Kehlet}, \citenamefont
  {Schulte-Herbr\"{u}ggen},\ and\ \citenamefont {Glaser}}]{Khaneja2005}%
  \BibitemOpen
  \bibfield  {author} {\bibinfo {author} {\bibfnamefont {N.}~\bibnamefont
  {Khaneja}}, \bibinfo {author} {\bibfnamefont {T.}~\bibnamefont {Reiss}},
  \bibinfo {author} {\bibfnamefont {C.}~\bibnamefont {Kehlet}}, \bibinfo
  {author} {\bibfnamefont {T.}~\bibnamefont {Schulte-Herbr\"{u}ggen}}, \ and\
  \bibinfo {author} {\bibfnamefont {S.~J.}\ \bibnamefont {Glaser}},\ }\bibinfo
  {title} {Optimal control of coupled spin dynamics: {D}esign of {NMR} pulse
  sequences by gradient ascent algorithms},\ \href {\doibase
  10.1016/j.jmr.2004.11.004} {\bibfield  {journal} {\bibinfo  {journal} {J.
  Magn. Reson.}\ }\textbf {\bibinfo {volume} {172}},\ \bibinfo {pages} {296}
  (\bibinfo {year} {2005})}\BibitemShut {NoStop}%
\bibitem [{\citenamefont {Lovett}\ \emph {et~al.}(2013)\citenamefont {Lovett},
  \citenamefont {Crosnier}, \citenamefont {Perarnau-Llobet},\ and\
  \citenamefont {Sanders}}]{PhysRevLett.110.220501}%
  \BibitemOpen
  \bibfield  {author} {\bibinfo {author} {\bibfnamefont {N.~B.}\ \bibnamefont
  {Lovett}}, \bibinfo {author} {\bibfnamefont {C.}~\bibnamefont {Crosnier}},
  \bibinfo {author} {\bibfnamefont {M.}~\bibnamefont {Perarnau-Llobet}}, \ and\
  \bibinfo {author} {\bibfnamefont {B.~C.}\ \bibnamefont {Sanders}},\ }\bibinfo
  {title} {Differential Evolution for Many-Particle Adaptive Quantum
  Metrology},\ \href {\doibase 10.1103/PhysRevLett.110.220501} {\bibfield
  {journal} {\bibinfo  {journal} {Phys. Rev. Lett.}\ }\textbf {\bibinfo
  {volume} {110}},\ \bibinfo {pages} {220501} (\bibinfo {year}
  {2013})}\BibitemShut {NoStop}%
\bibitem [{\citenamefont {Zahedinejad}\ \emph {et~al.}(2015)\citenamefont
  {Zahedinejad}, \citenamefont {Ghosh},\ and\ \citenamefont
  {Sanders}}]{PhysRevLett.114.200502}%
  \BibitemOpen
  \bibfield  {author} {\bibinfo {author} {\bibfnamefont {E.}~\bibnamefont
  {Zahedinejad}}, \bibinfo {author} {\bibfnamefont {J.}~\bibnamefont {Ghosh}},
  \ and\ \bibinfo {author} {\bibfnamefont {B.~C.}\ \bibnamefont {Sanders}},\
  }\bibinfo {title} {High-Fidelity Single-Shot Toffoli Gate via Quantum
  Control},\ \href {\doibase 10.1103/PhysRevLett.114.200502} {\bibfield
  {journal} {\bibinfo  {journal} {Phys. Rev. Lett.}\ }\textbf {\bibinfo
  {volume} {114}},\ \bibinfo {pages} {200502} (\bibinfo {year}
  {2015})}\BibitemShut {NoStop}%
\bibitem [{\citenamefont {Huang}\ \emph
  {et~al.}(2022{\natexlab{b}})\citenamefont {Huang}, \citenamefont {Li},
  \citenamefont {Hou}, \citenamefont {Wu}, \citenamefont {Yung}, \citenamefont
  {Bayat},\ and\ \citenamefont {Wang}}]{PhysRevA.105.052414}%
  \BibitemOpen
  \bibfield  {author} {\bibinfo {author} {\bibfnamefont {Y.}~\bibnamefont
  {Huang}}, \bibinfo {author} {\bibfnamefont {Q.}~\bibnamefont {Li}}, \bibinfo
  {author} {\bibfnamefont {X.}~\bibnamefont {Hou}}, \bibinfo {author}
  {\bibfnamefont {R.-B.}\ \bibnamefont {Wu}}, \bibinfo {author} {\bibfnamefont
  {M.-H.}\ \bibnamefont {Yung}}, \bibinfo {author} {\bibfnamefont
  {A.}~\bibnamefont {Bayat}}, \ and\ \bibinfo {author} {\bibfnamefont
  {X.}~\bibnamefont {Wang}},\ }\bibinfo {title} {Robust resource-efficient
  quantum variational ansatz through an evolutionary algorithm},\ \href
  {\doibase 10.1103/PhysRevA.105.052414} {\bibfield  {journal} {\bibinfo
  {journal} {Phys. Rev. A}\ }\textbf {\bibinfo {volume} {105}},\ \bibinfo
  {pages} {052414} (\bibinfo {year} {2022}{\natexlab{b}})}\BibitemShut
  {NoStop}%
\bibitem [{\citenamefont {Magann}\ \emph {et~al.}(2021)\citenamefont {Magann},
  \citenamefont {Arenz}, \citenamefont {Grace}, \citenamefont {Ho},
  \citenamefont {Kosut}, \citenamefont {McClean}, \citenamefont {Rabitz},\ and\
  \citenamefont {Sarovar}}]{PRXQuantum.2.010101}%
  \BibitemOpen
  \bibfield  {author} {\bibinfo {author} {\bibfnamefont {A.~B.}\ \bibnamefont
  {Magann}}, \bibinfo {author} {\bibfnamefont {C.}~\bibnamefont {Arenz}},
  \bibinfo {author} {\bibfnamefont {M.~D.}\ \bibnamefont {Grace}}, \bibinfo
  {author} {\bibfnamefont {T.-S.}\ \bibnamefont {Ho}}, \bibinfo {author}
  {\bibfnamefont {R.~L.}\ \bibnamefont {Kosut}}, \bibinfo {author}
  {\bibfnamefont {J.~R.}\ \bibnamefont {McClean}}, \bibinfo {author}
  {\bibfnamefont {H.~A.}\ \bibnamefont {Rabitz}}, \ and\ \bibinfo {author}
  {\bibfnamefont {M.}~\bibnamefont {Sarovar}},\ }\bibinfo {title} {From Pulses
  to Circuits and Back Again: A Quantum Optimal Control Perspective on
  Variational Quantum Algorithms},\ \href {\doibase
  10.1103/PRXQuantum.2.010101} {\bibfield  {journal} {\bibinfo  {journal} {PRX
  Quantum}\ }\textbf {\bibinfo {volume} {2}},\ \bibinfo {pages} {010101}
  (\bibinfo {year} {2021})}\BibitemShut {NoStop}%
\bibitem [{\citenamefont {Roslund}\ and\ \citenamefont
  {Rabitz}(2009)}]{PhysRevA.79.053417}%
  \BibitemOpen
  \bibfield  {author} {\bibinfo {author} {\bibfnamefont {J.}~\bibnamefont
  {Roslund}}\ and\ \bibinfo {author} {\bibfnamefont {H.}~\bibnamefont
  {Rabitz}},\ }\bibinfo {title} {Gradient algorithm applied to laboratory
  quantum control},\ \href {\doibase 10.1103/PhysRevA.79.053417} {\bibfield
  {journal} {\bibinfo  {journal} {Phys. Rev. A}\ }\textbf {\bibinfo {volume}
  {79}},\ \bibinfo {pages} {053417} (\bibinfo {year} {2009})}\BibitemShut
  {NoStop}%
\bibitem [{\citenamefont {Johansson}\ \emph {et~al.}(2013)\citenamefont
  {Johansson}, \citenamefont {Nation},\ and\ \citenamefont
  {Nori}}]{Johansson2013}%
  \BibitemOpen
  \bibfield  {author} {\bibinfo {author} {\bibfnamefont {J.}~\bibnamefont
  {Johansson}}, \bibinfo {author} {\bibfnamefont {P.}~\bibnamefont {Nation}}, \
  and\ \bibinfo {author} {\bibfnamefont {F.}~\bibnamefont {Nori}},\ }\bibinfo
  {title} {{QuTiP} 2: A Python framework for the dynamics of open quantum
  systems},\ \href {\doibase 10.1016/j.cpc.2012.11.019} {\bibfield  {journal}
  {\bibinfo  {journal} {Comp. Phys. Comm.}\ }\textbf {\bibinfo {volume}
  {184}},\ \bibinfo {pages} {1234} (\bibinfo {year} {2013})}\BibitemShut
  {NoStop}%
\bibitem [{\citenamefont {Wu}\ \emph {et~al.}(2019)\citenamefont {Wu},
  \citenamefont {Ding}, \citenamefont {Dong},\ and\ \citenamefont
  {Wang}}]{PhysRevA.99.042327}%
  \BibitemOpen
  \bibfield  {author} {\bibinfo {author} {\bibfnamefont {R.-B.}\ \bibnamefont
  {Wu}}, \bibinfo {author} {\bibfnamefont {H.}~\bibnamefont {Ding}}, \bibinfo
  {author} {\bibfnamefont {D.}~\bibnamefont {Dong}}, \ and\ \bibinfo {author}
  {\bibfnamefont {X.}~\bibnamefont {Wang}},\ }\bibinfo {title} {Learning robust
  and high-precision quantum controls},\ \href {\doibase
  10.1103/PhysRevA.99.042327} {\bibfield  {journal} {\bibinfo  {journal} {Phys.
  Rev. A}\ }\textbf {\bibinfo {volume} {99}},\ \bibinfo {pages} {042327}
  (\bibinfo {year} {2019})}\BibitemShut {NoStop}%
\bibitem [{\citenamefont {Ge}\ \emph {et~al.}(2020)\citenamefont {Ge},
  \citenamefont {Ding}, \citenamefont {Rabitz},\ and\ \citenamefont
  {Wu}}]{PhysRevA.101.052317}%
  \BibitemOpen
  \bibfield  {author} {\bibinfo {author} {\bibfnamefont {X.}~\bibnamefont
  {Ge}}, \bibinfo {author} {\bibfnamefont {H.}~\bibnamefont {Ding}}, \bibinfo
  {author} {\bibfnamefont {H.}~\bibnamefont {Rabitz}}, \ and\ \bibinfo {author}
  {\bibfnamefont {R.-B.}\ \bibnamefont {Wu}},\ }\bibinfo {title} {Robust
  quantum control in games: An adversarial learning approach},\ \href {\doibase
  10.1103/PhysRevA.101.052317} {\bibfield  {journal} {\bibinfo  {journal}
  {Phys. Rev. A}\ }\textbf {\bibinfo {volume} {101}},\ \bibinfo {pages}
  {052317} (\bibinfo {year} {2020})}\BibitemShut {NoStop}%
\bibitem [{\citenamefont {Ding}\ \emph
  {et~al.}(2021{\natexlab{b}})\citenamefont {Ding}, \citenamefont {Chu},
  \citenamefont {Qi},\ and\ \citenamefont {Wu}}]{PhysRevApplied.16.014056}%
  \BibitemOpen
  \bibfield  {author} {\bibinfo {author} {\bibfnamefont {H.-J.}\ \bibnamefont
  {Ding}}, \bibinfo {author} {\bibfnamefont {B.}~\bibnamefont {Chu}}, \bibinfo
  {author} {\bibfnamefont {B.}~\bibnamefont {Qi}}, \ and\ \bibinfo {author}
  {\bibfnamefont {R.-B.}\ \bibnamefont {Wu}},\ }\bibinfo {title} {Collaborative
  Learning of High-Precision Quantum Control and Tomography},\ \href {\doibase
  10.1103/PhysRevApplied.16.014056} {\bibfield  {journal} {\bibinfo  {journal}
  {Phys. Rev. Appl.}\ }\textbf {\bibinfo {volume} {16}},\ \bibinfo {pages}
  {014056} (\bibinfo {year} {2021}{\natexlab{b}})}\BibitemShut {NoStop}%
\bibitem [{\citenamefont {Li}\ \emph {et~al.}(2022)\citenamefont {Li},
  \citenamefont {Ahmed}, \citenamefont {Saraogi}, \citenamefont {Lambert},
  \citenamefont {Nori}, \citenamefont {Pitchford},\ and\ \citenamefont
  {Shammah}}]{Li2022quantum}%
  \BibitemOpen
  \bibfield  {author} {\bibinfo {author} {\bibfnamefont {B.}~\bibnamefont
  {Li}}, \bibinfo {author} {\bibfnamefont {S.}~\bibnamefont {Ahmed}}, \bibinfo
  {author} {\bibfnamefont {S.}~\bibnamefont {Saraogi}}, \bibinfo {author}
  {\bibfnamefont {N.}~\bibnamefont {Lambert}}, \bibinfo {author} {\bibfnamefont
  {F.}~\bibnamefont {Nori}}, \bibinfo {author} {\bibfnamefont {A.}~\bibnamefont
  {Pitchford}}, \ and\ \bibinfo {author} {\bibfnamefont {N.}~\bibnamefont
  {Shammah}},\ }\bibinfo {title} {Pulse-level noisy quantum circuits with
  {QuTiP}},\ \href {\doibase 10.22331/q-2022-01-24-630} {\bibfield  {journal}
  {\bibinfo  {journal} {Quantum}\ }\textbf {\bibinfo {volume} {6}},\ \bibinfo
  {pages} {630} (\bibinfo {year} {2022})}\BibitemShut {NoStop}%
\bibitem [{\citenamefont {Doria}\ \emph {et~al.}(2011)\citenamefont {Doria},
  \citenamefont {Calarco},\ and\ \citenamefont
  {Montangero}}]{PhysRevLett.106.190501}%
  \BibitemOpen
  \bibfield  {author} {\bibinfo {author} {\bibfnamefont {P.}~\bibnamefont
  {Doria}}, \bibinfo {author} {\bibfnamefont {T.}~\bibnamefont {Calarco}}, \
  and\ \bibinfo {author} {\bibfnamefont {S.}~\bibnamefont {Montangero}},\
  }\bibinfo {title} {Optimal Control Technique for Many-Body Quantum
  Dynamics},\ \href {\doibase 10.1103/PhysRevLett.106.190501} {\bibfield
  {journal} {\bibinfo  {journal} {Phys. Rev. Lett.}\ }\textbf {\bibinfo
  {volume} {106}},\ \bibinfo {pages} {190501} (\bibinfo {year}
  {2011})}\BibitemShut {NoStop}%
\bibitem [{\citenamefont {Caneva}\ \emph {et~al.}(2011)\citenamefont {Caneva},
  \citenamefont {Calarco},\ and\ \citenamefont
  {Montangero}}]{PhysRevA.84.022326}%
  \BibitemOpen
  \bibfield  {author} {\bibinfo {author} {\bibfnamefont {T.}~\bibnamefont
  {Caneva}}, \bibinfo {author} {\bibfnamefont {T.}~\bibnamefont {Calarco}}, \
  and\ \bibinfo {author} {\bibfnamefont {S.}~\bibnamefont {Montangero}},\
  }\bibinfo {title} {Chopped random-basis quantum optimization},\ \href
  {\doibase 10.1103/PhysRevA.84.022326} {\bibfield  {journal} {\bibinfo
  {journal} {Phys. Rev. A}\ }\textbf {\bibinfo {volume} {84}},\ \bibinfo
  {pages} {022326} (\bibinfo {year} {2011})}\BibitemShut {NoStop}%
\bibitem [{\citenamefont {Rach}\ \emph {et~al.}(2015)\citenamefont {Rach},
  \citenamefont {M\"uller}, \citenamefont {Calarco},\ and\ \citenamefont
  {Montangero}}]{PhysRevA.92.062343}%
  \BibitemOpen
  \bibfield  {author} {\bibinfo {author} {\bibfnamefont {N.}~\bibnamefont
  {Rach}}, \bibinfo {author} {\bibfnamefont {M.~M.}\ \bibnamefont {M\"uller}},
  \bibinfo {author} {\bibfnamefont {T.}~\bibnamefont {Calarco}}, \ and\
  \bibinfo {author} {\bibfnamefont {S.}~\bibnamefont {Montangero}},\ }\bibinfo
  {title} {Dressing the chopped-random-basis optimization: A bandwidth-limited
  access to the trap-free landscape},\ \href {\doibase
  10.1103/PhysRevA.92.062343} {\bibfield  {journal} {\bibinfo  {journal} {Phys.
  Rev. A}\ }\textbf {\bibinfo {volume} {92}},\ \bibinfo {pages} {062343}
  (\bibinfo {year} {2015})}\BibitemShut {NoStop}%
\bibitem [{\citenamefont {S\o{}rensen}\ \emph {et~al.}(2018)\citenamefont
  {S\o{}rensen}, \citenamefont {Aranburu}, \citenamefont {Heinzel},\ and\
  \citenamefont {Sherson}}]{PhysRevA.98.022119}%
  \BibitemOpen
  \bibfield  {author} {\bibinfo {author} {\bibfnamefont {J.~J. W.~H.}\
  \bibnamefont {S\o{}rensen}}, \bibinfo {author} {\bibfnamefont {M.~O.}\
  \bibnamefont {Aranburu}}, \bibinfo {author} {\bibfnamefont {T.}~\bibnamefont
  {Heinzel}}, \ and\ \bibinfo {author} {\bibfnamefont {J.~F.}\ \bibnamefont
  {Sherson}},\ }\bibinfo {title} {Quantum optimal control in a chopped basis:
  Applications in control of Bose-Einstein condensates},\ \href {\doibase
  10.1103/PhysRevA.98.022119} {\bibfield  {journal} {\bibinfo  {journal} {Phys.
  Rev. A}\ }\textbf {\bibinfo {volume} {98}},\ \bibinfo {pages} {022119}
  (\bibinfo {year} {2018})}\BibitemShut {NoStop}%
\bibitem [{\citenamefont {de~Fouquieres}\ \emph {et~al.}(2011)\citenamefont
  {de~Fouquieres}, \citenamefont {Schirmer}, \citenamefont {Glaser},\ and\
  \citenamefont {Kuprov}}]{deFouquieres2011}%
  \BibitemOpen
  \bibfield  {author} {\bibinfo {author} {\bibfnamefont {P.}~\bibnamefont
  {de~Fouquieres}}, \bibinfo {author} {\bibfnamefont {S.}~\bibnamefont
  {Schirmer}}, \bibinfo {author} {\bibfnamefont {S.}~\bibnamefont {Glaser}}, \
  and\ \bibinfo {author} {\bibfnamefont {I.}~\bibnamefont {Kuprov}},\ }\bibinfo
  {title} {Second order gradient ascent pulse engineering},\ \href {\doibase
  10.1016/j.jmr.2011.07.023} {\bibfield  {journal} {\bibinfo  {journal} {J.
  Magn. Reson.}\ }\textbf {\bibinfo {volume} {212}},\ \bibinfo {pages} {412}
  (\bibinfo {year} {2011})}\BibitemShut {NoStop}%
\bibitem [{\citenamefont {Nocedal}\ and\ \citenamefont
  {Wright}(2006)}]{Nocedal2006}%
  \BibitemOpen
  \bibfield  {author} {\bibinfo {author} {\bibfnamefont {J.}~\bibnamefont
  {Nocedal}}\ and\ \bibinfo {author} {\bibfnamefont {S.~J.}\ \bibnamefont
  {Wright}},\ }\href@noop {} {\emph {\bibinfo {title} {Numerical
  Optimization}}}\ (\bibinfo  {publisher} {Springer, Berlin},\ \bibinfo {year}
  {2006})\BibitemShut {NoStop}%
\bibitem [{\citenamefont {Torosov}\ \emph {et~al.}(2021)\citenamefont
  {Torosov}, \citenamefont {Shore},\ and\ \citenamefont
  {Vitanov}}]{PhysRevA.103.033110}%
  \BibitemOpen
  \bibfield  {author} {\bibinfo {author} {\bibfnamefont {B.~T.}\ \bibnamefont
  {Torosov}}, \bibinfo {author} {\bibfnamefont {B.~W.}\ \bibnamefont {Shore}},
  \ and\ \bibinfo {author} {\bibfnamefont {N.~V.}\ \bibnamefont {Vitanov}},\
  }\bibinfo {title} {Coherent control techniques for two-state quantum systems:
  A comparative study},\ \href {\doibase 10.1103/PhysRevA.103.033110}
  {\bibfield  {journal} {\bibinfo  {journal} {Phys. Rev. A}\ }\textbf {\bibinfo
  {volume} {103}},\ \bibinfo {pages} {033110} (\bibinfo {year}
  {2021})}\BibitemShut {NoStop}%
\bibitem [{\citenamefont {Wang}\ \emph {et~al.}(2012)\citenamefont {Wang},
  \citenamefont {Bishop}, \citenamefont {Kestner}, \citenamefont {Barnes},
  \citenamefont {Sun},\ and\ \citenamefont {Das~Sarma}}]{Wang2012}%
  \BibitemOpen
  \bibfield  {author} {\bibinfo {author} {\bibfnamefont {X.}~\bibnamefont
  {Wang}}, \bibinfo {author} {\bibfnamefont {L.~S.}\ \bibnamefont {Bishop}},
  \bibinfo {author} {\bibfnamefont {J.}~\bibnamefont {Kestner}}, \bibinfo
  {author} {\bibfnamefont {E.}~\bibnamefont {Barnes}}, \bibinfo {author}
  {\bibfnamefont {K.}~\bibnamefont {Sun}}, \ and\ \bibinfo {author}
  {\bibfnamefont {S.}~\bibnamefont {Das~Sarma}},\ }\bibinfo {title} {Composite
  pulses for robust universal control of singlet{\textendash}triplet qubits},\
  \href {\doibase 10.1038/ncomms2003} {\bibfield  {journal} {\bibinfo
  {journal} {Nat. Commun.}\ }\textbf {\bibinfo {volume} {3}},\ \bibinfo {pages}
  {997} (\bibinfo {year} {2012})}\BibitemShut {NoStop}%
\bibitem [{\citenamefont {Wang}\ \emph
  {et~al.}(2014{\natexlab{a}})\citenamefont {Wang}, \citenamefont {Bishop},
  \citenamefont {Barnes}, \citenamefont {Kestner},\ and\ \citenamefont
  {Das~Sarma}}]{PhysRevA.89.022310}%
  \BibitemOpen
  \bibfield  {author} {\bibinfo {author} {\bibfnamefont {X.}~\bibnamefont
  {Wang}}, \bibinfo {author} {\bibfnamefont {L.~S.}\ \bibnamefont {Bishop}},
  \bibinfo {author} {\bibfnamefont {E.}~\bibnamefont {Barnes}}, \bibinfo
  {author} {\bibfnamefont {J.~P.}\ \bibnamefont {Kestner}}, \ and\ \bibinfo
  {author} {\bibfnamefont {S.}~\bibnamefont {Das~Sarma}},\ }\bibinfo {title}
  {Robust quantum gates for singlet-triplet spin qubits using composite
  pulses},\ \href {\doibase 10.1103/PhysRevA.89.022310} {\bibfield  {journal}
  {\bibinfo  {journal} {Phys. Rev. A}\ }\textbf {\bibinfo {volume} {89}},\
  \bibinfo {pages} {022310} (\bibinfo {year} {2014}{\natexlab{a}})}\BibitemShut
  {NoStop}%
\bibitem [{\citenamefont {Scully}\ and\ \citenamefont
  {Zubairy}(1997)}]{scully97}%
  \BibitemOpen
  \bibfield  {author} {\bibinfo {author} {\bibfnamefont {M.~O.}\ \bibnamefont
  {Scully}}\ and\ \bibinfo {author} {\bibfnamefont {M.~S.}\ \bibnamefont
  {Zubairy}},\ }\href@noop {} {\emph {\bibinfo {title} {Quantum Optics}}}\
  (\bibinfo  {publisher} {Cambridge University Press, Cambridge, UK},\ \bibinfo
  {year} {1997})\BibitemShut {NoStop}%
\bibitem [{\citenamefont {Genov}\ \emph {et~al.}(2017)\citenamefont {Genov},
  \citenamefont {Schraft}, \citenamefont {Vitanov},\ and\ \citenamefont
  {Halfmann}}]{PhysRevLett.118.133202}%
  \BibitemOpen
  \bibfield  {author} {\bibinfo {author} {\bibfnamefont {G.~T.}\ \bibnamefont
  {Genov}}, \bibinfo {author} {\bibfnamefont {D.}~\bibnamefont {Schraft}},
  \bibinfo {author} {\bibfnamefont {N.~V.}\ \bibnamefont {Vitanov}}, \ and\
  \bibinfo {author} {\bibfnamefont {T.}~\bibnamefont {Halfmann}},\ }\bibinfo
  {title} {Arbitrarily Accurate Pulse Sequences for Robust Dynamical
  Decoupling},\ \href {\doibase 10.1103/PhysRevLett.118.133202} {\bibfield
  {journal} {\bibinfo  {journal} {Phys. Rev. Lett.}\ }\textbf {\bibinfo
  {volume} {118}},\ \bibinfo {pages} {133202} (\bibinfo {year}
  {2017})}\BibitemShut {NoStop}%
\bibitem [{\citenamefont {Genov}\ \emph {et~al.}(2020)\citenamefont {Genov},
  \citenamefont {Hain}, \citenamefont {Vitanov},\ and\ \citenamefont
  {Halfmann}}]{PhysRevA.101.013827}%
  \BibitemOpen
  \bibfield  {author} {\bibinfo {author} {\bibfnamefont {G.~T.}\ \bibnamefont
  {Genov}}, \bibinfo {author} {\bibfnamefont {M.}~\bibnamefont {Hain}},
  \bibinfo {author} {\bibfnamefont {N.~V.}\ \bibnamefont {Vitanov}}, \ and\
  \bibinfo {author} {\bibfnamefont {T.}~\bibnamefont {Halfmann}},\ }\bibinfo
  {title} {Universal composite pulses for efficient population inversion with
  an arbitrary excitation profile},\ \href {\doibase
  10.1103/PhysRevA.101.013827} {\bibfield  {journal} {\bibinfo  {journal}
  {Phys. Rev. A}\ }\textbf {\bibinfo {volume} {101}},\ \bibinfo {pages}
  {013827} (\bibinfo {year} {2020})}\BibitemShut {NoStop}%
\bibitem [{\citenamefont {Kyoseva}\ \emph {et~al.}(2019)\citenamefont
  {Kyoseva}, \citenamefont {Greener},\ and\ \citenamefont
  {Suchowski}}]{PhysRevA.100.032333}%
  \BibitemOpen
  \bibfield  {author} {\bibinfo {author} {\bibfnamefont {E.}~\bibnamefont
  {Kyoseva}}, \bibinfo {author} {\bibfnamefont {H.}~\bibnamefont {Greener}}, \
  and\ \bibinfo {author} {\bibfnamefont {H.}~\bibnamefont {Suchowski}},\
  }\bibinfo {title} {Detuning-modulated composite pulses for high-fidelity
  robust quantum control},\ \href {\doibase 10.1103/PhysRevA.100.032333}
  {\bibfield  {journal} {\bibinfo  {journal} {Phys. Rev. A}\ }\textbf {\bibinfo
  {volume} {100}},\ \bibinfo {pages} {032333} (\bibinfo {year}
  {2019})}\BibitemShut {NoStop}%
\bibitem [{\citenamefont {Kaplan}\ \emph {et~al.}(2023)\citenamefont {Kaplan},
  \citenamefont {Erew}, \citenamefont {Piasetzky}, \citenamefont {Goldstein},
  \citenamefont {Oz},\ and\ \citenamefont {Suchowski}}]{PhysRevA.108.042401}%
  \BibitemOpen
  \bibfield  {author} {\bibinfo {author} {\bibfnamefont {I.}~\bibnamefont
  {Kaplan}}, \bibinfo {author} {\bibfnamefont {M.}~\bibnamefont {Erew}},
  \bibinfo {author} {\bibfnamefont {Y.}~\bibnamefont {Piasetzky}}, \bibinfo
  {author} {\bibfnamefont {M.}~\bibnamefont {Goldstein}}, \bibinfo {author}
  {\bibfnamefont {Y.}~\bibnamefont {Oz}}, \ and\ \bibinfo {author}
  {\bibfnamefont {H.}~\bibnamefont {Suchowski}},\ }\bibinfo {title} {Segmented
  composite design of robust single-qubit quantum gates},\ \href {\doibase
  10.1103/PhysRevA.108.042401} {\bibfield  {journal} {\bibinfo  {journal}
  {Phys. Rev. A}\ }\textbf {\bibinfo {volume} {108}},\ \bibinfo {pages}
  {042401} (\bibinfo {year} {2023})}\BibitemShut {NoStop}%
\bibitem [{\citenamefont {Levitt}\ and\ \citenamefont
  {Freeman}(1979)}]{Levitt1979}%
  \BibitemOpen
  \bibfield  {author} {\bibinfo {author} {\bibfnamefont {M.~H.}\ \bibnamefont
  {Levitt}}\ and\ \bibinfo {author} {\bibfnamefont {R.}~\bibnamefont
  {Freeman}},\ }\bibinfo {title} {{NMR} population inversion using a composite
  pulse},\ \href {\doibase 10.1016/0022-2364(79)90265-8} {\bibfield  {journal}
  {\bibinfo  {journal} {J. Magn. Reson.}\ }\textbf {\bibinfo {volume} {33}},\
  \bibinfo {pages} {473} (\bibinfo {year} {1979})}\BibitemShut {NoStop}%
\bibitem [{\citenamefont {Levitt}(1986)}]{Levitt1986}%
  \BibitemOpen
  \bibfield  {author} {\bibinfo {author} {\bibfnamefont {M.~H.}\ \bibnamefont
  {Levitt}},\ }\bibinfo {title} {Composite pulses},\ \href {\doibase
  10.1016/0079-6565(86)80005-x} {\bibfield  {journal} {\bibinfo  {journal}
  {Prog. NMR Spectrosc.}\ }\textbf {\bibinfo {volume} {18}},\ \bibinfo {pages}
  {61} (\bibinfo {year} {1986})}\BibitemShut {NoStop}%
\bibitem [{\citenamefont {Wimperis}(1994)}]{Wimperis1994}%
  \BibitemOpen
  \bibfield  {author} {\bibinfo {author} {\bibfnamefont {S.}~\bibnamefont
  {Wimperis}},\ }\bibinfo {title} {Broadband, Narrowband, and Passband
  Composite Pulses for Use in Advanced NMR Experiments},\ \href {\doibase
  https://doi.org/10.1006/jmra.1994.1159} {\bibfield  {journal} {\bibinfo
  {journal} {J. Magn. Reson.}\ }\textbf {\bibinfo {volume} {109}},\ \bibinfo
  {pages} {221} (\bibinfo {year} {1994})}\BibitemShut {NoStop}%
\bibitem [{\citenamefont {Cummins}\ \emph {et~al.}(2003)\citenamefont
  {Cummins}, \citenamefont {Llewellyn},\ and\ \citenamefont
  {Jones}}]{PhysRevA.67.042308}%
  \BibitemOpen
  \bibfield  {author} {\bibinfo {author} {\bibfnamefont {H.~K.}\ \bibnamefont
  {Cummins}}, \bibinfo {author} {\bibfnamefont {G.}~\bibnamefont {Llewellyn}},
  \ and\ \bibinfo {author} {\bibfnamefont {J.~A.}\ \bibnamefont {Jones}},\
  }\bibinfo {title} {Tackling systematic errors in quantum logic gates with
  composite rotations},\ \href {\doibase 10.1103/PhysRevA.67.042308} {\bibfield
   {journal} {\bibinfo  {journal} {Phys. Rev. A}\ }\textbf {\bibinfo {volume}
  {67}},\ \bibinfo {pages} {042308} (\bibinfo {year} {2003})}\BibitemShut
  {NoStop}%
\bibitem [{\citenamefont {Brown}\ \emph {et~al.}(2004)\citenamefont {Brown},
  \citenamefont {Harrow},\ and\ \citenamefont {Chuang}}]{PhysRevA.70.052318}%
  \BibitemOpen
  \bibfield  {author} {\bibinfo {author} {\bibfnamefont {K.~R.}\ \bibnamefont
  {Brown}}, \bibinfo {author} {\bibfnamefont {A.~W.}\ \bibnamefont {Harrow}}, \
  and\ \bibinfo {author} {\bibfnamefont {I.~L.}\ \bibnamefont {Chuang}},\
  }\bibinfo {title} {Arbitrarily accurate composite pulse sequences},\ \href
  {\doibase 10.1103/PhysRevA.70.052318} {\bibfield  {journal} {\bibinfo
  {journal} {Phys. Rev. A}\ }\textbf {\bibinfo {volume} {70}},\ \bibinfo
  {pages} {052318} (\bibinfo {year} {2004})}\BibitemShut {NoStop}%
\bibitem [{\citenamefont {Ichikawa}\ \emph {et~al.}(2011)\citenamefont
  {Ichikawa}, \citenamefont {Bando}, \citenamefont {Kondo},\ and\ \citenamefont
  {Nakahara}}]{PhysRevA.84.062311}%
  \BibitemOpen
  \bibfield  {author} {\bibinfo {author} {\bibfnamefont {T.}~\bibnamefont
  {Ichikawa}}, \bibinfo {author} {\bibfnamefont {M.}~\bibnamefont {Bando}},
  \bibinfo {author} {\bibfnamefont {Y.}~\bibnamefont {Kondo}}, \ and\ \bibinfo
  {author} {\bibfnamefont {M.}~\bibnamefont {Nakahara}},\ }\bibinfo {title}
  {Designing robust unitary gates: Application to concatenated composite
  pulses},\ \href {\doibase 10.1103/PhysRevA.84.062311} {\bibfield  {journal}
  {\bibinfo  {journal} {Phys. Rev. A}\ }\textbf {\bibinfo {volume} {84}},\
  \bibinfo {pages} {062311} (\bibinfo {year} {2011})}\BibitemShut {NoStop}%
\bibitem [{\citenamefont {Jones}(2013)}]{PhysRevA.87.052317}%
  \BibitemOpen
  \bibfield  {author} {\bibinfo {author} {\bibfnamefont {J.~A.}\ \bibnamefont
  {Jones}},\ }\bibinfo {title} {Designing short robust {NOT} gates for quantum
  computation},\ \href {\doibase 10.1103/PhysRevA.87.052317} {\bibfield
  {journal} {\bibinfo  {journal} {Phys. Rev. A}\ }\textbf {\bibinfo {volume}
  {87}},\ \bibinfo {pages} {052317} (\bibinfo {year} {2013})}\BibitemShut
  {NoStop}%
\bibitem [{\citenamefont {Cohen}\ \emph {et~al.}(2016)\citenamefont {Cohen},
  \citenamefont {Rotem},\ and\ \citenamefont {Retzker}}]{PhysRevA.93.032340}%
  \BibitemOpen
  \bibfield  {author} {\bibinfo {author} {\bibfnamefont {I.}~\bibnamefont
  {Cohen}}, \bibinfo {author} {\bibfnamefont {A.}~\bibnamefont {Rotem}}, \ and\
  \bibinfo {author} {\bibfnamefont {A.}~\bibnamefont {Retzker}},\ }\bibinfo
  {title} {Refocusing two-qubit-gate noise for trapped ions by composite
  pulses},\ \href {\doibase 10.1103/PhysRevA.93.032340} {\bibfield  {journal}
  {\bibinfo  {journal} {Phys. Rev. A}\ }\textbf {\bibinfo {volume} {93}},\
  \bibinfo {pages} {032340} (\bibinfo {year} {2016})}\BibitemShut {NoStop}%
\bibitem [{\citenamefont {Dridi}\ \emph
  {et~al.}(2020{\natexlab{b}})\citenamefont {Dridi}, \citenamefont {Mejatty},
  \citenamefont {Glaser},\ and\ \citenamefont {Sugny}}]{PhysRevA.101.012321}%
  \BibitemOpen
  \bibfield  {author} {\bibinfo {author} {\bibfnamefont {G.}~\bibnamefont
  {Dridi}}, \bibinfo {author} {\bibfnamefont {M.}~\bibnamefont {Mejatty}},
  \bibinfo {author} {\bibfnamefont {S.~J.}\ \bibnamefont {Glaser}}, \ and\
  \bibinfo {author} {\bibfnamefont {D.}~\bibnamefont {Sugny}},\ }\bibinfo
  {title} {Robust control of a {NOT} gate by composite pulses},\ \href
  {\doibase 10.1103/PhysRevA.101.012321} {\bibfield  {journal} {\bibinfo
  {journal} {Phys. Rev. A}\ }\textbf {\bibinfo {volume} {101}},\ \bibinfo
  {pages} {012321} (\bibinfo {year} {2020}{\natexlab{b}})}\BibitemShut
  {NoStop}%
\bibitem [{\citenamefont {Torosov}\ and\ \citenamefont
  {Vitanov}(2020)}]{PhysRevResearch.2.043194}%
  \BibitemOpen
  \bibfield  {author} {\bibinfo {author} {\bibfnamefont {B.~T.}\ \bibnamefont
  {Torosov}}\ and\ \bibinfo {author} {\bibfnamefont {N.~V.}\ \bibnamefont
  {Vitanov}},\ }\bibinfo {title} {High-fidelity composite quantum gates for
  {R}aman qubits},\ \href {\doibase 10.1103/PhysRevResearch.2.043194}
  {\bibfield  {journal} {\bibinfo  {journal} {Phys. Rev. Research}\ }\textbf
  {\bibinfo {volume} {2}},\ \bibinfo {pages} {043194} (\bibinfo {year}
  {2020})}\BibitemShut {NoStop}%
\bibitem [{\citenamefont {Bulmer}\ \emph {et~al.}(2020)\citenamefont {Bulmer},
  \citenamefont {Jones},\ and\ \citenamefont {Walmsley}}]{Bulmer2020}%
  \BibitemOpen
  \bibfield  {author} {\bibinfo {author} {\bibfnamefont {J.~F.~F.}\
  \bibnamefont {Bulmer}}, \bibinfo {author} {\bibfnamefont {J.~A.}\
  \bibnamefont {Jones}}, \ and\ \bibinfo {author} {\bibfnamefont {I.~A.}\
  \bibnamefont {Walmsley}},\ }\bibinfo {title} {Drive-noise tolerant optical
  switching inspired by composite pulses},\ \href {\doibase 10.1364/oe.378469}
  {\bibfield  {journal} {\bibinfo  {journal} {Opt. Express}\ }\textbf {\bibinfo
  {volume} {28}},\ \bibinfo {pages} {8646} (\bibinfo {year}
  {2020})}\BibitemShut {NoStop}%
\bibitem [{\citenamefont {Rong}\ \emph {et~al.}(2015)\citenamefont {Rong},
  \citenamefont {Geng}, \citenamefont {Shi}, \citenamefont {Liu}, \citenamefont
  {Xu}, \citenamefont {Ma}, \citenamefont {Kong}, \citenamefont {Jiang},
  \citenamefont {Wu},\ and\ \citenamefont {Du}}]{Rong2015nc}%
  \BibitemOpen
  \bibfield  {author} {\bibinfo {author} {\bibfnamefont {X.}~\bibnamefont
  {Rong}}, \bibinfo {author} {\bibfnamefont {J.}~\bibnamefont {Geng}}, \bibinfo
  {author} {\bibfnamefont {F.}~\bibnamefont {Shi}}, \bibinfo {author}
  {\bibfnamefont {Y.}~\bibnamefont {Liu}}, \bibinfo {author} {\bibfnamefont
  {K.}~\bibnamefont {Xu}}, \bibinfo {author} {\bibfnamefont {W.}~\bibnamefont
  {Ma}}, \bibinfo {author} {\bibfnamefont {F.}~\bibnamefont {Kong}}, \bibinfo
  {author} {\bibfnamefont {Z.}~\bibnamefont {Jiang}}, \bibinfo {author}
  {\bibfnamefont {Y.}~\bibnamefont {Wu}}, \ and\ \bibinfo {author}
  {\bibfnamefont {J.~F.}\ \bibnamefont {Du}},\ }\bibinfo {title} {Experimental
  fault-tolerant universal quantum gates with solid-state spins under ambient
  conditions},\ \href {\doibase 10.1038/ncomms9748} {\bibfield  {journal}
  {\bibinfo  {journal} {Nat. Commun.}\ }\textbf {\bibinfo {volume} {6}},\
  \bibinfo {pages} {8748} (\bibinfo {year} {2015})}\BibitemShut {NoStop}%
\bibitem [{\citenamefont {Mount}\ \emph {et~al.}(2015)\citenamefont {Mount},
  \citenamefont {Kabytayev}, \citenamefont {Crain}, \citenamefont {Harper},
  \citenamefont {Baek}, \citenamefont {Vrijsen}, \citenamefont {Flammia},
  \citenamefont {Brown}, \citenamefont {Maunz},\ and\ \citenamefont
  {Kim}}]{PhysRevA.92.060301}%
  \BibitemOpen
  \bibfield  {author} {\bibinfo {author} {\bibfnamefont {E.}~\bibnamefont
  {Mount}}, \bibinfo {author} {\bibfnamefont {C.}~\bibnamefont {Kabytayev}},
  \bibinfo {author} {\bibfnamefont {S.}~\bibnamefont {Crain}}, \bibinfo
  {author} {\bibfnamefont {R.}~\bibnamefont {Harper}}, \bibinfo {author}
  {\bibfnamefont {S.-Y.}\ \bibnamefont {Baek}}, \bibinfo {author}
  {\bibfnamefont {G.}~\bibnamefont {Vrijsen}}, \bibinfo {author} {\bibfnamefont
  {S.~T.}\ \bibnamefont {Flammia}}, \bibinfo {author} {\bibfnamefont {K.~R.}\
  \bibnamefont {Brown}}, \bibinfo {author} {\bibfnamefont {P.}~\bibnamefont
  {Maunz}}, \ and\ \bibinfo {author} {\bibfnamefont {J.}~\bibnamefont {Kim}},\
  }\bibinfo {title} {Error compensation of single-qubit gates in a
  surface-electrode ion trap using composite pulses},\ \href {\doibase
  10.1103/PhysRevA.92.060301} {\bibfield  {journal} {\bibinfo  {journal} {Phys.
  Rev. A}\ }\textbf {\bibinfo {volume} {92}},\ \bibinfo {pages} {060301}
  (\bibinfo {year} {2015})}\BibitemShut {NoStop}%
\bibitem [{\citenamefont {Ivanov}\ \emph {et~al.}(2022)\citenamefont {Ivanov},
  \citenamefont {Torosov},\ and\ \citenamefont
  {Vitanov}}]{PhysRevLett.129.240505}%
  \BibitemOpen
  \bibfield  {author} {\bibinfo {author} {\bibfnamefont {S.~S.}\ \bibnamefont
  {Ivanov}}, \bibinfo {author} {\bibfnamefont {B.~T.}\ \bibnamefont {Torosov}},
  \ and\ \bibinfo {author} {\bibfnamefont {N.~V.}\ \bibnamefont {Vitanov}},\
  }\bibinfo {title} {High-Fidelity Quantum Control by Polychromatic Pulse
  Trains},\ \href {\doibase 10.1103/PhysRevLett.129.240505} {\bibfield
  {journal} {\bibinfo  {journal} {Phys. Rev. Lett.}\ }\textbf {\bibinfo
  {volume} {129}},\ \bibinfo {pages} {240505} (\bibinfo {year}
  {2022})}\BibitemShut {NoStop}%
\bibitem [{\citenamefont {Torosov}\ and\ \citenamefont
  {Vitanov}(2022)}]{PhysRevApplied.18.034062}%
  \BibitemOpen
  \bibfield  {author} {\bibinfo {author} {\bibfnamefont {B.~T.}\ \bibnamefont
  {Torosov}}\ and\ \bibinfo {author} {\bibfnamefont {N.~V.}\ \bibnamefont
  {Vitanov}},\ }\bibinfo {title} {Experimental Demonstration of Composite
  Pulses on IBM's Quantum Computer},\ \href {\doibase
  10.1103/PhysRevApplied.18.034062} {\bibfield  {journal} {\bibinfo  {journal}
  {Phys. Rev. Appl.}\ }\textbf {\bibinfo {volume} {18}},\ \bibinfo {pages}
  {034062} (\bibinfo {year} {2022})}\BibitemShut {NoStop}%
\bibitem [{\citenamefont {Torosov}\ and\ \citenamefont
  {Vitanov}(2011)}]{PhysRevA.83.053420}%
  \BibitemOpen
  \bibfield  {author} {\bibinfo {author} {\bibfnamefont {B.~T.}\ \bibnamefont
  {Torosov}}\ and\ \bibinfo {author} {\bibfnamefont {N.~V.}\ \bibnamefont
  {Vitanov}},\ }\bibinfo {title} {Smooth composite pulses for high-fidelity
  quantum information processing},\ \href {\doibase 10.1103/PhysRevA.83.053420}
  {\bibfield  {journal} {\bibinfo  {journal} {Phys. Rev. A}\ }\textbf {\bibinfo
  {volume} {83}},\ \bibinfo {pages} {053420} (\bibinfo {year}
  {2011})}\BibitemShut {NoStop}%
\bibitem [{\citenamefont {Genov}\ \emph {et~al.}(2014)\citenamefont {Genov},
  \citenamefont {Schraft}, \citenamefont {Halfmann},\ and\ \citenamefont
  {Vitanov}}]{PhysRevLett.113.043001}%
  \BibitemOpen
  \bibfield  {author} {\bibinfo {author} {\bibfnamefont {G.~T.}\ \bibnamefont
  {Genov}}, \bibinfo {author} {\bibfnamefont {D.}~\bibnamefont {Schraft}},
  \bibinfo {author} {\bibfnamefont {T.}~\bibnamefont {Halfmann}}, \ and\
  \bibinfo {author} {\bibfnamefont {N.~V.}\ \bibnamefont {Vitanov}},\ }\bibinfo
  {title} {Correction of Arbitrary Field Errors in Population Inversion of
  Quantum Systems by Universal Composite Pulses},\ \href {\doibase
  10.1103/PhysRevLett.113.043001} {\bibfield  {journal} {\bibinfo  {journal}
  {Phys. Rev. Lett.}\ }\textbf {\bibinfo {volume} {113}},\ \bibinfo {pages}
  {043001} (\bibinfo {year} {2014})}\BibitemShut {NoStop}%
\bibitem [{\citenamefont {Low}\ \emph {et~al.}(2014)\citenamefont {Low},
  \citenamefont {Yoder},\ and\ \citenamefont {Chuang}}]{PhysRevA.89.022341}%
  \BibitemOpen
  \bibfield  {author} {\bibinfo {author} {\bibfnamefont {G.~H.}\ \bibnamefont
  {Low}}, \bibinfo {author} {\bibfnamefont {T.~J.}\ \bibnamefont {Yoder}}, \
  and\ \bibinfo {author} {\bibfnamefont {I.~L.}\ \bibnamefont {Chuang}},\
  }\bibinfo {title} {Optimal arbitrarily accurate composite pulse sequences},\
  \href {\doibase 10.1103/PhysRevA.89.022341} {\bibfield  {journal} {\bibinfo
  {journal} {Phys. Rev. A}\ }\textbf {\bibinfo {volume} {89}},\ \bibinfo
  {pages} {022341} (\bibinfo {year} {2014})}\BibitemShut {NoStop}%
\bibitem [{\citenamefont {Torosov}\ and\ \citenamefont
  {Vitanov}(2018)}]{PhysRevA.97.043408}%
  \BibitemOpen
  \bibfield  {author} {\bibinfo {author} {\bibfnamefont {B.~T.}\ \bibnamefont
  {Torosov}}\ and\ \bibinfo {author} {\bibfnamefont {N.~V.}\ \bibnamefont
  {Vitanov}},\ }\bibinfo {title} {Arbitrarily accurate twin composite
  $\ensuremath{\pi}$-pulse sequences},\ \href {\doibase
  10.1103/PhysRevA.97.043408} {\bibfield  {journal} {\bibinfo  {journal} {Phys.
  Rev. A}\ }\textbf {\bibinfo {volume} {97}},\ \bibinfo {pages} {043408}
  (\bibinfo {year} {2018})}\BibitemShut {NoStop}%
\bibitem [{\citenamefont {Demeter}(2016)}]{PhysRevA.93.023830}%
  \BibitemOpen
  \bibfield  {author} {\bibinfo {author} {\bibfnamefont {G.}~\bibnamefont
  {Demeter}},\ }\bibinfo {title} {Composite pulses for high-fidelity population
  inversion in optically dense, inhomogeneously broadened atomic ensembles},\
  \href {\doibase 10.1103/PhysRevA.93.023830} {\bibfield  {journal} {\bibinfo
  {journal} {Phys. Rev. A}\ }\textbf {\bibinfo {volume} {93}},\ \bibinfo
  {pages} {023830} (\bibinfo {year} {2016})}\BibitemShut {NoStop}%
\bibitem [{\citenamefont {Shi}\ \emph {et~al.}(2021)\citenamefont {Shi},
  \citenamefont {Wu}, \citenamefont {Shen}, \citenamefont {Song}, \citenamefont
  {Xia}, \citenamefont {Yi},\ and\ \citenamefont
  {Zheng}}]{PhysRevA.103.052612}%
  \BibitemOpen
  \bibfield  {author} {\bibinfo {author} {\bibfnamefont {Z.-C.}\ \bibnamefont
  {Shi}}, \bibinfo {author} {\bibfnamefont {H.-N.}\ \bibnamefont {Wu}},
  \bibinfo {author} {\bibfnamefont {L.-T.}\ \bibnamefont {Shen}}, \bibinfo
  {author} {\bibfnamefont {J.}~\bibnamefont {Song}}, \bibinfo {author}
  {\bibfnamefont {Y.}~\bibnamefont {Xia}}, \bibinfo {author} {\bibfnamefont
  {X.~X.}\ \bibnamefont {Yi}}, \ and\ \bibinfo {author} {\bibfnamefont {S.-B.}\
  \bibnamefont {Zheng}},\ }\bibinfo {title} {Robust single-qubit gates by
  composite pulses in three-level systems},\ \href {\doibase
  10.1103/PhysRevA.103.052612} {\bibfield  {journal} {\bibinfo  {journal}
  {Phys. Rev. A}\ }\textbf {\bibinfo {volume} {103}},\ \bibinfo {pages}
  {052612} (\bibinfo {year} {2021})}\BibitemShut {NoStop}%
\bibitem [{\citenamefont {Gevorgyan}\ and\ \citenamefont
  {Vitanov}(2021)}]{PhysRevA.104.012609}%
  \BibitemOpen
  \bibfield  {author} {\bibinfo {author} {\bibfnamefont {H.~L.}\ \bibnamefont
  {Gevorgyan}}\ and\ \bibinfo {author} {\bibfnamefont {N.~V.}\ \bibnamefont
  {Vitanov}},\ }\bibinfo {title} {Ultrahigh-fidelity composite rotational
  quantum gates},\ \href {\doibase 10.1103/PhysRevA.104.012609} {\bibfield
  {journal} {\bibinfo  {journal} {Phys. Rev. A}\ }\textbf {\bibinfo {volume}
  {104}},\ \bibinfo {pages} {012609} (\bibinfo {year} {2021})}\BibitemShut
  {NoStop}%
\bibitem [{\citenamefont {Torosov}\ \emph {et~al.}(2020)\citenamefont
  {Torosov}, \citenamefont {Drewsen},\ and\ \citenamefont
  {Vitanov}}]{PhysRevResearch.2.043235}%
  \BibitemOpen
  \bibfield  {author} {\bibinfo {author} {\bibfnamefont {B.~T.}\ \bibnamefont
  {Torosov}}, \bibinfo {author} {\bibfnamefont {M.}~\bibnamefont {Drewsen}}, \
  and\ \bibinfo {author} {\bibfnamefont {N.~V.}\ \bibnamefont {Vitanov}},\
  }\bibinfo {title} {Chiral resolution by composite {R}aman pulses},\ \href
  {\doibase 10.1103/PhysRevResearch.2.043235} {\bibfield  {journal} {\bibinfo
  {journal} {Phys. Rev. Research}\ }\textbf {\bibinfo {volume} {2}},\ \bibinfo
  {pages} {043235} (\bibinfo {year} {2020})}\BibitemShut {NoStop}%
\bibitem [{\citenamefont {Shi}\ \emph {et~al.}(2022)\citenamefont {Shi},
  \citenamefont {Zhang}, \citenamefont {Ran}, \citenamefont {Xia},
  \citenamefont {Ianconescu}, \citenamefont {Friedman}, \citenamefont {Yi},\
  and\ \citenamefont {Zheng}}]{Shi2022}%
  \BibitemOpen
  \bibfield  {author} {\bibinfo {author} {\bibfnamefont {Z.-C.}\ \bibnamefont
  {Shi}}, \bibinfo {author} {\bibfnamefont {C.}~\bibnamefont {Zhang}}, \bibinfo
  {author} {\bibfnamefont {D.}~\bibnamefont {Ran}}, \bibinfo {author}
  {\bibfnamefont {Y.}~\bibnamefont {Xia}}, \bibinfo {author} {\bibfnamefont
  {R.}~\bibnamefont {Ianconescu}}, \bibinfo {author} {\bibfnamefont
  {A.}~\bibnamefont {Friedman}}, \bibinfo {author} {\bibfnamefont {X.~X.}\
  \bibnamefont {Yi}}, \ and\ \bibinfo {author} {\bibfnamefont {S.-B.}\
  \bibnamefont {Zheng}},\ }\bibinfo {title} {Composite pulses for high fidelity
  population transfer in three-level systems},\ \href {\doibase
  10.1088/1367-2630/ac48e7} {\bibfield  {journal} {\bibinfo  {journal} {New J.
  Phys.}\ }\textbf {\bibinfo {volume} {24}},\ \bibinfo {pages} {023014}
  (\bibinfo {year} {2022})}\BibitemShut {NoStop}%
\bibitem [{\citenamefont {Wu}\ \emph {et~al.}(2023)\citenamefont {Wu},
  \citenamefont {Zhang}, \citenamefont {Song}, \citenamefont {Xia},\ and\
  \citenamefont {Shi}}]{PhysRevA.107.023103}%
  \BibitemOpen
  \bibfield  {author} {\bibinfo {author} {\bibfnamefont {H.-N.}\ \bibnamefont
  {Wu}}, \bibinfo {author} {\bibfnamefont {C.}~\bibnamefont {Zhang}}, \bibinfo
  {author} {\bibfnamefont {J.}~\bibnamefont {Song}}, \bibinfo {author}
  {\bibfnamefont {Y.}~\bibnamefont {Xia}}, \ and\ \bibinfo {author}
  {\bibfnamefont {Z.-C.}\ \bibnamefont {Shi}},\ }\bibinfo {title} {Composite
  pulses for optimal robust control in two-level systems},\ \href {\doibase
  10.1103/PhysRevA.107.023103} {\bibfield  {journal} {\bibinfo  {journal}
  {Phys. Rev. A}\ }\textbf {\bibinfo {volume} {107}},\ \bibinfo {pages}
  {023103} (\bibinfo {year} {2023})}\BibitemShut {NoStop}%
\bibitem [{\citenamefont {D'Arrigo}\ \emph {et~al.}(2016)\citenamefont
  {D'Arrigo}, \citenamefont {Falci},\ and\ \citenamefont
  {Paladino}}]{PhysRevA.94.022303}%
  \BibitemOpen
  \bibfield  {author} {\bibinfo {author} {\bibfnamefont {A.}~\bibnamefont
  {D'Arrigo}}, \bibinfo {author} {\bibfnamefont {G.}~\bibnamefont {Falci}}, \
  and\ \bibinfo {author} {\bibfnamefont {E.}~\bibnamefont {Paladino}},\
  }\bibinfo {title} {High-fidelity two-qubit gates via dynamical decoupling of
  local $1/f$ noise at the optimal point},\ \href {\doibase
  10.1103/PhysRevA.94.022303} {\bibfield  {journal} {\bibinfo  {journal} {Phys.
  Rev. A}\ }\textbf {\bibinfo {volume} {94}},\ \bibinfo {pages} {022303}
  (\bibinfo {year} {2016})}\BibitemShut {NoStop}%
\bibitem [{\citenamefont {Huang}\ and\ \citenamefont
  {Goan}(2017)}]{PhysRevA.95.062325}%
  \BibitemOpen
  \bibfield  {author} {\bibinfo {author} {\bibfnamefont {C.-H.}\ \bibnamefont
  {Huang}}\ and\ \bibinfo {author} {\bibfnamefont {H.-S.}\ \bibnamefont
  {Goan}},\ }\bibinfo {title} {Robust quantum gates for stochastic time-varying
  noise},\ \href {\doibase 10.1103/PhysRevA.95.062325} {\bibfield  {journal}
  {\bibinfo  {journal} {Phys. Rev. A}\ }\textbf {\bibinfo {volume} {95}},\
  \bibinfo {pages} {062325} (\bibinfo {year} {2017})}\BibitemShut {NoStop}%
\bibitem [{\citenamefont {Kestner}\ \emph {et~al.}(2013)\citenamefont
  {Kestner}, \citenamefont {Wang}, \citenamefont {Bishop}, \citenamefont
  {Barnes},\ and\ \citenamefont {Das~Sarma}}]{PhysRevLett.110.140502}%
  \BibitemOpen
  \bibfield  {author} {\bibinfo {author} {\bibfnamefont {J.~P.}\ \bibnamefont
  {Kestner}}, \bibinfo {author} {\bibfnamefont {X.}~\bibnamefont {Wang}},
  \bibinfo {author} {\bibfnamefont {L.~S.}\ \bibnamefont {Bishop}}, \bibinfo
  {author} {\bibfnamefont {E.}~\bibnamefont {Barnes}}, \ and\ \bibinfo {author}
  {\bibfnamefont {S.}~\bibnamefont {Das~Sarma}},\ }\bibinfo {title}
  {Noise-Resistant Control for a Spin Qubit Array},\ \href {\doibase
  10.1103/PhysRevLett.110.140502} {\bibfield  {journal} {\bibinfo  {journal}
  {Phys. Rev. Lett.}\ }\textbf {\bibinfo {volume} {110}},\ \bibinfo {pages}
  {140502} (\bibinfo {year} {2013})}\BibitemShut {NoStop}%
\bibitem [{\citenamefont {Wang}\ \emph
  {et~al.}(2014{\natexlab{b}})\citenamefont {Wang}, \citenamefont
  {Calderon-Vargas}, \citenamefont {Rana}, \citenamefont {Kestner},
  \citenamefont {Barnes},\ and\ \citenamefont
  {Das~Sarma}}]{PhysRevB.90.155306}%
  \BibitemOpen
  \bibfield  {author} {\bibinfo {author} {\bibfnamefont {X.}~\bibnamefont
  {Wang}}, \bibinfo {author} {\bibfnamefont {F.~A.}\ \bibnamefont
  {Calderon-Vargas}}, \bibinfo {author} {\bibfnamefont {M.~S.}\ \bibnamefont
  {Rana}}, \bibinfo {author} {\bibfnamefont {J.~P.}\ \bibnamefont {Kestner}},
  \bibinfo {author} {\bibfnamefont {E.}~\bibnamefont {Barnes}}, \ and\ \bibinfo
  {author} {\bibfnamefont {S.}~\bibnamefont {Das~Sarma}},\ }\bibinfo {title}
  {Noise-compensating pulses for electrostatically controlled silicon spin
  qubits},\ \href {\doibase 10.1103/PhysRevB.90.155306} {\bibfield  {journal}
  {\bibinfo  {journal} {Phys. Rev. B}\ }\textbf {\bibinfo {volume} {90}},\
  \bibinfo {pages} {155306} (\bibinfo {year} {2014}{\natexlab{b}})}\BibitemShut
  {NoStop}%
\bibitem [{\citenamefont {Colmenar}\ and\ \citenamefont
  {Kestner}(2022)}]{PhysRevA.106.032611}%
  \BibitemOpen
  \bibfield  {author} {\bibinfo {author} {\bibfnamefont {R.~K.~L.}\
  \bibnamefont {Colmenar}}\ and\ \bibinfo {author} {\bibfnamefont {J.~P.}\
  \bibnamefont {Kestner}},\ }\bibinfo {title} {Reverse engineering of one-qubit
  filter functions with dynamical invariants},\ \href {\doibase
  10.1103/PhysRevA.106.032611} {\bibfield  {journal} {\bibinfo  {journal}
  {Phys. Rev. A}\ }\textbf {\bibinfo {volume} {106}},\ \bibinfo {pages}
  {032611} (\bibinfo {year} {2022})}\BibitemShut {NoStop}%
\bibitem [{\citenamefont {Milne}\ \emph {et~al.}(2020)\citenamefont {Milne},
  \citenamefont {Edmunds}, \citenamefont {Hempel}, \citenamefont {Roy},
  \citenamefont {Mavadia},\ and\ \citenamefont
  {Biercuk}}]{PhysRevApplied.13.024022}%
  \BibitemOpen
  \bibfield  {author} {\bibinfo {author} {\bibfnamefont {A.~R.}\ \bibnamefont
  {Milne}}, \bibinfo {author} {\bibfnamefont {C.~L.}\ \bibnamefont {Edmunds}},
  \bibinfo {author} {\bibfnamefont {C.}~\bibnamefont {Hempel}}, \bibinfo
  {author} {\bibfnamefont {F.}~\bibnamefont {Roy}}, \bibinfo {author}
  {\bibfnamefont {S.}~\bibnamefont {Mavadia}}, \ and\ \bibinfo {author}
  {\bibfnamefont {M.~J.}\ \bibnamefont {Biercuk}},\ }\bibinfo {title}
  {Phase-Modulated Entangling Gates Robust to Static and Time-Varying Errors},\
  \href {\doibase 10.1103/PhysRevApplied.13.024022} {\bibfield  {journal}
  {\bibinfo  {journal} {Phys. Rev. Appl.}\ }\textbf {\bibinfo {volume} {13}},\
  \bibinfo {pages} {024022} (\bibinfo {year} {2020})}\BibitemShut {NoStop}%
\bibitem [{\citenamefont {Kang}\ \emph {et~al.}(2023)\citenamefont {Kang},
  \citenamefont {Wang}, \citenamefont {Fang}, \citenamefont {Zhang},
  \citenamefont {Khosravani}, \citenamefont {Kim},\ and\ \citenamefont
  {Brown}}]{PhysRevApplied.19.014014}%
  \BibitemOpen
  \bibfield  {author} {\bibinfo {author} {\bibfnamefont {M.}~\bibnamefont
  {Kang}}, \bibinfo {author} {\bibfnamefont {Y.}~\bibnamefont {Wang}}, \bibinfo
  {author} {\bibfnamefont {C.}~\bibnamefont {Fang}}, \bibinfo {author}
  {\bibfnamefont {B.}~\bibnamefont {Zhang}}, \bibinfo {author} {\bibfnamefont
  {O.}~\bibnamefont {Khosravani}}, \bibinfo {author} {\bibfnamefont
  {J.}~\bibnamefont {Kim}}, \ and\ \bibinfo {author} {\bibfnamefont {K.~R.}\
  \bibnamefont {Brown}},\ }\bibinfo {title} {Designing Filter Functions of
  Frequency-Modulated Pulses for High-Fidelity Two-Qubit Gates in Ion Chains},\
  \href {\doibase 10.1103/PhysRevApplied.19.014014} {\bibfield  {journal}
  {\bibinfo  {journal} {Phys. Rev. Appl.}\ }\textbf {\bibinfo {volume} {19}},\
  \bibinfo {pages} {014014} (\bibinfo {year} {2023})}\BibitemShut {NoStop}%
\bibitem [{\citenamefont {Araki}\ \emph {et~al.}(2023)\citenamefont {Araki},
  \citenamefont {Nori},\ and\ \citenamefont {Gneiting}}]{PhysRevA.107.032609}%
  \BibitemOpen
  \bibfield  {author} {\bibinfo {author} {\bibfnamefont {T.}~\bibnamefont
  {Araki}}, \bibinfo {author} {\bibfnamefont {F.}~\bibnamefont {Nori}}, \ and\
  \bibinfo {author} {\bibfnamefont {C.}~\bibnamefont {Gneiting}},\ }\bibinfo
  {title} {Robust quantum control with disorder-dressed evolution},\ \href
  {\doibase 10.1103/PhysRevA.107.032609} {\bibfield  {journal} {\bibinfo
  {journal} {Phys. Rev. A}\ }\textbf {\bibinfo {volume} {107}},\ \bibinfo
  {pages} {032609} (\bibinfo {year} {2023})}\BibitemShut {NoStop}%
\bibitem [{\citenamefont {Khalid}\ \emph {et~al.}(2023)\citenamefont {Khalid},
  \citenamefont {Weidner}, \citenamefont {Jonckheere}, \citenamefont
  {Shermer},\ and\ \citenamefont {Langbein}}]{PhysRevA.107.032606}%
  \BibitemOpen
  \bibfield  {author} {\bibinfo {author} {\bibfnamefont {I.}~\bibnamefont
  {Khalid}}, \bibinfo {author} {\bibfnamefont {C.~A.}\ \bibnamefont {Weidner}},
  \bibinfo {author} {\bibfnamefont {E.~A.}\ \bibnamefont {Jonckheere}},
  \bibinfo {author} {\bibfnamefont {S.~G.}\ \bibnamefont {Shermer}}, \ and\
  \bibinfo {author} {\bibfnamefont {F.~C.}\ \bibnamefont {Langbein}},\
  }\bibinfo {title} {Statistically characterizing robustness and fidelity of
  quantum controls and quantum control algorithms},\ \href {\doibase
  10.1103/PhysRevA.107.032606} {\bibfield  {journal} {\bibinfo  {journal}
  {Phys. Rev. A}\ }\textbf {\bibinfo {volume} {107}},\ \bibinfo {pages}
  {032606} (\bibinfo {year} {2023})}\BibitemShut {NoStop}%
\bibitem [{\citenamefont {Bhole}(2020)}]{bhole2020a}%
  \BibitemOpen
  \bibfield  {author} {\bibinfo {author} {\bibfnamefont {G.}~\bibnamefont
  {Bhole}},\ }\emph {\bibinfo {title} {Coherent control for quantum information
  processing}},\ \href@noop {} {Ph.D. thesis},\ \bibinfo  {school} {University
  of Oxford} (\bibinfo {year} {2020})\BibitemShut {NoStop}%
\bibitem [{\citenamefont {Ram}\ \emph {et~al.}(2022)\citenamefont {Ram},
  \citenamefont {Krithika}, \citenamefont {Batra},\ and\ \citenamefont
  {Mahesh}}]{PhysRevA.105.042437}%
  \BibitemOpen
  \bibfield  {author} {\bibinfo {author} {\bibfnamefont {M.~H.}\ \bibnamefont
  {Ram}}, \bibinfo {author} {\bibfnamefont {V.~R.}\ \bibnamefont {Krithika}},
  \bibinfo {author} {\bibfnamefont {P.}~\bibnamefont {Batra}}, \ and\ \bibinfo
  {author} {\bibfnamefont {T.~S.}\ \bibnamefont {Mahesh}},\ }\bibinfo {title}
  {Robust quantum control using hybrid pulse engineering},\ \href {\doibase
  10.1103/PhysRevA.105.042437} {\bibfield  {journal} {\bibinfo  {journal}
  {Phys. Rev. A}\ }\textbf {\bibinfo {volume} {105}},\ \bibinfo {pages}
  {042437} (\bibinfo {year} {2022})}\BibitemShut {NoStop}%
\bibitem [{\citenamefont {Mahesh}\ \emph {et~al.}(2023)\citenamefont {Mahesh},
  \citenamefont {Batra},\ and\ \citenamefont {Ram}}]{Mahesh2022}%
  \BibitemOpen
  \bibfield  {author} {\bibinfo {author} {\bibfnamefont {T.~S.}\ \bibnamefont
  {Mahesh}}, \bibinfo {author} {\bibfnamefont {P.}~\bibnamefont {Batra}}, \
  and\ \bibinfo {author} {\bibfnamefont {M.~H.}\ \bibnamefont {Ram}},\
  }\bibinfo {title} {Quantum Optimal Control: Practical Aspects and Diverse
  Methods},\ \href {\doibase 10.1007/s41745-022-00311-2} {\bibfield  {journal}
  {\bibinfo  {journal} {J. Indian Inst. Sci.}\ }\textbf {\bibinfo {volume}
  {103}},\ \bibinfo {pages} {591} (\bibinfo {year} {2023})}\BibitemShut
  {NoStop}%
\bibitem [{\citenamefont {Bishop}(2006)}]{Bishop2006}%
  \BibitemOpen
  \bibfield  {author} {\bibinfo {author} {\bibfnamefont {C.~M.}\ \bibnamefont
  {Bishop}},\ }\href@noop {} {\emph {\bibinfo {title} {Pattern {R}ecognition
  and {M}achine {L}earning}}}\ (\bibinfo  {publisher} {Springer, Singapore},\
  \bibinfo {year} {2006})\BibitemShut {NoStop}%
\bibitem [{\citenamefont {Freeman}(1997)}]{Freeman97}%
  \BibitemOpen
  \bibfield  {author} {\bibinfo {author} {\bibfnamefont {R.}~\bibnamefont
  {Freeman}},\ }\href@noop {} {\emph {\bibinfo {title} {Spin {C}horeography}}}\
  (\bibinfo  {publisher} {Spektrum, Oxford},\ \bibinfo {year}
  {1997})\BibitemShut {NoStop}%
\bibitem [{\citenamefont {Green}\ \emph {et~al.}(2013)\citenamefont {Green},
  \citenamefont {Sastrawan}, \citenamefont {Uys},\ and\ \citenamefont
  {Biercuk}}]{Green2013}%
  \BibitemOpen
  \bibfield  {author} {\bibinfo {author} {\bibfnamefont {T.~J.}\ \bibnamefont
  {Green}}, \bibinfo {author} {\bibfnamefont {J.}~\bibnamefont {Sastrawan}},
  \bibinfo {author} {\bibfnamefont {H.}~\bibnamefont {Uys}}, \ and\ \bibinfo
  {author} {\bibfnamefont {M.~J.}\ \bibnamefont {Biercuk}},\ }\bibinfo {title}
  {Arbitrary quantum control of qubits in the presence of universal noise},\
  \href {\doibase 10.1088/1367-2630/15/9/095004} {\bibfield  {journal}
  {\bibinfo  {journal} {New J. Phys.}\ }\textbf {\bibinfo {volume} {15}},\
  \bibinfo {pages} {095004} (\bibinfo {year} {2013})}\BibitemShut {NoStop}%
\bibitem [{\citenamefont {Kabytayev}\ \emph {et~al.}(2014)\citenamefont
  {Kabytayev}, \citenamefont {Green}, \citenamefont {Khodjasteh}, \citenamefont
  {Biercuk}, \citenamefont {Viola},\ and\ \citenamefont
  {Brown}}]{PhysRevA.90.012316}%
  \BibitemOpen
  \bibfield  {author} {\bibinfo {author} {\bibfnamefont {C.}~\bibnamefont
  {Kabytayev}}, \bibinfo {author} {\bibfnamefont {T.~J.}\ \bibnamefont
  {Green}}, \bibinfo {author} {\bibfnamefont {K.}~\bibnamefont {Khodjasteh}},
  \bibinfo {author} {\bibfnamefont {M.~J.}\ \bibnamefont {Biercuk}}, \bibinfo
  {author} {\bibfnamefont {L.}~\bibnamefont {Viola}}, \ and\ \bibinfo {author}
  {\bibfnamefont {K.~R.}\ \bibnamefont {Brown}},\ }\bibinfo {title} {Robustness
  of composite pulses to time-dependent control noise},\ \href {\doibase
  10.1103/PhysRevA.90.012316} {\bibfield  {journal} {\bibinfo  {journal} {Phys.
  Rev. A}\ }\textbf {\bibinfo {volume} {90}},\ \bibinfo {pages} {012316}
  (\bibinfo {year} {2014})}\BibitemShut {NoStop}%
\bibitem [{\citenamefont {Paz-Silva}\ and\ \citenamefont
  {Viola}(2014)}]{PhysRevLett.113.250501}%
  \BibitemOpen
  \bibfield  {author} {\bibinfo {author} {\bibfnamefont {G.~A.}\ \bibnamefont
  {Paz-Silva}}\ and\ \bibinfo {author} {\bibfnamefont {L.}~\bibnamefont
  {Viola}},\ }\bibinfo {title} {General Transfer-Function Approach to Noise
  Filtering in Open-Loop Quantum Control},\ \href {\doibase
  10.1103/PhysRevLett.113.250501} {\bibfield  {journal} {\bibinfo  {journal}
  {Phys. Rev. Lett.}\ }\textbf {\bibinfo {volume} {113}},\ \bibinfo {pages}
  {250501} (\bibinfo {year} {2014})}\BibitemShut {NoStop}%
\bibitem [{\citenamefont {Zhen}\ \emph {et~al.}(2016)\citenamefont {Zhen},
  \citenamefont {Xin}, \citenamefont {Zhang},\ and\ \citenamefont
  {Long}}]{Zhen2016}%
  \BibitemOpen
  \bibfield  {author} {\bibinfo {author} {\bibfnamefont {X.-L.}\ \bibnamefont
  {Zhen}}, \bibinfo {author} {\bibfnamefont {T.}~\bibnamefont {Xin}}, \bibinfo
  {author} {\bibfnamefont {F.-H.}\ \bibnamefont {Zhang}}, \ and\ \bibinfo
  {author} {\bibfnamefont {G.-L.}\ \bibnamefont {Long}},\ }\bibinfo {title}
  {Experimental demonstration of concatenated composite pulses robustness to
  non-static errors},\ \href {\doibase 10.1007/s11433-016-0208-7} {\bibfield
  {journal} {\bibinfo  {journal} {Sci. China Phys. Mech. Astron.}\ }\textbf
  {\bibinfo {volume} {59}},\ \bibinfo {pages} {103011} (\bibinfo {year}
  {2016})}\BibitemShut {NoStop}%
\bibitem [{\citenamefont {Haas}\ \emph {et~al.}(2019)\citenamefont {Haas},
  \citenamefont {Puzzuoli}, \citenamefont {Zhang},\ and\ \citenamefont
  {Cory}}]{Haas2019}%
  \BibitemOpen
  \bibfield  {author} {\bibinfo {author} {\bibfnamefont {H.}~\bibnamefont
  {Haas}}, \bibinfo {author} {\bibfnamefont {D.}~\bibnamefont {Puzzuoli}},
  \bibinfo {author} {\bibfnamefont {F.}~\bibnamefont {Zhang}}, \ and\ \bibinfo
  {author} {\bibfnamefont {D.~G.}\ \bibnamefont {Cory}},\ }\bibinfo {title}
  {Engineering effective Hamiltonians},\ \href {\doibase
  10.1088/1367-2630/ab4525} {\bibfield  {journal} {\bibinfo  {journal} {New J.
  Phys.}\ }\textbf {\bibinfo {volume} {21}},\ \bibinfo {pages} {103011}
  (\bibinfo {year} {2019})}\BibitemShut {NoStop}%
\bibitem [{\citenamefont {Li}\ \emph {et~al.}(2021)\citenamefont {Li},
  \citenamefont {Calderon-Vargas}, \citenamefont {Zeng},\ and\ \citenamefont
  {Barnes}}]{Li2021}%
  \BibitemOpen
  \bibfield  {author} {\bibinfo {author} {\bibfnamefont {B.}~\bibnamefont
  {Li}}, \bibinfo {author} {\bibfnamefont {F.~A.}\ \bibnamefont
  {Calderon-Vargas}}, \bibinfo {author} {\bibfnamefont {J.}~\bibnamefont
  {Zeng}}, \ and\ \bibinfo {author} {\bibfnamefont {E.}~\bibnamefont
  {Barnes}},\ }\bibinfo {title} {Designing arbitrary single-axis rotations
  robust against perpendicular time-dependent noise},\ \href {\doibase
  10.1088/1367-2630/ac22ea} {\bibfield  {journal} {\bibinfo  {journal} {New J.
  Phys.}\ }\textbf {\bibinfo {volume} {23}},\ \bibinfo {pages} {093032}
  (\bibinfo {year} {2021})}\BibitemShut {NoStop}%
\bibitem [{\citenamefont {Su}\ \emph {et~al.}(2021)\citenamefont {Su},
  \citenamefont {Bruinsma},\ and\ \citenamefont
  {Campbell}}]{PhysRevA.104.052625}%
  \BibitemOpen
  \bibfield  {author} {\bibinfo {author} {\bibfnamefont {Q.~D.}\ \bibnamefont
  {Su}}, \bibinfo {author} {\bibfnamefont {R.}~\bibnamefont {Bruinsma}}, \ and\
  \bibinfo {author} {\bibfnamefont {W.~C.}\ \bibnamefont {Campbell}},\
  }\bibinfo {title} {Quantum gates robust to secular amplitude drifts},\ \href
  {\doibase 10.1103/PhysRevA.104.052625} {\bibfield  {journal} {\bibinfo
  {journal} {Phys. Rev. A}\ }\textbf {\bibinfo {volume} {104}},\ \bibinfo
  {pages} {052625} (\bibinfo {year} {2021})}\BibitemShut {NoStop}%
\bibitem [{\citenamefont {Gefen}\ \emph {et~al.}(2017)\citenamefont {Gefen},
  \citenamefont {Cohen}, \citenamefont {Cohen},\ and\ \citenamefont
  {Retzker}}]{PhysRevA.95.032314}%
  \BibitemOpen
  \bibfield  {author} {\bibinfo {author} {\bibfnamefont {T.}~\bibnamefont
  {Gefen}}, \bibinfo {author} {\bibfnamefont {D.}~\bibnamefont {Cohen}},
  \bibinfo {author} {\bibfnamefont {I.}~\bibnamefont {Cohen}}, \ and\ \bibinfo
  {author} {\bibfnamefont {A.}~\bibnamefont {Retzker}},\ }\bibinfo {title}
  {Enhancing the fidelity of two-qubit gates by measurements},\ \href {\doibase
  10.1103/PhysRevA.95.032314} {\bibfield  {journal} {\bibinfo  {journal} {Phys.
  Rev. A}\ }\textbf {\bibinfo {volume} {95}},\ \bibinfo {pages} {032314}
  (\bibinfo {year} {2017})}\BibitemShut {NoStop}%
\end{thebibliography}%

\end{document}